\documentclass[11pt,a4paper]{article}

\usepackage{jheppub}
\usepackage{lipsum}
\usepackage{braket}
\usepackage{tensor}
\usepackage{dsfont}
\usepackage{tcolorbox}
\usepackage{caption}
\usepackage{subcaption}
\usepackage{tikz}
\usetikzlibrary{decorations.pathreplacing,calligraphy}
\usetikzlibrary{decorations.pathreplacing}
\usetikzlibrary{decorations.pathmorphing, patterns,shapes}
\usetikzlibrary{positioning,arrows,calc}
\usetikzlibrary{arrows.meta}
\usetikzlibrary{decorations.markings}
\usetikzlibrary{shapes.geometric}
\usetikzlibrary{patterns}
\tikzset{
  snake it/.style={
    decorate, 
    decoration=snake,
    segment length=3
  }
}
\usepackage{bm}
\usepackage{caption}
\usepackage{multirow}
\usepackage{subcaption}
\usepackage{ragged2e}
\usepackage{mathtools}
\usepackage{cancel}
\usepackage{comment}
\DeclareCaptionJustification{justified}{\justifying}
\usepackage{hyperref}
\usepackage{amssymb}
\usepackage{tikz}
\usetikzlibrary{arrows,decorations.markings,shapes.geometric,decorations.pathmorphing}
\tikzset{snake it/.style={decorate, decoration=snake}}
\usetikzlibrary{decorations.pathreplacing,calligraphy}
\usetikzlibrary{decorations.pathreplacing}
\usetikzlibrary{decorations.pathmorphing, patterns,shapes}
\usepackage{bm}
\definecolor{DarkBlueGrey}{RGB}{76,94,107}
\definecolor{MediumBlueGrey}{RGB}{110,135,153}
\definecolor{LightBlueGrey}{RGB}{134,163,184}
\definecolor{WCOrange}{RGB}{242,146,29}
\definecolor{SCRed}{RGB}{179,48,48}
\definecolor{VertexColor}{RGB}{242,146,29}
\definecolor{GluonColor}{RGB}{255,172,172}
\definecolor{SEColor}{RGB}{134,163,184}

\definecolor{BGBox}{RGB}{255,254,230}
\definecolor{PlaneColor}{RGB}{230,230,230}
\definecolor{BlobColor}{RGB}{190,180,230}

\DeclareMathOperator{\Disc}{Disc}

\newcommand{\Li}{{\normalfont\text{Li}}}

\newcommand{\vev}[1]{\langle\, #1 \, \rangle}

\def\veps{\varepsilon}

\newcommand{\hb}{\bar{h}}

\newcommand{\op}{\boldsymbol}

\def\Nm{{\mathcal{N}}}
\def\Om{{\mathcal{O}}}

\usepackage{bbm}

\def\eps{\epsilon}
\def\veps{\varepsilon}

\newcommand\zb{{\bar{z}}}

\newif\ifstartcompletesineup
\newif\ifendcompletesineup
\pgfkeys{
    /pgf/decoration/.cd,
    start up/.is if=startcompletesineup,
    start up=true,
    start up/.default=true,
    start down/.style={/pgf/decoration/start up=false},
    end up/.is if=endcompletesineup,
    end up=true,
    end up/.default=true,
    end down/.style={/pgf/decoration/end up=false}
}
\pgfdeclaredecoration{complete sines}{initial}
{
    \state{initial}[
        width=+0pt,
        next state=upsine,
        persistent precomputation={
            \ifstartcompletesineup
                \pgfkeys{/pgf/decoration automaton/next state=upsine}
                \ifendcompletesineup
                    \pgfmathsetmacro\matchinglength{
                        0.5*\pgfdecoratedinputsegmentlength / (ceil(0.5* \pgfdecoratedinputsegmentlength / \pgfdecorationsegmentlength) )
                    }
                \else
                    \pgfmathsetmacro\matchinglength{
                        0.5 * \pgfdecoratedinputsegmentlength / (ceil(0.5 * \pgfdecoratedinputsegmentlength / \pgfdecorationsegmentlength ) - 0.499)
                    }
                \fi
            \else
                \pgfkeys{/pgf/decoration automaton/next state=downsine}
                \ifendcompletesineup
                    \pgfmathsetmacro\matchinglength{
                        0.5* \pgfdecoratedinputsegmentlength / (ceil(0.5 * \pgfdecoratedinputsegmentlength / \pgfdecorationsegmentlength ) - 0.4999)
                    }
                \else
                    \pgfmathsetmacro\matchinglength{
                        0.5 * \pgfdecoratedinputsegmentlength / (ceil(0.5 * \pgfdecoratedinputsegmentlength / \pgfdecorationsegmentlength ) )
                    }
                \fi
            \fi
            \setlength{\pgfdecorationsegmentlength}{\matchinglength pt}
        }] {}
    \state{downsine}[width=\pgfdecorationsegmentlength,next state=upsine]{
        \pgfpathsine{\pgfpoint{0.5\pgfdecorationsegmentlength}{0.5\pgfdecorationsegmentamplitude}}
        \pgfpathcosine{\pgfpoint{0.5\pgfdecorationsegmentlength}{-0.5\pgfdecorationsegmentamplitude}}
    }
    \state{upsine}[width=\pgfdecorationsegmentlength,next state=downsine]{
        \pgfpathsine{\pgfpoint{0.5\pgfdecorationsegmentlength}{-0.5\pgfdecorationsegmentamplitude}}
        \pgfpathcosine{\pgfpoint{0.5\pgfdecorationsegmentlength}{0.5\pgfdecorationsegmentamplitude}}
}
    \state{final}{}
}

\tikzset{
corner/.style={line width=1pt,dashed,draw=black,dash pattern=on 6pt off 4pt},
scalar/.style={line width=1pt,draw=black},
gluon/.style={line width=1pt,decorate, draw=GluonColor,
    decoration={complete sines,aspect=0,amplitude=1.25mm,segment length=1.5mm,start up,end up}},
gluontwo/.style={line width=1pt,decorate, draw=GluonColor,
    decoration={complete sines,aspect=0,amplitude=.7mm,segment length=1mm,start up,end up}},
ghost/.style={line width=1pt,loosely dotted,draw=black},
wilson/.style={line width=2pt,draw=black},
 }
\NewDocumentCommand\semiloop{O{black}mmmO{}O{above}}
{%
\draw[#1] let \p1 = ($(#3)-(#2)$) in (#3) arc (#4:({#4+180}):({0.5*veclen(\x1,\y1)})node[midway, #6] {#5};)
}

\newcommand\x{.75}

\newcommand\vertexsize{.07}

\newcommand{\DiagramB}{\begin{tikzpicture}[baseline={([yshift=-.5ex]current bounding box.center)}]
\draw[scalar] (0,0) -- (1.5,0);
\draw[scalar] (.75,.25) circle (.25);
\draw[fill=white] (0,0) circle (.075);
\draw[fill=white] (1.5,0) circle (.075);
\draw[fill=VertexColor,VertexColor] (.75,0) circle (\vertexsize);
\end{tikzpicture}}

\newcommand{\GMIFigOne}{\begin{tikzpicture}[baseline={([yshift=-.5ex]current bounding box.center)}]
\node[] at (3.25,1.5) {$\text{Im}\ x$};
\node[] at (6.3,-1) {$\text{Re}\ x$};
\node[] at (6,-4.75) {$\tau$};
\draw[scalar,-Latex,opacity=.7] (2-1.25,-1+1.25) -- (4.5+1.25,-3.5-1.25);
\draw[scalar,-Latex,opacity=.7] (2-1.25,-.25-4+.75) -- (4.5+1.25,2.25-4+.75);
\draw[scalar,-Latex,opacity=.7] (3.25,-5) -- (3.25,1);
\fill[LightBlueGrey,opacity=.25] (0.5,-.25) -- (3.5,1.25) -- (3.5,-1.75) -- (0.5,-3.25);
\fill[LightBlueGrey,opacity=.25] (0.5+1.25,-.25-1.25) -- (3.5+1.25,1.25-1.25) -- (3.5+1.25,-1.75-1.25) -- (0.5+1.25,-3.25-1.25);
\fill[LightBlueGrey,opacity=.25] (0.5+2.5,-.25-2.5) -- (3.5+2.5,1.25-2.5) -- (3.5+2.5,-1.75-2.5) -- (0.5+2.5,-3.25-2.5);
\draw[very thick,SEColor,opacity=1] (1.5,.25) -- (5,-4.75);
\draw[very thick,SEColor,opacity=1] (1,-.9) -- (5.5,-3.6);
\draw[-, decorate, decoration={zigzag, segment length=6, amplitude=3},line width=1,SCRed] (3.25,-3.75) -- (3.25,-.75);
\draw[-, decorate, decoration={zigzag, segment length=6, amplitude=3},line width=1,SCRed] (3.25+1.25,-3.75-1.25+1.5+.5) -- (3.25+1.25,-.75-1.25);
\draw[-, decorate, decoration={zigzag, segment length=6, amplitude=3},line width=1,SCRed] (3.25+1.25,-3.75-1.25) -- (3.25+1.25,-.75-1.25-1.5-.5);
\draw[-, decorate, decoration={zigzag, segment length=6, amplitude=3},line width=1,SCRed] (3.25-1.25,-3.75+1.25+1.5+.5) -- (3.25-1.25,-.75+1.25);
\draw[-, decorate, decoration={zigzag, segment length=6, amplitude=3},line width=1,SCRed] (3.25-1.25,-3.75+1.25) -- (3.25-1.25,-.75-1.25+1.5-.5-.5);
\draw[fill=DarkBlueGrey] (3.25,-2.25) circle (.08);
\draw[fill=DarkBlueGrey] (3.25+1.25,-2.25-.75) circle (.08);
\draw[fill=DarkBlueGrey] (3.25+1.25,-2.25-1.75) circle (.08);
\draw[fill=DarkBlueGrey] (3.25-1.25,-2.25+.75) circle (.08);
\draw[fill=DarkBlueGrey] (3.25-1.25,-2.25+1.75) circle (.08);
\end{tikzpicture}}

\newcommand{\GMIFigTwo}{\begin{tikzpicture}[scale = 1, baseline={([yshift=-.5ex]current bounding box.center)}]
\draw[scalar] (5.4+.7,1.9+1) -- (5.4+.7,1.45+1);
\draw[scalar] (5.4+.7,1.45+1) -- (6+.7,1.45+1);
\node[] at (5.7+.7,1.7+1) {$w$};
\draw[scalar,-Latex] (-.5,0) -- (6.7,0);
\draw[scalar,-Latex] (3,-3) -- (3,3);
\fill [WCOrange,even odd rule,opacity=.25] (3,0) circle[radius=2.5cm] circle[radius=1.75cm];
\fill [LightBlueGrey,even odd rule,opacity=.25] (3,0) circle[radius=2cm] circle[radius=1cm];
\draw[wilson,LightBlueGrey,opacity=.5] (3,0) circle (2);
\draw[wilson,LightBlueGrey,opacity=.5] (3,0) circle (1);
\draw[wilson,dashed,WCOrange,opacity=1] (3,0) circle (2.5);
\draw[wilson,dashed,WCOrange,opacity=1] (3,0) circle (1.75);
\draw[-, decorate, decoration={zigzag, segment length=6, amplitude=3},line width=1,SCRed] (-.5,0) -- (1,0);
\draw[fill=SCRed,SCRed] (1,0) circle (.075);
\draw[fill=SCRed,SCRed] (2,0) circle (.075);
\draw[-, decorate, decoration={zigzag, segment length=6, amplitude=3},line width=1,SCRed] (2,0) -- (4,0);
\draw[fill=SCRed,SCRed] (4,0) circle (.075);
\draw[-, decorate, decoration={zigzag, segment length=6, amplitude=3},line width=1,SCRed] (5,0) -- (6.5,0);
\draw[fill=SCRed,SCRed] (5,0) circle (.075);
\end{tikzpicture}}

\newcommand{\AnalyticStructureThermal}{\begin{tikzpicture}[scale = 1, baseline={([yshift=-.5ex]current bounding box.center)}]
\draw[scalar,-Latex] (0,0) -- (6.2,0);
\draw[scalar,-Latex] (3,-2) -- (3,2);
\node[] at (5.7,1.7) {$w'$};
\draw[scalar] (5.4,1.9) -- (5.4,1.45);
\draw[scalar] (5.4,1.45) -- (6,1.45);
\draw[fill=MediumBlueGrey,MediumBlueGrey] (4,1.25) circle (.075);
\node[MediumBlueGrey] at (4.7,1.25) {$w$};
\draw[scalar] (4,1.25) circle (.4);
\draw[scalar,-Latex] (3.62,1.09) -- (3.62,1.08);
\draw[-, decorate, decoration={zigzag, segment length=6, amplitude=3},line width=1,SCRed] (0,0) -- (1,0);
\draw[fill=SCRed,SCRed] (1,0) circle (.075);
\node[SCRed] at (1,-.5) {$-1/r$};
\draw[fill=SCRed,SCRed] (2,0) circle (.075);
\node[SCRed] at (2,-.5) {$-r$};
\draw[-, decorate, decoration={zigzag, segment length=6, amplitude=3},line width=1,SCRed] (2,0) -- (4,0);
\draw[fill=SCRed,SCRed] (4,0) circle (.075);
\node[SCRed] at (4,-.5) {$r$};
\draw[-, decorate, decoration={zigzag, segment length=6, amplitude=3},line width=1,SCRed] (5,0) -- (6,0);
\draw[fill=SCRed,SCRed] (5,0) circle (.075);
\node[SCRed] at (5,-.5) {$1/r$};
\end{tikzpicture}}

\newcommand{\ContourThermal}{\begin{tikzpicture}[scale=1, baseline={([yshift=-.5ex]current bounding box.center)}]
\draw[scalar,-Latex] (0,0) -- (6.2,0);
\draw[scalar,-Latex] (3,-2) -- (3,2);
\node[] at (5.75,1.75) {$w'$};
\draw[scalar] (5.4,1.9) -- (5.4,1.45);
\draw[scalar] (5.4,1.45) -- (6,1.45);
\draw[-, decorate, decoration={zigzag, segment length=6, amplitude=3},line width=1,SCRed] (0,0) -- (1,0);
\draw[fill=SCRed,SCRed] (1,0) circle (.075);
\draw[fill=SCRed,SCRed] (2,0) circle (.075);
\draw[-, decorate, decoration={zigzag, segment length=6, amplitude=3},line width=1,SCRed] (2,0) -- (4,0);
\draw[fill=SCRed,SCRed] (4,0) circle (.075);
\draw[-, decorate, decoration={zigzag, segment length=6, amplitude=3},line width=1,SCRed] (5,0) -- (6,0);
\draw[fill=SCRed,SCRed] (5,0) circle (.075);
\draw[scalar,-Latex] (.6,.3) -- (.7,.3);
\draw[scalar,-Latex] (3.5,.3) -- (3.6,.3);
\draw[scalar,-Latex] (5.6,.3) -- (5.7,.3);
\draw[scalar] (0,.3) -- (1,.3);
\draw[scalar] (0,-.3) -- (1,-.3);
\draw[scalar] (1,-.3) arc (270:450:.3);
\draw[scalar] (2,.3) -- (4,.3);
\draw[scalar] (2,-.3) -- (4,-.3);
\draw[scalar] (2,.3) arc (90:270:.3);
\draw[scalar] (4,-.3) arc (270:450:.3);
\draw[scalar] (5,.3) -- (6,.3);
\draw[scalar] (5,-.3) -- (6,-.3);
\draw[scalar] (5,.3) arc (90:270:.3);
\end{tikzpicture}}

\newcommand{\ThermaltauAnal}{\begin{tikzpicture}[scale = 1, baseline={([yshift=-.4ex]current bounding box.center)}]
\draw[scalar,-Latex] (0,0) -- (6.2,0);
\draw[scalar,-Latex] (3,-2) -- (3,2);
\node[] at (5.75,1.8) {$\tau'$};
\draw[scalar] (5.4,1.9) -- (5.4,1.45);
\draw[scalar] (5.4,1.45) -- (6,1.45);
\draw[fill=MediumBlueGrey,MediumBlueGrey] (4,1.25) circle (.075);
\node[MediumBlueGrey] at (4.7,1.25) {$\tau$};
\draw[scalar] (4,1.25) circle (.4);
\draw[scalar,-Latex] (3.62,1.09) -- (3.62,1.08);
\draw[-, decorate, decoration={zigzag, segment length=6, amplitude=3},line width=1,SCRed] (1,-2) -- (1,0);
\draw[fill=SCRed,SCRed] (1,0) circle (.075);
\draw[-, decorate, decoration={zigzag, segment length=6, amplitude=3},line width=1,SCRed] (3,-2) -- (3,0);
\draw[fill=SCRed,SCRed] (3,0) circle (.075);
\draw[-, decorate, decoration={zigzag, segment length=6, amplitude=3},line width=1,SCRed] (2,-2) -- (2,0);
\draw[fill=SCRed,SCRed] (2,0) circle (.075);
\draw[-, decorate, decoration={zigzag, segment length=6, amplitude=3},line width=1,SCRed] (4,-2) -- (4,0);
\draw[fill=SCRed,SCRed] (4,0) circle (.075);
\draw[-, decorate, decoration={zigzag, segment length=6, amplitude=3},line width=1,SCRed] (5,-2) -- (5,0);
\draw[fill=SCRed,SCRed] (5,0) circle (.075);
\end{tikzpicture}}

\newcommand{\ContourThermaltau}{\begin{tikzpicture}[scale = 1, baseline={([yshift=-.4ex]current bounding box.center)}]
\draw[scalar,-Latex] (0,0) -- (6.2,0);
\draw[scalar,-Latex] (3,-2) -- (3,2);
\node[] at (5.75,1.8) {$\tau'$};
\draw[scalar] (5.4,1.9) -- (5.4,1.45);
\draw[scalar] (5.4,1.45) -- (6,1.45);
\draw[fill=MediumBlueGrey,MediumBlueGrey] (4,1.25) circle (.075);
\node[MediumBlueGrey] at (4.35,1.25) {$\tau$};
\draw[scalar,-Latex] (1.1,.395) -- (1.13,.395);
\draw[scalar] (0.7,0.1) -- (0.7,-2);
\draw[scalar] (1.3,0.1) -- (1.3,-2);
\draw[scalar] (0.7,.1) arc (180:0:.3);
\draw[scalar,-Latex] (2.1,.395) -- (2.13,.395);
\draw[scalar] (1.7,0.1) -- (1.7,-2);
\draw[scalar] (2.3,0.1) -- (2.3,-2);
\draw[scalar] (1.7,.1) arc (180:0:.3);
\draw[scalar,-Latex] (3.1,.395) -- (3.13,.395);
\draw[scalar] (2.7,0.1) -- (2.7,-2);
\draw[scalar] (3.3,0.1) -- (3.3,-2);
\draw[scalar] (2.7,.1) arc (180:0:.3);
\draw[scalar,-Latex] (4.1,.395) -- (4.13,.395);
\draw[scalar] (3.7,0.1) -- (3.7,-2);
\draw[scalar] (4.3,0.1) -- (4.3,-2);
\draw[scalar] (3.7,.1) arc (180:0:.3);
\draw[scalar,-Latex] (5.1,.395) -- (5.13,.395);
\draw[scalar] (4.7,0.1) -- (4.7,-2);
\draw[scalar] (5.3,0.1) -- (5.3,-2);
\draw[scalar] (4.7,.1) arc (180:0:.3);
\draw[-, decorate, decoration={zigzag, segment length=6, amplitude=3},line width=1,SCRed] (1,-2) -- (1,0);
\draw[fill=SCRed,SCRed] (1,0) circle (.075);
\draw[-, decorate, decoration={zigzag, segment length=6, amplitude=3},line width=1,SCRed] (3,-2) -- (3,0);
\draw[fill=SCRed,SCRed] (3,0) circle (.075);
\draw[-, decorate, decoration={zigzag, segment length=6, amplitude=3},line width=1,SCRed] (2,-2) -- (2,0);
\draw[fill=SCRed,SCRed] (2,0) circle (.075);
\draw[-, decorate, decoration={zigzag, segment length=6, amplitude=3},line width=1,SCRed] (4,-2) -- (4,0);
\draw[fill=SCRed,SCRed] (4,0) circle (.075);
\draw[-, decorate, decoration={zigzag, segment length=6, amplitude=3},line width=1,SCRed] (5,-2) -- (5,0);
\draw[fill=SCRed,SCRed] (5,0) circle (.075);
\end{tikzpicture}}

\newcommand{\ScalarPropagator}{\begin{tikzpicture}[baseline={([yshift=1.5ex]current bounding box.center)}]
\draw[scalar] (0,.5) -- (1,.5);
\draw[fill=white] (0,.5) circle (.075);
\draw[fill=white] (1,.5) circle (.075);
\node[] at (0,0) {\footnotesize \makebox[\widthof{$\smash{i}$}][c]{$\smash{i}$}};
\node[] at (1,0) {\footnotesize \makebox[\widthof{$\smash{j}$}][c]{$\smash{j}$}};
\end{tikzpicture}}
\usepackage{dsfont}
\usepackage{float}
\usepackage{pgfplots}
\pgfplotsset{compat=1.14}

\let\oldbfseries=\bfseries
\let\oldmdseries=\mdseries
\let\oldnormalfont=\normalfont
\renewcommand{\bfseries}{\oldbfseries\boldmath}
\renewcommand{\mdseries}{\oldmdseries\unboldmath}
\renewcommand{\normalfont}{\oldnormalfont\unboldmath}

\makeatletter
\newlength{\apb@width}
\newcommand{\autoparbox}[2][c]{\settowidth{\apb@width}{#2}\parbox[#1]{\apb@width}{#2}}

\makeatother

\def\Nm{{\mathcal{N}}}
\def\Om{{\mathcal{O}}}

\def\zb{{\bar{z}}}

\def\eps{\epsilon}
\def\veps{\varepsilon}

\newcommand{\beq}{\begin{equation}}
\newcommand{\eeq}{\end{equation}}

\definecolor{nicegreen}{rgb}{0.1,0.6,0.1}

\newmuskip\pFqskip
\mathchardef\pFcomma=\mathcode`,

\makeatletter
\renewcommand*\env@matrix[1][\arraystretch]{%
  \edef\arraystretch{#1}%
  \hskip -\arraycolsep
  \let\@ifnextchar\new@ifnextchar
  \array{*\c@MaxMatrixCols c}}
\makeatother

\graphicspath{{./figures/}}

\begin{document}

\vspace*{-.6in} \thispagestyle{empty}
\begin{flushright}
DESY-25-078
\end{flushright}
\vspace{1cm} {\large
\begin{center}
{\Large \bf The analytic bootstrap at finite temperature
}\\
\end{center}}
\vspace{0.5cm}
\begin{center}
{Julien Barrat,$^{a,}\footnote{julien.barrat@desy.de}$ Deniz N. Bozkurt,$^{b,}\footnote{deniz.bozkurt@desy.de}$ Enrico Marchetto,$^{a,}\footnote{enrico.marchetto@desy.de}$ \\ Alessio Miscioscia,$^{a,}\footnote{alessio.miscioscia@desy.de}$ and Elli Pomoni$^{a,}\footnote{elli.pomoni@desy.de}$}\\[0.5cm] 
{ \small
 $^{a}$Deutsches Elektronen-Synchrotron DESY, Notkestr. 85, 22607 Hamburg, Germany\\
\small $^{b}$ Institut f\"ur Theoretische Physik, Universit\"at Hamburg, Luruper Chaussee 149,
22607 Hamburg, Germany
}
\vspace{1cm} 

   \bf Abstract
\end{center}
\begin{abstract}
\noindent
We propose  new universal formulae for thermal two-point functions of scalar operators based on their analytic structure, constructed to manifestly satisfy all the bootstrap conditions.
We derive a dispersion relation in the complexified time plane, which fixes the correlator up to an additive constant and theory-dependent dynamical information.
At non-zero spatial separation we introduce a formula for the thermal two-point function obtained by summing over images of the dispersion relation result obtained in the OPE regime.
This construction satisfies all thermal bootstrap conditions, with the exception of clustering at infinite distance, which must be verified on a case-by-case basis.
We test our results both in weakly and strongly-coupled theories.
In particular, we show that the asymptotic behavior for the heavy sector proposed in~\cite{Marchetto:2023xap} and its correction can be explicitly derived from the dispersion relation.
We combine analytical and numerical results to compute the thermal two-point function of the energy operator in the $3d$ Ising model and find agreement with Monte Carlo simulations.
\end{abstract}

\newpage

\setlength{\parindent}{0pt}

{
\setcounter{tocdepth}{3}
\tableofcontents
}

\setlength{\parskip}{0.1in}

\section{Introduction}
\label{sec:Introduction}

Conformal field theories (CFTs) play a distinguished role in the space of quantum field theories (QFTs).
These scale-invariant models correspond to fixed points in the renormalization group (RG) flow and serve as landmarks in the classification of QFTs.
In statistical physics, they describe a wide range of critical phenomena, from classical phase transitions such as the liquid-gas critical point to quantum critical points in strongly-correlated systems \cite{cardy1996scaling,Sachdev:2011fcc,Mussardo:2020rxh,pelissetto2002critical,Rong:2023fdy}.
CFTs also play a central role in the understanding of quantum gravity, particularly through the holographic principle: gravitational theories in Anti-de Sitter (AdS) space are conjectured to be dual to CFTs living in one fewer spacetime dimension \cite{Maldacena:1997re}.

The study of conformal field theories (CFTs) can be extended beyond the vacuum by placing the system in a non-trivial state.
Among the possible choices, the thermal state plays a particularly central role.
In real-world experiments, quantum critical behavior always occurs at non-zero temperature, making it crucial to understand thermal effects in CFTs to enable meaningful comparisons with experimental data.
In the context of the AdS/CFT correspondence, black holes and black branes in Anti-de Sitter space are holographically dual to CFTs at finite temperature, in finite and infinite spatial volume, respectively \cite{Witten:1998zw}.
Thermal effects in quantum field theory—and thus in CFTs—can be incorporated via the Matsubara formalism \cite{Matsubara:1955ws}, in which time is compactified on a circle of circumference $\beta = 1/T$, corresponding to the Euclidean manifold $S^1_\beta \times \mathbb{R}^{d-1}$.
This setting provides the simplest non-flat manifold in which conformality is broken, making it a natural and tractable arena to study thermal CFTs.
Despite the loss of full conformal symmetry, the theory remains under strong analytic control, and powerful constraints continue to govern its observables \cite{Cardy:1986ie,El-Showk:2011yvt,Benjamin:2023qsc,Benjamin:2024kdg,Diatlyk:2024qpr}.

In recent years, the study of thermal effects in CFTs has gained significant attention, driven by the realization that key tools — such as the operator product expansion (OPE) — remain applicable even at finite temperature~\cite{Katz:2014rla}, despite the explicit breaking of conformal symmetry~\cite{Marchetto:2023fcw}.
One immediate consequence of this symmetry breaking is the emergence of non-vanishing one-point functions for local operators. Moreover, the success of the conformal bootstrap program in solving strongly-coupled theories at zero temperature — most notably the three-dimensional Ising model (see~\cite{Poland:2018epd,Rychkov:2023wsd} for recent reviews and references therein) — has inspired efforts to extend bootstrap techniques to thermal settings.
In ~\cite{El-Showk:2011yvt}, it was proposed that the Kubo–Martin–Schwinger (KMS) condition, which encodes the periodicity along the thermal circle $S_\beta^1$, could serve as a consistency constraint on finite-temperature dynamics.
The first concrete realization of this idea appeared in the seminal work~\cite{Iliesiu:2018fao}, where the bootstrap framework was adapted to reconcile the OPE of a thermal two-point function with the KMS condition.
This approach was subsequently applied to the three-dimensional Ising model, leading to the numerical determination of several thermal observables \cite{Iliesiu:2018zlz}.
The program has since been further developed, yielding a plethora of new results \cite{Petkou:2018ynm,Parisini:2022wkb,Marchetto:2023fcw,Marchetto:2023xap,David:2023uya,David:2024naf,Barrat:2024aoa,David:2024pir,Kumar:2025txh,Buric:2024kxo,Buric:2025anb}.
Notably, in~\cite{Barrat:2024fwq}, numerical values for thermal one-point functions and the free energy of the Ising, XY, and Heisenberg models were obtained within a bootstrap framework that relies on accurately approximating heavy operators. 

Another powerful approach to the bootstrap problem is to exploit its constraints analytically rather than numerically.
This strategy builds on the universal behavior of CFTs at large spin, which imposes systematic and model-independent constraints on the CFT data \cite{Fitzpatrick:2012yx,Komargodski:2012ek,Alday:2015eya}.
A major breakthrough in this direction was the development of the Lorentzian inversion formula \cite{Caron-Huot:2017vep} and dispersion relations \cite{Carmi:2019cub,Bissi:2019kkx,Carmi:2025mtx} — equivalent techniques that reconstruct CFT data from the discontinuities of four-point functions at zero temperature.
This method, known as the \emph{analytic bootstrap}, is particularly effective in perturbative settings, such as holographic theories, where the relevant discontinuities can be computed from the gravitational dual \cite{Aharony:2016dwx,Rastelli:2016nze,Alday:2017xua,Rastelli:2017udc,Jafferis:2017zna,Caron-Huot:2018kta,Mukhametzhanov:2018zja,Bissi:2022mrs,Hartman:2022zik}.
Analytic bootstrap techniques offer a robust framework for studying strongly-coupled systems, complementing numerical methods and extending naturally to situations where conformal symmetry is partially broken \cite{Lemos:2017vnx,Liendo:2019jpu,Barrat:2022psm,Bianchi:2022ppi,Bonomi:2024lky,Carmi:2024tmp}.
In particular, an inversion formula for thermal two-point functions was derived in \cite{Iliesiu:2018zlz}, and a thermal dispersion relation was proposed in \cite{Alday:2020eua}.

In this work, we revisit the dispersion relation\footnote{We emphasize that, throughout this work, the term “dispersion relation” is used in a broad sense.
Since the discontinuity of thermal correlators lacks definite positivity properties, it cannot be interpreted as the absorptive part of a physical process.} introduced in~\cite{Alday:2020eua} and present a new formulation applicable to thermal two-point functions evaluated at zero spatial separation.
We study the interplay between dispersion relations and the OPE, deriving explicit expressions for thermal correlators in terms of their discontinuities.
Our framework is tested in both exactly solvable models and perturbative regimes.
We also examine the connection between analytic and numerical approaches to finite-temperature CFTs.
In particular, we demonstrate that the asymptotic behavior of thermal OPE coefficients for heavy operators, derived in~\cite{Marchetto:2023xap}, can be reproduced from the dispersion relation, including all subleading corrections.
Finally, we apply numerically determined OPE data to our analytic framework to compute the thermal correlator $\langle \epsilon(0) \epsilon (\tau) \rangle_\beta$ in the three-dimensional Ising model, and find good agreement with Monte Carlo simulations.

\subsection{Summary of the results}
\label{subsec:SummaryOfTheResults}

\paragraph{Dispersion relation at zero spatial separation.}
In Section \ref{sec:AnalyticToolsForThermalCorrelators} we derive a novel dispersion relation for thermal two-point functions of identical scalar operators evaluated in the complex $\tau$-plane.
We argue that the two-point function $g(\tau)$ is fully determined by its discontinuity, up to an additive constant $\kappa$:
\begin{equation}
    g(\tau) = \sum_{m = -\infty}^{\infty} \frac{1}{2\pi i} \int_{-i\infty}^{0} \mathrm{d}\tau' \, \frac{\operatorname{Disc} g(\tau')}{\tau' + m \beta - \tau} + \frac{\kappa}{\beta^{2\Delta_\phi}} \ .
\end{equation}
By combining this formula with the OPE, we find that all thermal two-point functions admit an expansion in terms of generalized free field correlators:
\begin{equation} \label{eq: summ1}
    g(\tau) = \sum_{\Delta} \frac{a_{\Delta}}{\beta^{2\Delta_\phi}} \left[ \zeta_H\left(2\Delta_\phi - \Delta, 1 - \frac{\tau}{\beta}\right) + \zeta_H\left(2\Delta_\phi - \Delta, \frac{\tau}{\beta}\right) \right] + \frac{\kappa}{\beta^{2\Delta_\phi}}  \,,
\end{equation}
where the coefficients $a_\Delta$ are the thermal OPE coefficients contributing to the discontinuity.
This formulation has the advantage to encode the consistency conditions that the correlator should satisfy.
In particular, it is manifestly KMS-invariant and is consistent with the OPE by construction.

\paragraph{Non-zero spatial separation and generalized method of images.}
We revisit the dispersion relation derived in~\cite{Alday:2020eua} for thermal two-point functions at non-zero spatial separation, $g(\tau,x)$, and combine it with the OPE.
When the discontinuity acts on individual thermal blocks, the resulting expression captures only the large-spin approximation of the thermal correlator; subleading corrections of order $O(2^{-J})$ are not accounted for.
In addition, the dispersion relation does not capture the thermal OPE coefficients associated with operators of spin $J \leq J^\star$, where $J^\star$ characterizes the Regge behavior of the correlator.
These contributions are encoded in the integral over the arc at infinity in the complex plane.
As a result, the correlator obtained from the dispersion relation is not KMS invariant and must be appropriately reconstructed.

We address these issues by presenting two distinct solutions.
The first one involves introducing a correction term to OPE coefficients, which are then fixed by imposing the KMS condition.
This approach is tested both for the free scalar theory and the $\mathrm{O}(N)$ model at the Wilson-Fisher fixed point, as presented in Appendix~\ref{app:KMScompensator}.
The second method is motivated by the complex $\tau$-plane formula and consists of reinstating KMS invariance by summing over the thermal images of the dispersion relation:
\begin{equation} \label{eq: summ2}
    g(\tau, x) = \frac{1}{2} \sum_{m = -\infty}^\infty g_\text{dr}(\tau - m\beta, x) + g_\text{arcs}(\tau, x) \ .
\end{equation}
We call this formula the \textit{generalized method of images}.
Here $g_\text{arcs}(\tau, x)$ captures the contribution of the low-spin operators with $J \leq J^\star$ and should satisfy the KMS condition on its own.
We argue that this formula satisfies all the thermal bootstrap axioms, apart from clustering at infinite distance, which must be tested on a case-by-case basis.
If the sum over images already encodes the correct behavior at infinite distance, the arc contributions can be shown to vanish and the generalized method of images simplifies to only the first term in \eqref{eq: summ2}.

\paragraph{Connection to momentum space.}
We establish a connection between the generalized method of images and the Fourier transform of thermal two-point functions.
We conclude that the block-by-block contributions to the dispersion relation provide a non-perturbative extension of the Fourier thermal blocks derived in~\cite{Manenti:2019wxs}.
Concretely, momentum-space thermal blocks associated to an operator $\Om$, neglecting arcs, are given by 
\begin{equation}
    \widetilde{f}_{\Delta,J}(\omega_n, \vec{k}) = \int_0^1 \mathrm{d}\tau \, e^{i \omega_n \tau} \int_{\mathbb{R}^{d-1}} \hspace{-1em} \mathrm{d} \vec{x} \, e^{i \vec{k} \cdot \vec{x}} \left( \frac{1}{2} \sum_{m = -\infty}^\infty g_\text{dr}^{(\mathcal{O})}(\tau - m, x)\right) \,.
\end{equation}
In perturbative theories, this expression matches the momentum-space OPE obtained via standard diagrammatic calculations.

\paragraph{Applications.}
Section~\ref{sec:applications} is devoted to applications of the formulae~\eqref{eq: summ1} and~\eqref{eq: summ2} to both perturbative and non-perturbative settings.
We study generalized free fields, the $\mathrm{O}(N)$ model, at first order in $\veps$, and in the large $N$ limit, as well as two-point functions of Virasoro primaries in two dimensions.

\paragraph{Asymptotic corrections and  \texorpdfstring{$3d$}{\texttwoinferior} Ising.}
In Section~\ref{eq:hyb} we explore the synergistic use of the analytic methods developed in this work together with the numerical techniques introduced in previous studies, particularly in~\cite{Barrat:2024fwq}. 
We show that the asymptotic behavior of thermal OPE coefficients for heavy operators, originally derived in~\cite{Marchetto:2023xap}, can be recovered analytically from the complex $\tau$-plane framework.
The dispersion relation gives access to the full tower of polynomial corrections to the leading asymptotic behavior — corrections that were previously unknown.
All coefficients in this expansion are expressed in terms of the conformal dimensions $\lbrace 0, \Delta_1, \Delta_2,\dots\rbrace $ and of the OPE coefficients $\lbrace 1,a_{\Delta_1}, a_{\Delta_2},\dots\rbrace $  of light operators in the OPE:
\begin{equation}
         a_\Delta \overset{\Delta \gg 1}{\sim}  \frac{2\Delta^{2\Delta_\phi-1}}{\Gamma(2\Delta_\phi)}\left(1+\frac{\Delta_\phi (2\Delta_\phi-1)}{\Delta }+a_{\Delta_1}\frac{\Gamma(2\Delta_\phi)}{\Gamma(2\Delta_\phi-\Delta_1)}\frac{1}{\Delta^{\Delta_1}}+\ldots \right) \ .
\end{equation}

We then apply the dispersion relation in the complex $\tau$-plane to the $3d$ Ising model, using as input the numerical OPE coefficients bootstrapped in~\cite{Barrat:2024fwq}.
We produce for the first time an approximation of the thermal correlator $\langle\epsilon(0) \epsilon(\tau)\rangle_\beta$.
In both cases, our results show agreement with Monte Carlo simulations.

We conclude in Section~\ref{sec:conclusions}, where we discuss several outlooks, including potential applications to holographic theories, improvements to numerical methods, and further developments of bootstrap approaches to thermal effects in CFTs.

\section{Analytic tools for thermal correlators}
\label{sec:AnalyticToolsForThermalCorrelators}

We present in this section analytic tools to bootstrap two-point functions of identical scalar operators at finite temperature.

After a review on thermal correlators, we study the limit in which the operators have zero separation in the spatial directions.
In this setup, we derive a novel dispersion relation to reconstruct the two-point function from its discontinuity.
We show that the arc contributions are at most a constant and obtain a formula that encompasses all the consistency conditions in this limit.

We then address correlators with non-zero spatial separation, starting with a review of the derivation of a dispersion relation that was first obtained in \cite{Alday:2020eua}.
We discuss the arc contributions, as well as two methods to implement the consistency conditions of the bootstrap problem.
The first one is numerical and relies on the form of corrections at large spin.
The second one is analytical and consists of summing the result of the dispersion relation over images.
The result satisfies by construction most of the consistency conditions, in particular periodicity.

We conclude this section by connecting to momentum-space correlators. Notably, the operators contributing to the momentum-space OPE coincide with those relevant to the discontinuity of the position-space OPE. However, while the momentum-space OPE misses certain non-perturbative effects, the method of images applied to the dispersion relation is expected to successfully capture these corrections.

\subsection{Thermal two-point functions}
\label{subsec:ThermalTwoPointFunctions}

We start by reviewing the essential properties of correlation functions at finite temperature.

\subsubsection{Kinematics}
\label{subsubsec:Kinematics}

We denote the two-point function of identical scalars at finite temperature by
\begin{equation}
    g(\tau, x) = \vev{\phi(\tau,x) \phi(0,0)}_\beta \ ,
    \label{eq: gen 2pt}
\end{equation}
where we use the shorthand notation $x = | \vec x|$ for the spatial dependence.
An important part of this work focuses on the study of these correlators at zero spatial coordinates, in which case we write
\begin{equation}
    g(\tau) = g(\tau, 0) \ .
    \label{eq:ZeroSpatialCorrelator_ShorthandNotation}
\end{equation}
When the spatial dependence is preserved, it is convenient to use the variables
\begin{equation}
    z = \tau + i x\,, \qquad \zb = \tau - i x\,.
    \label{eq:zzbVariables}
\end{equation}
For compactness, we often write $g(z,\zb)=g(\tau, x)$ as a slight abuse of notation.
Another useful change of variables is given by
\begin{equation}
    r^2=z \zb \ , \qquad w^2=\frac{z}{\zb}  \ .
    \label{eq:rwVariables}
\end{equation}
Both sets of variables are frequent in the literature and reminiscent of four-point functions at zero temperature.

In CFTs at finite temperature, the correlators can be expanded into thermal conformal blocks in the regime of convergence of the OPE ($\sqrt{z \zb} = \sqrt{\tau^2+\vec{x}^2} < \beta$):
\begin{equation}
    g(z,\zb)
    =
    \sum_{\mathcal O} a_{\mathcal O}\, f_{\Delta, J} (z,\zb)\,,
    \label{eq:ThermalBlockExpansion}
\end{equation}
where the sum runs over the operators exchanged in the OPE $\phi \times \phi$ with quantum numbers $\Delta$ and $J$ corresponding, respectively, to their scaling dimensions and spins.
The blocks depend implicitly on $\Delta_\phi$, and are given by
\begin{equation}
    f_{\Delta, J} (z,\zb)
    =
    (z \zb)^{\Delta/2-\Delta_\phi} C_J^{(\nu)}\left(\frac{z+\zb}{2\sqrt{z \zb}}\right) \ ,
    \label{eq:ThermalBlocks}
\end{equation}
with $C_J^{(\nu)} (x)$ the Gegenbauer polynomial and $\nu = (d-2)/2$.
Meanwhile, the OPE coefficients $a_\Om$ are defined as
\begin{equation}
    a_{\mathcal O}
    =
    \frac{J!}{2^J (\nu)_J}
    \frac{f_{\phi \phi \mathcal O} b_{\mathcal O}}{c_{\mathcal O}}\ ,
    \label{eq:DefOPECoefficients}
\end{equation}
where $f_{\phi \phi \mathcal O}$ is the three-point function OPE coefficient at zero temperature, $c_{\mathcal O}$ is the normalization of the operator $\mathcal O$, and $b_{\mathcal O}$ is the thermal one-point function coefficient defined through
\begin{equation}
    \vev{\mathcal O^{\mu_1\ldots \mu_J}}_\beta
    =
    \frac{b_{\mathcal O}}{\beta^{\Delta}}
    \left(e^{\mu_1}\ldots e^{\mu_J} - \text{traces}\right)\ .
    \label{eq:ThermalOnePointFunctions}
\end{equation}
where $e^{\mu} = \delta^{0\mu}$.
The notation $(\nu)_J$ refers to the Pochhammer symbol. In the case of the stress energy tensor, $c_{\mathcal O}$ coincide with the $c$-anomaly coefficient  $C_T$ (the ``central charge'').

At zero spatial distance, the OPE becomes:
\begin{equation}
    g(\tau)
    =
    \frac{1}{\tau^{2\Delta_\phi}}
    \sum_{\Delta}
    a_\Delta \left( \frac{\tau}{\beta} \right)^\Delta \ ,
    \label{eq:OPEintau}
\end{equation}
reflecting an effective one-dimensional system, insensitive to the spin $J$.
Here we defined the reduced OPE coefficients
\begin{equation}
    a_\Delta = \sum_{\mathcal O, \Delta_{\mathcal O} = \Delta}
    a_{\mathcal O}\, C_J^{(\nu)}(1) \ .
    \label{eq:aDeltadef}
\end{equation}
When there are several operators with equal scaling dimension but different spins, $a_\Delta$ is said to be degenerate.
This situation often occurs in perturbative settings or highly symmetric models.

For the rest of this paper (with the exception of Section \ref{subsec:CorrelatorsInThe3dIsingModel}), we set $\beta=1$ without loss of generality.

\subsubsection{Periodicity}
\label{subsubsec:Periodicity}

The thermal geometry imposes stringent constraints on the thermal one-point functions.
In particular, the periodicity along the thermal circle results in a crossing-like equation, known as \textit{KMS condition} \cite{Kubo:1957mj,Martin:1959jp}, which in our conventions is given by\footnote{In Eq.~\eqref{eq:KMSCondition} we combined the KMS condition with parity ($\tau \to -\tau$) to obtain an equation in which, if $z \zb < \beta^2$, both side can be expanded using the OPE.}
\begin{equation}
    g(z,\zb) = g(1-\zb,1-z)\,.
    \label{eq:KMSCondition}
\end{equation}
Each side of this equation can be expanded into thermal blocks following \eqref{eq:ThermalBlockExpansion}.
In the following, we call \textit{$s$-channel} the left-hand side and \textit{$t$-channel} the right-hand one, by analogy to crossing symmetry at zero temperature.
The strength of the KMS relation is that it holds outside of the OPE regime, and using it to reconstruct full correlators is the main goal of the thermal bootstrap program.

Note also that the correlator is \textit{real}, which results into the relations
\begin{equation}
    g(z,\zb) = g(\zb,z)
    \quad \longleftrightarrow \quad
    g( r,w) = g( r,w^{-1})\,,
    \label{eq:RealityCondition}
\end{equation}
and that \textit{parity} imposes\footnote{In general, a theory may not be parity invariant. However, it is easy to see that the transformation $\tau \to -\tau$ is always a symmetry of any CFT~\cite{Iliesiu:2018fao,Alday:2020eua}. Combining this fact with the reality of the correlator, one obtains the condition~\eqref{eq:ParityCondition} even if the theory itself is not parity invariant.}
\begin{equation}
    g(z,\zb) = g(-z,-\zb)
    \quad \longleftrightarrow \quad
    g( r,w) = g( r,-w)\,.
    \label{eq:ParityCondition}
\end{equation}

\subsection{Reconstructing correlators at zero spatial separation}
\label{subsec:ReconstructingCorrelatorsAtZeroSpatialSeparation}

In this section, we derive a novel dispersion relation for thermal correlators with zero spatial separation and discuss how to reconstruct the correlators through an expansion in generalized free field correlators.

\subsubsection{Dispersion relation}
\label{subsubsec:DispersionRelation1}

\begin{figure}[t]
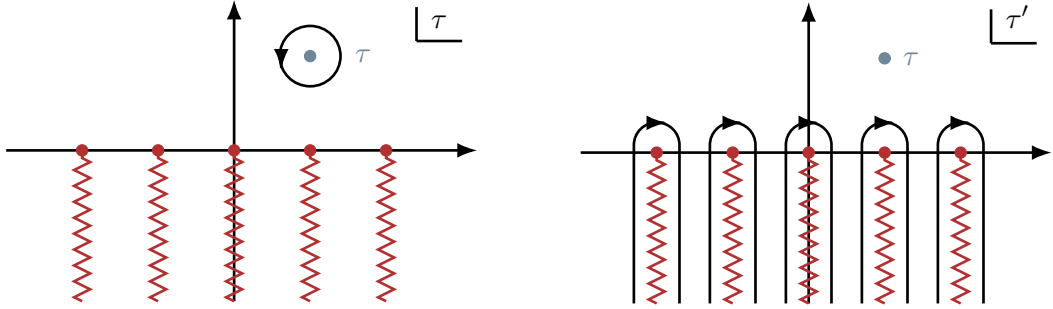

    \centering
    \begin{minipage}{.5\textwidth}
    \centering
    \ThermaltauAnal
    \end{minipage}%
    \begin{minipage}{.5\textwidth}
    \centering
    \ContourThermaltau
    \end{minipage}
    \caption{\textbf{Left panel:} Analytic structure in the complex $\tau$-plane of the thermal two-point function at zero spatial separation, $g(\tau)$. \textbf{Right panel:} The contours considered to compute the correlation function in $\tau$ via a dispersion relation.
    }
    \label{fig:collanalytic}
\end{figure}

The zero spatial distance limit introduces interesting features.
First, the OPE is convergent everywhere on the thermal circle.
Second, the function $g(\tau)$ depends solely on the time coordinate, making the problem effectively one-dimensional.\footnote{We are grateful to Dalimil Mazáč for interesting discussions and insights connected to this section.}

\paragraph{Analytic structure in the complex  \texorpdfstring{$\tau$}{\texttwoinferior}-plane.}
We now consider $g(\tau)$ as a function of the complex variable $\tau \in \mathbb{C}$.
The analytic structure is strongly constrained by the KMS condition: the function has simple poles on the real axis located at
\begin{equation}
    \tau = n \beta \ , \quad n \in \mathbb{Z} \ .
    \label{eq:LocationSimplePoles}
\end{equation}
Branch cuts might additionally be present.
To determine this, we focus on the region $|\tau| < \beta$, where we can expand the two-point function using the OPE \eqref{eq:OPEintau}.
As mentioned in Section \ref{subsubsec:Kinematics}, this system is effectively one-dimensional and the OPE is not sensitive to the spin of the operators.
The branching points in \eqref{eq:LocationSimplePoles} would depend on the conformal dimensions of the operators present in the OPE.
Because of periodicity ($\tau \sim \tau+n \beta$) and time reversal ($\tau \sim -\tau$), such branch cuts cannot be placed in arbitrary directions but are forced to lie along the imaginary time direction, as shown in Fig.~\ref{fig:collanalytic}.
It should also be noticed that if such branch cuts exist, they must extend all the way to infinity.

\paragraph{Dispersion relation.}
Knowledge of the analytic structure in the complex $\tau$-plane allows us to write an explicit formula for $g(\tau)$ in terms of the discontinuity along the branch cuts.
To do this, we compute the correlation function around a point $\tau$ by using the Cauchy integral formula
\begin{equation} \label{eq:Chauc1}
    g(\tau)=\frac{1}{2 \pi i}\oint_{\mathcal{C}} \text{d}  \tau' \frac{g(\tau')}{\tau'-\tau}\ .
\end{equation}
The next step is to deform the contour $\mathcal{C}$.
We must consider two contributions: the first comes from the infinite set of discontinuities along the cuts, while the second one arises from the arcs at $|\tau| \to \infty$.
We denote the latter as $g_\text{arcs}(\tau)$, and the two-point function becomes
\begin{equation}
    g(\tau)
    =
    g_{\text{dr}}(\tau) + g_\text{arcs}(\tau)\ ,
    \label{eq:DRAndArcs}
\end{equation}
with
\begin{equation}
    g_{\text{dr}}(\tau)
    =
    \sum_{m=-\infty}^{\infty}
    \int_{-i \infty}^{0} \frac{\text{d}\tau'}{2 \pi i} \frac{\operatorname{Disc} g(\tau')}{\tau'+m - \tau}\,,
    \label{eq:disp}
\end{equation}
where the discontinuity is defined as 
\begin{equation}
    \operatorname{Disc} g(\tau') = g(\tau'  + 0) - g(\tau' -  0)\,.
\end{equation}
In fact, the branch cuts of the two-point function extend along $\tau \in (-i\infty + n\beta,\, i\infty + n\beta)$, as a consequence of the OPE terms of the form $\left(x^2 + \tau^2\right)^s$ in the limit $x \to 0$. The correct analytic structure is instead depicted in Fig.~\ref{fig:GMIFigOne}, while in Fig.~\ref{fig:collanalytic} we depict the analytic structure corresponding to terms as $\tau^{s}$, appearing directly after setting $x = 0$. Nevertheless, computing the discontinuity as illustrated in Fig.~\ref{fig:collanalytic} is equivalent to evaluating it for the cuts in Fig.~\ref{fig:GMIFigOne} in the limit $x \to 0$.

Note that the discontinuity in equation \eqref{eq:disp} is taken on the imaginary axis ($\tau' \in i \mathbb R$ as shown in Fig. \ref{fig:collanalytic}).
Here the sum runs over all the locations of the poles, automatically encoding the KMS condition. Note that \eqref{eq:disp} is independent of $d$, the number of spacetime dimensions.

\subsubsection{Arc contributions}
\label{subsubsec:ArcContributions1}

We now discuss the arc contributions in the dispersion relation \eqref{eq:disp}, and show that they correspond at most to a constant.

The starting point is to notice that $g_\text{dr}(\tau)$ captures the whole analytic structure of the full correlator $g(\tau)$, which implies that $g_\text{arcs}(\tau)$ has to be an entire function.
Similarly, $g_\text{dr}(\tau)$ is periodic by construction and thus
\begin{equation}
    g_\text{arcs}(\tau)
    =
    g_\text{arcs}(1-\tau)\,.
    \label{eq:CocircularKMS}
\end{equation}
We now have to discuss the boundedness of the function $g_\text{arcs}(\tau)$.
On the real axis, the absence of finite poles and the periodicity ensure that the function is bounded.
Hence, we only need to focus on the behavior in the upper complex $\tau$-plane; the lower counterpart would be analogous.

First of all, we prove that the function $g(\tau)$ is bounded as $|\tau| \to \infty$. This can be shown by adapting one of the arguments in \cite{Iliesiu:2018fao}. We can quantize the thermal theory by identifying spatial slices $\mathbb{R}^{d-2}\times S^1_{\beta}$ on the thermal geometry $\mathbb{R}^{d-1}\times S^1_{\beta}$, where the quantum states are defined; the time variable is represented by the missing spatial direction. The circle compactification is now interpreted as a Kaluza--Klein compactification. The two-point function can be rewritten as 
\begin{equation}
    g(\tau) = \langle \phi(\tau) \phi(0)  \rangle_{\beta}=  \langle \Psi | e^{i \tau \op P_{\text{KK}} } |\Psi \rangle  \ ,
\end{equation}
where $ |\Psi \rangle=\phi(0) | 0 \rangle_{\mathbb{R}^{d-2}\times S^1_{\beta}}$.
The complex time evolution operator can be rewritten in terms of two operators $V$ and $U$ defined as \begin{equation}
    V = e^{-\operatorname{Im}(\tau) \op P_{\text{KK}}} \ , \quad U = e^{i\operatorname{Re}(\tau) \op P_{\text{KK}}}\ , \quad e^{i \tau \op P_{\text{KK}} } = V^{\frac{1}{2}}U V^{\frac{1}{2}} \ .
\end{equation}
Observe that $V$ is a positive Hermitian operator while $U$ is a unitary operator. By Cauchy--Schwartz inequality we have \begin{equation}
    |g(\tau)|^2 \le \langle \Psi| V^{\frac{1}{2}}V^{\frac{1}{2}}|\Psi \rangle  \langle \Psi| V^{\frac{1}{2}}U^\dagger U V^{\frac{1}{2}}|\Psi \rangle  = \langle \Psi | V |\Psi \rangle^2 \ .
\end{equation}
Therefore we have 
\begin{equation} \label{eq: bound t}
    |g(\tau)|\le |g(i \operatorname{Im}(\tau))|   \ .
\end{equation}
 Eq.~\eqref{eq: bound t} shows that the problem of bounding the function $g(\tau)$ in the full upper complex $\tau$-plane can be reduced to the problem of bounding it on the imaginary axis, i.e., of bounding the real time correlator. The real time correlator is not affected by poles at finite real time and it is bounded at infinity by the clustering decomposition of the correlator for $|t| \to \infty$.
 
Given the absence of finite poles on the imaginary axis, we conclude that $g(\tau)$ is bounded by a constant on the imaginary axis and then on the entire upper complex $\tau$-plane. Secondly, we need to discuss the boundedness of the function $g_{\text{dr}}(\tau)$ in the upper complex $\tau$-plane. By construction, this function is \emph{polynomially} bounded. Hence, we conclude that $g_{\text{arcs}}(\tau)$ is at least polynomially bounded, but for the purposes of our discussion it is enough to say
\begin{equation} \label{eq: exp bd}
    |g_{\text{arcs}}(\tau)| < C e^{|\tau|} \ , \qquad |\tau| \to \infty\,,
\end{equation}
for any constant real number $C$, which is a weaker assumption then the polynomial bound. 

We conclude that the function $g_{\text{arcs}}(\tau)$ is entire, periodic on the real axis, and surely exponentially bounded as in \eqref{eq: exp bd}. As we show in Appendix \ref{app:pefconst}, these conditions lead to the conclusion
\begin{equation} \label{eq: kappa}
    g_{\text{arcs}}(\tau)=\kappa \ , \qquad \kappa \in \mathbb{R} \ .
\end{equation}

\subsubsection{A formula for  \texorpdfstring{$g(\tau)$}{\texttwoinferior}}
\label{subsubsec:AFormulaForg}

We can now combine the insights of Sections \ref{subsubsec:DispersionRelation1} and \ref{subsubsec:ArcContributions1} to derive a formula for $g(\tau)$ that is manifestly KMS invariant.

Our strategy is to insert the OPE \eqref{eq:OPEintau} in the dispersion relation \eqref{eq:disp} and to commute the discontinuity with the sum.
Following the same arguments as in Section \ref{subsubsec:ArcContributions1}, this procedure misses at most a constant which can be reabsorbed in $\kappa$. This may be interpreted as a consequence of the existence of a conformal map between the strip and the disc, which makes the OPE effectively converge everywhere. 
We then compute the discontinuity of single thermal blocks and find
\begin{equation}
     \Disc_{\operatorname{Im}(\tau) < 0 } \tau^{\Delta - 2\Delta_\phi}
     =
     \tau^{\Delta - 2\Delta_\phi} e^{\frac{3}{2}\pi i (\Delta - 2 \Delta_\phi)}\left[1 - e^{-2 \pi i (\Delta - 2\Delta_\phi)} \right] \ ,
     \label{eq: disc tau}
\end{equation}
where the overall phase reflects the fact that we are parametrizing a contour surrounding the lower imaginary axis.
Resumming the discontinuities, we obtain an alternative expansion for the correlator:
\begin{equation}
    g(\tau) = \sum_{\Delta} a_{\Delta} \Big[ \zeta_{H}(2\Delta_\phi - \Delta, \tau)+\zeta_{H}(2\Delta_\phi - \Delta, 1 - \tau)  \Big] + \kappa\ ,
    \label{eq:dispcomplextau}
\end{equation}
where in each term we find a kernel written in terms of Hurwitz $\zeta$-functions multiplied by the OPE coefficients $a_\Delta$ that encode the dynamical information. The dependence on $\beta$ can be reintroduced by dimensional analysis: the complete formula is given in Eq.~\eqref{eq: summ1}, where it is also straightforward to check that the limit $\beta \to \infty$ gives the correct zero-temperature correlator. 
A similar equation appeared before in \cite{Parisini:2023nbd}, where it was employed in the study of holographic correlators through a different setup: in our work, this formula is derived by inverting operators via dispersion relation.

In the rest of this section, we discuss the implications of \eqref{eq:dispcomplextau}.
This expression is powerful, as KMS invariance is built in term by term, as opposed to the thermal block expansion \eqref{eq:OPEintau}.
Furthermore, we note that \eqref{eq:dispcomplextau} can be suggestively interpreted as an infinite sum over two-point function of fundamental scalars in generalized free field (GFF) theory, which is given by~\cite{Marchetto:2023xap}
\begin{equation}
    g_\text{GFF} (\tau, \Delta)
    =
    \zeta_H(2\Delta, \tau) + \zeta_H(2\Delta, 1 - \tau) \,.
\end{equation}
In words, we conclude that \textit{thermal two-point functions at zero spatial separation admit an expansion in terms of GFF correlators}, with coefficients given by the thermal OPE data.\footnote{Note that correlators produced with the dispersion relation can escape this statement if they are not consistent with the boundedness condition discussed in Section \ref{subsubsec:ArcContributions1}. The arc contributions are non-trivial in this case.}
At zero temperature, a similar structure was proposed for one-dimensional systems in~\cite{Ghosh:2025sic}, in which solutions to the crossing equation are expressed in terms of \textit{extremal solutions} — minimal solutions that individually satisfy crossing.
In the same spirit, \eqref{eq:dispcomplextau} implies that GFF correlators form a basis for thermal two-point functions.

This formulation can be made immediately useful in perturbative settings or in cases where the spectrum is protected by additional symmetries (for instance free or supersymmetric theories).
The kernel $g_\text{GFF} (\tau, \Delta)$ vanishes identically when $\Delta - 2\Delta_\phi \in 2\mathbb{Z}^{>0}$, leading to considerable simplifications in favorable circumstances.
For example, as we will see in more detail in Section \ref{subsec:GeneralizedFreeTheories}, the double-twist operators $[\phi \phi]_{n,J}$ do not contribute in a generalized free field theory, and the correlator $g(\tau)$ is fully reconstructed by the term corresponding to the identity operator.
In a general interacting theory, the double-twist operators acquire anomalous dimensions, and in principle their contribution needs to be considered.
Nonetheless, a classical result from analytic conformal bootstrap \cite{Alday:2015eya,Fitzpatrick:2012yx,Komargodski:2012ek,Simmons-Duffin:2016gjk} gives \footnote{In specific theories, such as gauge theories (when $\Gamma_\text{cusp} \neq 0$), the anomalous dimensions of operators may grow as $\sqrt{J}$ rather than $1/J$. This behavior is well understood in the literature \cite{Alday:2007mf,Alday:2013cwa,Caron-Huot:2017vep}. In such cases, it is necessary to properly account for the contribution of these operators in the discontinuity.}
\begin{equation}
    \Delta= 2\Delta_\phi + 2n + J + \gamma \ , \qquad \gamma \sim O\left(\frac{1}{J}\right) \quad \text{as } J \to \infty \ .
\end{equation}
This means that in the large spin limit, the OPE spectrum approaches that of a generalized free field.
Therefore the double-twist operators do not contribute to the discontinuity, since
\begin{equation}
   \Disc_{\operatorname{Im}(\tau) < 0 } \tau^{\Delta- 2\Delta_\phi} \propto  \frac{1}{J} \ .
\end{equation}
Clearly, this conclusion is correct only if $a_\Delta$ does not grow as $J$ in the large spin limit. This property is theory dependent.
The contribution of these operators to the discontinuity was discussed from a different perspective in~\cite{Manenti:2019wxs}; we return to this connection in Section \ref{subsec:ConnectionToMomentumSpaceCorrelators}.

\subsection{Reconstructing correlators at non-zero spatial separation}
\label{subsec:ReconstructingCorrelatorsAtNonZeroSpatialSeparation}

We now extend our analysis of the previous section to the case in which the spatial distance between the two external operators is kept non-vanishing.
We first review the derivation of the dispersion relation presented in \cite{Alday:2020eua} and discuss the arc contributions.
We then lay out our strategy for reconstructing correlators, using the OPE and a set of consistency conditions.
This is applied through two different methods: the first one is numerical and relies on the form of the large-spin expansion for OPE coefficients.
The second one is analytical, where correlators are obtained through promoting the dispersion relation of \cite{Alday:2020eua} to a version that is KMS invariant by construction.

\subsubsection{Dispersion relation}
\label{subsubsec:DispersionRelation2}

\begin{figure}[t]
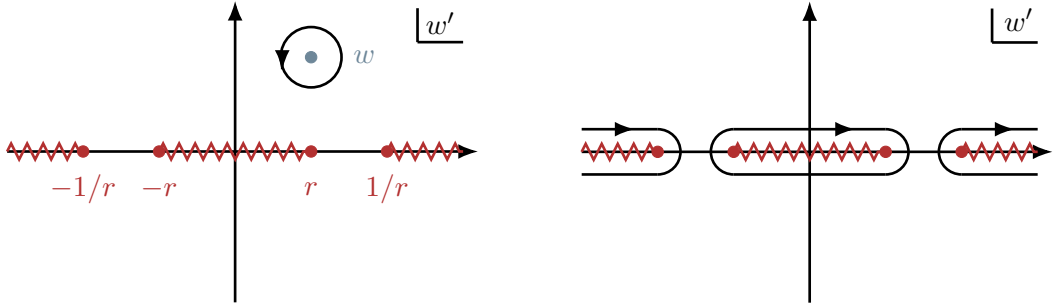

    \centering
    \begin{minipage}{.5\textwidth}
    \centering
    \AnalyticStructureThermal
    \end{minipage}%
    \begin{minipage}{.5\textwidth}
    \centering
    \ContourThermal
    \end{minipage} 
    \caption{\textbf{Left panel}: Analytic structure of a thermal correlator in the complex $w$-plane. \textbf{Right panel}: Contours used in the derivation of the dispersion relation. 
    }
    \label{fig:AnalyticStructurethermal}
\end{figure}

For non-zero spatial coordinates, a dispersion relation was already derived in \cite{Alday:2020eua}; we review here the derivation from the analytic structure and the formula.

\paragraph{Analytic structure in the \texorpdfstring{$w$}{\texttwoinferior}-plane.}
The procedure to compute the dispersion relation is similar to the one described above for the case of the $\tau$-plane, but using the analytic structure in the complex $w$-plane.
It was argued in \cite{Iliesiu:2018fao} that the corresponding analytic structure is the one depicted in Fig.~\ref{fig:AnalyticStructurethermal}.
This was shown by employing the OPE and suitable positivity assumptions on a combination of the momentum and Hamiltonian operators in the Kaluza--Klein formalism.

\paragraph{Dispersion relation.}
As always, the starting point is to consider the two-point function at a generic point $w$ and make use of Cauchy's formula to rewrite it as a contour integration:
\begin{equation}
    g(r,w)=
    \oint_{\mathcal{C}} \frac{\text{d} w'}{2 \pi i} \frac{g(r,w')}{w' -w} \ .
\end{equation}
We can then deform the contour to go around the cuts: this turns into an integral over the discontinuity of $g( r,w)$, and dropping the arcs we obtain\footnote{The precise derivation involves making use of the reality and parity conditions given in \eqref{eq:RealityCondition} and \eqref{eq:ParityCondition} to combine the contours into one expression. The steps are in one-to-one agreement with the derivation provided in \cite{Barrat:2022psm,Bianchi:2022ppi} for the defect case, which shares the same analytic structure, apart from the fact that parity has to be imposed.
Note that, in Eq.~(5.5) of~\cite{Alday:2020eua}, there is an additional term with a residue.
We do not keep track of this term here, as its main purpose is to reconstruct the identity contribution.}
\begin{equation}
    g_\text{dr} ( r,w) = \int_{0}^{ r} \frac{\text{d} w'}{2 \pi i} \
    \frac{w^2 (1-w'^4)}{w' (w'-w)(w'+w) (1-w^2 w'^2)}
    \operatorname{Disc} g( r,w')\,,
    \label{eq:DRw}
\end{equation}
with
\begin{equation}
    \operatorname{Disc} g( r,w') = g( r,w'+ i 0) - g( r,w'-i 0)\,.
\end{equation}
The full correlator is then in principle given by\footnote{It should be emphasized at this point that the dispersion relation \eqref{eq:DRw} is fully equivalent to the Lorentzian inversion formula of \cite{Iliesiu:2018fao} (see Section~5.2 of~\cite{Alday:2020eua} for details). The fact that it relates the discontinuity of the correlator to itself simply provides a shortcut to calculating the OPE coefficients individually.}
\begin{equation}\label{eq:DRfull}
    g ( r,w)
    =
    g_\text{dr} ( r,w)
    +
    g_\text{arcs} ( r,w)\,,
\end{equation}
where $g_\text{arcs}(r,w)$ comes from the contribution of the integral in the limit $|w|\to \infty$.

\subsubsection{OPE inversion and consistency conditions}
\label{subsubsec:OPEInversionAndConsistencyConditions}

In this section, we lay out a general strategy and summarize the consistency conditions that the true correlator needs to satisfy.
We ignore the arc contributions for now, which will be discussed in Section \ref{subsubsec:ArcContributions2}.

\paragraph{Inverting the OPE.}
The main idea is to insert the $t$-channel of the OPE \eqref{eq:ThermalBlockExpansion} in the dispersion relation \eqref{eq:DRw} and to commute the discontinuity with the sum:
\begin{equation}
    \operatorname{Disc}_{\operatorname{Re}(\zb) > 0} g(z,\zb)
    =
    \sum_{\mathcal O} a_{\mathcal O} \operatorname{Disc}_{\operatorname{Re}(\zb) > 0} f_{\Delta, J} (1-z,1-\zb)\,.
\end{equation}
Similarly to what we already discussed in Section \ref{subsubsec:AFormulaForg}, the blocks do not contribute when $\Delta - 2\Delta_\phi \in 2\mathbb{Z}^{>0}$, and in favorable settings the correlator can be reconstructed from a finite number of contributions.
This happens because the thermal blocks behave schematically near $\zb=1$ as
\begin{equation}
    f_{\Delta,J} (1-z,1-\zb) 
    \sim \hspace{-0.5em}
    \sum_{k=1}^{\tiny{\Delta_\phi-(\Delta-J)/2}} \hspace{-0.5em}\frac{1}{(1-\zb)^{k}} + \text{regular}\,,
\end{equation}
while the discontinuity vanishes for integer powers (the regular terms):
\begin{equation}
    \Disc_{\operatorname{Re}(\zb) > 0} (1-\zb)^\alpha
    = 0 \ ,
    \quad
    \alpha \in \mathbb{Z}^{>0}
    \label{eq:vanishingdisc}\ .
\end{equation}
For the blocks that survive the discontinuity, the following identity holds both when $\alpha \in \mathbb{Z}^{<0}$ or when $\alpha$ is not an integer:
\begin{equation}\label{eq:disc}
    \Disc_{\operatorname{Re}(\zb) > 0} (1-\zb)^\alpha
    = 2 i \sin(\alpha \pi)\, (\zb - 1)^\alpha\, \Theta(\text{Re}(\zb) - 1) \ , \quad \alpha \notin \mathbb{Z}^{>0}\,.
\end{equation}

\paragraph{OPE and out-of-OPE contributions.}
At zero temperature, the procedure described above typically already reproduces the correlator up to the arc contributions.
An additional obstacle is introduced at finite temperature, which stems from the fact that the OPE is not convergent everywhere.
Compared to the complex $\tau$-plane case, the prescription to exchange the discontinuity operation with the OPE expansion requires additional consideration.
This can be seen directly from \eqref{eq:DRw}, since the integral runs over $w \in (0, r)$, with $r = \sqrt{x^2 + \tau^2}$, and the OPE only converges for $r < 1$.
One can indeed verify that the function $g_{\text{dr}}(r,w)$ is in this case \textit{not KMS invariant}.
This discrepancy can be understood by employing the Lorentzian inversion formula of~\cite{Iliesiu:2018fao}, which returns a function $a(\Delta, J)$, whose poles correspond to the OPE coefficients of our correlator of interest.
As already identified in~\cite{Iliesiu:2018fao},\footnote{See Section 6 for additional detail.} we need to distinguish two types of contributions:
\begin{itemize}
    \item[$\star$] \emph{OPE contributions}. This type of contribution schematically appears in the inversion formula through integrals of the following form:
    \begin{equation}
       \int_1^\infty \frac{\text{d}\zb}{\zb} \zb^{\Delta_\phi - \hb - m} \Disc(1 - \zb)^c  \sim J^{-c-1}+\dots\ ,
    \end{equation}
    where $\hb = (\Delta + J)/2$ and $m\in \mathbb{Z}^{\geq 0}$
    This highlights the fact that contributions from the OPE region to the coefficients are \emph{polynomially} suppressed in the spin $J$; 
    \item[$\star$] \emph{Out-of-OPE contributions}. This type of contribution is schematically represented instead as integrals of the form
    \begin{equation}\label{eq:notOPELIF}
        \int_2^\infty \frac{\text{d}\zb}{\zb} \zb^{\Delta_\phi - \hb - m} f(\zb) \sim 2^{-J}+\dots \ .
    \end{equation}
    In this case, we learn instead that the out-of-OPE region encodes terms that are \emph{exponentially} suppressed in the spin $J$.
    These terms are missing from the Lorentzian inversion formula, and similarly they are not captured by the dispersion relation \eqref{eq:DRw}.
\end{itemize}
This analysis suggests that the thermal correlator $g(z,\zb)$ can be split into two parts, each associated to one type of contribution:
\begin{equation}
    g(z,\zb) = g_\text{OPE}(z,\zb) + g_{{\text{out-of-OPE}}}(z,\zb) \ ,
    \label{eq: gout}
\end{equation}
where $g_\text{OPE}(\tau,x)$ can be identified with the dispersion relation and its arc completion. 

\paragraph{Consistency conditions.}
We now discuss the consistency conditions that can be imposed on the correlator in order to reconstruct the function $g_{{\text{out-of-OPE}}}(z,\zb)$:
\begin{enumerate}
    \item \emph{KMS condition.}
    A fundamental property of thermal two-point functions is the KMS condition \eqref{eq:KMSCondition};
    
    \item \emph{Analytic structure.}
    The analytic structure of thermal two-point functions must match the one described above and illustrated in Fig.~\ref{fig:AnalyticStructurethermal}.
    In particular, the arc terms must have vanishing discontinuity;
    
    \item \emph{Consistency with the OPE.}
    The thermal two-point function must be consistent with the operator product expansion.
    The dispersion relation is expected to reproduce the correct OPE spectrum up to low-spin operators, which would then be encoded in $g_\text{arcs}$;
    
    \item \emph{Regge boundedness.}
    In the variables $(r, w)$, the thermal two-point function is expected to be \emph{polynomially} bounded in the Regge limit
    \begin{equation}
        w \to \infty \,, \quad  r \text{ fixed};
    \end{equation}
    \item \emph{Clustering at large distances.}
    At large spatial separation, the two-point function of a unitary theory must satisfy the clustering condition
    \begin{equation}
        g(z,\zb) \xrightarrow{x \to \infty} \langle \phi \rangle_\beta^2.
    \end{equation}
    The precise approach to this asymptotic value depends on the theory. Note that the clustering decomposition may be violated if the theory is not unitary and in particular if there are operators with negative anomalous dimensions (see Section \ref{subsubsec:VirasoroPrimariesLeeYang}).
    In general, for theories without an extensive symmetry structure,\footnote{A simple counterexample is the $4d$ free scalar theory, where the correlator between fundamental scalars decays as a power law: $g(\tau,x) \sim 1/x$.} the expected behavior is an exponential decay
    \begin{equation}
        g(z,\zb) \xrightarrow{x \to \infty} \langle \phi \rangle_\beta^2 + O(e^{-m_\text{th} x}),
    \end{equation}
    where $m_\text{th} \propto 1/\beta$ is the thermal mass.
\end{enumerate}

\subsubsection{A numerical approach}
\label{subsubsec:ANumericalApproach}

Our first approach to reconstruct $g_{{\text{out-of-OPE}}}(z,\zb)$ is based on the fact that the corrections in spin are organised into a set of \textit{exponential corrections} to the large-spin expansion of the OPE coefficients:\footnote{Strictly speaking, Eq.~\eqref{eq:expope} is reliable for the spin of the operator being large enough. This is discussed in further detail in Section \ref{subsubsec:ArcContributions2}. For the sake of the discussion, we ignore the arc contributions in this Section.}
\begin{equation}
    a_{\mathcal O}
    =a_{\Om}^{\text{(dr)}}
    \left(\textcolor{blue}{1}+  \textcolor{red}{\sum_{n = 1}^\infty \frac{c_n}{n^J}}\right) \ .
    \label{eq:expope}
\end{equation}
where $c_1 = 1$ and the \textcolor{blue}{blue term} is the contribution to $a_{\mathcal O}$ given by the dispersion relation applied to the $t$-channel expansion, and contained in $g_\text{OPE}(z,\zb)$. 
$a_{\mathcal O}^\text{(dr)}$ has at most a polynomial dependence on $J$.
The \textcolor{red}{red terms} instead are part of $g_{{\text{out-of-OPE}}}(z,\zb)$ and they encode the exponential dependence on $J$. Intriguingly enough this expansion is reminiscent of expansion of one-point functions in terms of polylogarithms \cite{Petkou:2021zhg,Karydas:2023ufs}.
Each exponential correction in the spin comes with an unknown coefficient $c_n$ to be fixed.

These coefficients can be fixed numerically by requiring that the candidate correlator satisfies KMS invariance
\begin{equation}
    g_{\text{cand}}(z,\zb) \overset{!}{=} g_{\text{cand}}(1 - z,1-\zb) \ .
\end{equation}
In practice, this procedure requires truncating the sum over exponential corrections to a finite order $n_\text{max}$.
One can then solve for a finite number of coefficients $c_n$ numerically and verify the stability and convergence of the procedure by varying $n_\text{max}$.
This procedure is shown explicitly in Appendix \ref{app:KMScompensator} for the $4d$ scalar free theory and for the $\mathrm{O}(N)$ model at first order in the $\varepsilon$-expansion.
In both of these cases, the exact correlator is recovered (up to a constant that the KMS condition cannot fix).

\subsubsection{The generalized method of images}
\label{subsubsec:TheGeneralizedMethodOfImages}

\begin{figure}
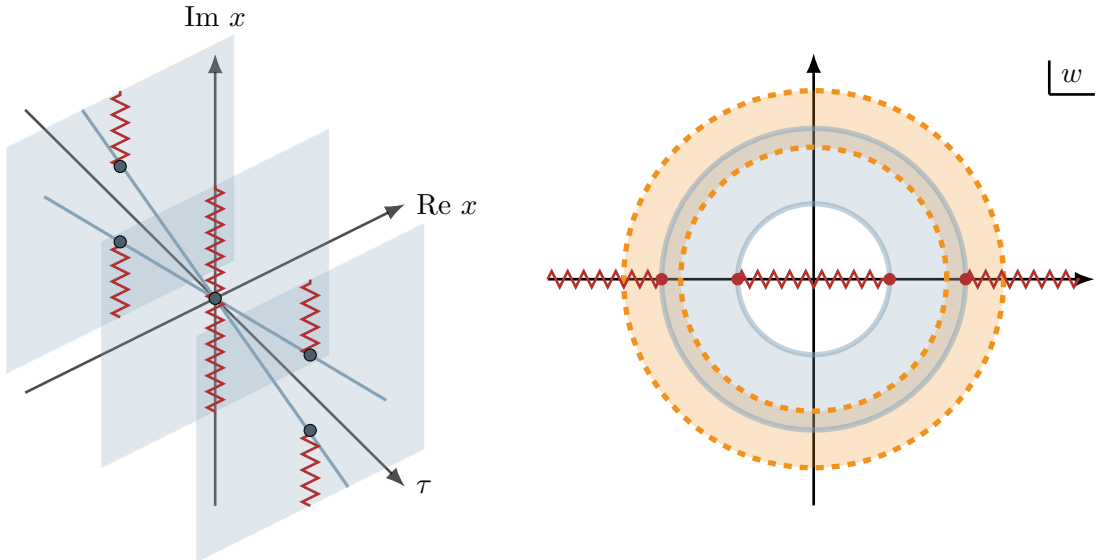

    \centering
    \begin{minipage}{.5\textwidth}
    \centering
    \GMIFigOne
    \end{minipage}%
    \begin{minipage}{.5\textwidth}
    \centering
    \GMIFigTwo
    \end{minipage} 
    \caption{\textbf{Left panel}:
    Illustration of the analytic structure of thermal two-point function in the complex $x$-plane, following \cite{Manenti:2019wxs}.
    The different planes represent integer shifts in $\tau$.
    The lines crossing the end points of the cuts show the lightcone $\tau^2 = (i x)^2$.
    \textbf{Right panel}:
    The blue-shaded annulus (with the solid contours) represent the region of convergence of the OPE in the $s$-channel in the $w$-plane, as shown in \cite{Iliesiu:2018fao}.
    The orange-shaded annulus (with the dashed contours) schematically depicts how the region of convergence of the OPE shifts with one image.
    This illustrates how \eqref{eq:GeneralizedMI} captures the full correlator by starting with one OPE region (in this case the $t$-channel) and summing over its images.}
    \label{fig:GMIFigOne}
\end{figure}

The numerical method discussed above is conceptually straightforward, but it can become technically challenging when many operators contribute to the dispersion relation.
For this reason, it is useful to explore an alternative approach and impose KMS invariance \emph{explicitly}, starting directly from the output of the dispersion relation.
To motivate this, let us recall the dispersion relation in the complex $\tau$-plane \eqref{eq:disp}.
In that case, each operator in the OPE contributes via a KMS-invariant function, constructed by summing over all its thermal images. Fig. \ref{fig:GMIFigOne} shows that this is also the case in the complex $x$-plane.

One might wonder whether a similar construction can be generalized to the case of non-zero spatial distance.
The key observation is that any non-periodic function $f(\tau)$ can be used to build a function $\widetilde{f}(\tau)$ periodic by construction 
\begin{equation}\label{eq:poissonresumm}
    \widetilde f(\tau) = \sum_{m = -\infty}^\infty f(\tau - m)\ ,
\end{equation}
and, conversely, any periodic function admits such a decomposition through Poisson resummation.
This suggests a natural Ansatz for the correlator: since $g_{\text{dr}}(\tau,x)$ is not KMS-invariant, periodicity can be reinstated by construction as follows\footnote{The factor $1/2$ comes from the fact that by summing in $m\in \mathbb Z$ we are over-counting the images, since when we computed the dispersion relation we already used the symmetry $w \to -w$.}
\begin{equation}
    g(z,\zb)
    =
    \frac{1}{2} \sum_{m = -\infty}^\infty g_\text{dr}(z - m, \zb-m) + g_\text{arcs}(z,\zb) \ .
    \label{eq:GeneralizedMI}
\end{equation}
We refer to Eq.~\eqref{eq:GeneralizedMI} as the \emph{generalized method of images}.
This conjecture can be seen to satisfy most of the consistency requirements listed in Section \ref{subsubsec:OPEInversionAndConsistencyConditions}: 
\begin{enumerate}
    \item\emph{KMS condition.}
    This condition is automatically satisfied by construction in \eqref{eq:GeneralizedMI}, as the sum over images ensures periodicity;

    \item\emph{Analytic structure.}
    This requirement is also satisfied, since each image is built out of the dispersion relation, which by definition implements the correct analytic behavior; 
    
    \item\emph{Consistency with the OPE.}
    In Eq. \eqref{eq:notOPELIF}, we showed that the image terms contribute only $1/n^J$ corrections to the thermal OPE coefficients, and thus \eqref{eq:GeneralizedMI} is consistent with the OPE structure by construction. In the right panel of Fig. \ref{fig:GMIFigOne} we show the OPE contribution (blue shaded region) and the same contribution shifting $\tau$ of one unit of $\beta$ (orange shaded region), which therefore is an out of OPE contribution in the sense of Eq. \eqref{eq: gout}. The figure provides a pictorial demonstration of the mechanism underlying the generalized method of images;

    \item\emph{Regge boundedness.}
    This assumption plays a crucial role in the dispersion relation. In fact one has that for the term $m = 0$ in the Regge limit the result of the dispersion relation is polynomially bounded.
    If $m \neq 0$, this is less trivial to conclude since under $z \to z-m$ and $\zb \to \zb-m$ we have
    \begin{equation}
        r \to \tilde r=\sqrt{r^2- m z - m \zb+ m^2} \ ,
    \end{equation}
    and therefore for $w \to \infty$ and $r$ fixed for $m \neq 0$ we have $\tilde r \to \infty$ in the images.
    Nonetheless the images of the dispersion relation can be defined by considering \eqref{eq:DRw} and making the change of variable $z \to z-m$ and $\zb \to \zb-m$.
    One can notice that if the integral is polynomally bounded for $m =0$ this is also the case for $m \neq 0$. 
    Therefore every $g^{(\mathcal O)}(z-m,\zb-m)$ is bounded and thus we expect the full sum over the generalized images to converge;

    \item\emph{Clustering at large distances.} This condition is \emph{not} automatically built into \eqref{eq:GeneralizedMI}, and must be checked on a case-by-case basis.
    If the sum over images does not reproduce the correct large-distance behavior, the term $g_\text{arcs}$ must compensate for it.
\end{enumerate}
One may now insert the OPE inside \eqref{eq:GeneralizedMI} to produce a formula in the gist of \eqref{eq:dispcomplextau}.
We then obtain the following convenient expression:
\begin{equation}
    g(z,\zb)
    =
    \sum_{\Om} a_\Om \sum_{m=-\infty}^\infty f_{\Delta,J} (z+m,\zb+m) + g_\text{arcs} (z,\zb)\,,
\end{equation}
i.e., the correlator can be reconstructed by summing the thermal blocks over images.

\subsubsection{Arc contributions}
\label{subsubsec:ArcContributions2}

In the case of the complex $\tau$-plane, we were able to show that the arc terms are given at most by a constant.
When spatial separation is non-zero, we need to distinguish between the arcs of the dispersion relation, i.e. the function $g_\text{arcs}$ in \eqref{eq:DRfull} , and the arcs after generalized images defined in \eqref{eq:GeneralizedMI}.
In this Section  we bound the piece with arc contributions.
Here we should expect operators of low spin $J < J^*$ to appear in $g_\text{arcs} (\tau,x)$:
\begin{align}
    g_\text{arcs}(\tau,x)
    &=
    \sum_{\Delta}\sum_{J< J_\star} a_{\Om}^{\text{(arcs)}} \left(x^2+\tau^2\right)^{\frac{\Delta}{2}-\Delta_\phi} C_J^{(\nu)}\left(\frac{\tau}{\sqrt{x^2+\tau^2}}\right) \notag \\
    &=  \sum_{J <J_{\star}} F_{J}(\tau,x) \, C_J^{(\nu)}\left(\frac{\tau}{\sqrt{x^2+\tau^2}} \right)  \ ,
\end{align}
where $J^*$ is an assumption in practical cases.
If the generalized method of images suggested in Section \ref{subsubsec:TheGeneralizedMethodOfImages} produces the correct clustering decomposition behavior at large distances, then we must have
\begin{equation}
    \lim_{x \to \infty }g_\text{arcs}(\tau, x ) = 0 \ \quad  \Rightarrow \quad  \lim_{x \to \infty }F_{J}(\tau,x)= 0 \ ,
    \label{eq:clusteringcomp}
\end{equation}
as given in \cite{Alday:2020eua}.
However the only function $F_{J}(\tau,x)$ satisfying this property with $g_\text{arcs}$ being KMS invariant is $F_{J}(\tau,x) = 0$.
To see this, use $\tau \to \tau+m$ and the limit $m \to \infty$ while considering \eqref{eq:clusteringcomp}.

Clearly, this argument fails in cases where the generalized images do not exhibit the correct behavior as $x \to \infty$. The contribution from $g_{\text{arcs}}$ is therefore theory-dependent and must be analyzed on a case-by-case basis. We will present explicit examples in Section~\ref{sec:applications}.

The argument presented above supports the uniqueness of the function \( g_{\text{arcs}} \) within the given framework. Specifically, for any candidate function that satisfies the axioms of the thermal bootstrap --- most notably, the correct clustering decomposition of the correlator --- the reasoning can be repeated to confirm that no additional contributions are required. While this does not amount to a full uniqueness theorem for the thermal bootstrap in the sense of~\cite{Alday:2020eua,Marchetto:2023fcw}, as it relies on prior knowledge of the correlator's discontinuity, it does imply that the generalized method of images may omit, at most, a uniquely determined and well-characterized function. As such, the correlator remains unambiguous within this approach.

\subsection{Connection to momentum-space correlators}
\label{subsec:ConnectionToMomentumSpaceCorrelators}

A complementary approach to computing the generalized method of images for the dispersion relation in position space involves transitioning to a momentum-space description. In this context, one may ask whether there is a notion of OPE in momentum space.\footnote{Similar questions for zero-temperature four-point functions are discussed in~\cite{Bzowski:2013sza,Bzowski:2020kfw,Marotta:2022jrp}.}  

\subsubsection{OPE in momentum space}
The main obstacle to translating the OPE expansion into momentum space is that the OPE converges only within a finite region determined by the size of the thermal circle, specifically \(\tau^2 + x^2 < \beta^2\).  
As a result, the Fourier transform does not commute with the OPE.  
Nevertheless, the concept of OPE can still be applied in the regime of large momenta, which corresponds to small distances. In fact, a well-defined OPE has been shown to exist in momentum space in the limit \(\omega \to \infty\), with \(\xi = |\vec{k}| / \omega\) held fixed~\cite{Caron-Huot:2009ypo}.

Only in this limit one can claim that the two-point function in momentum space is approximated by its OPE
\begin{equation}
    g(\omega,k) =\langle \phi(\omega,k) \phi(-\omega,-k)\rangle_\beta \overset{\omega \to \infty}{\sim} \sum_{\mathcal O} a_{\mathcal O} \tilde f_{\Delta,J}(\omega_n,k)\label{eq:momentumblocks} \ ,
\end{equation}
where $\tilde f_{\Delta,J}$ represents the Fourier transform of the thermal OPE block.
The blocks have been determined in \cite{Manenti:2019wxs}: 
\begin{equation}
    \tilde f_{\Delta,J}(\omega_n,k) = \sum_{j = 0}^{J/2} c_{J,j}^{(\nu)}\frac{2^{\beta_-}\pi^{\frac{d}{2}}\Gamma(\beta_+)}{\Gamma(\alpha)} \frac{ {}_2F_1\left(\beta_-,1-\beta_-;1-\beta_+;\frac{k+i \omega_n}{2k}\right)}{k^{\beta_-}(k+i \omega_n)^{\beta_+}}+ O(e^{-k})\ ,
    \label{eq:momentumcb}
\end{equation}
where we denote $k = |\vec k|$ and $\omega_n=2\pi n$ are the Matsubara modes.
We also define
\begin{equation}
    \tilde j = \frac{J}{2}-j , \hspace{0.5 cm} \tilde \Delta = \frac{\Delta}{2}-\Delta_\phi  \ , \hspace{0.5 cm} \alpha= \tilde j-\tilde \Delta\ , \hspace{0.5 cm} \beta_\pm = \frac{d}{2}+\tilde \Delta \pm \tilde j \ ,
\end{equation}
and
\begin{equation}
    c_{J,j}^{(\nu)} = (-1)^j \frac{\Gamma(J - j + \nu)}{\Gamma(\nu)\, j!\, (J - 2j)!} \, 2^{J - 2j} \ .
\end{equation}
Notice that all the blocks corresponding to the operators $[\phi\phi]_{n,J}$ with dimensions $\Delta = 2\Delta_\phi+2n+J$ do not appear in the momentum space OPE, and these are the operators which also do not contribute to the dispersion relation as given in \eqref{eq:vanishingdisc}.

In general, up to exponentially suppressed corrections at small momenta, the momentum-space correlator can be transformed back to position space by performing an inverse Fourier transform in flat space, followed by a Poisson resummation to lift the result to the thermal manifold:
\begin{align}
    g_{\text{pert}}(\tau,x) = \sum_{m=-\infty}^\infty \frac{1}{(2\pi)^{d}}  \int_{\mathbb{R}^d} \text{d}\omega\, \text{d}^{d-1}\vec{k}\  g(\omega,\vec{k})\ e^{i(\tau+m)\omega + i\vec{k}\cdot\vec{x}}\ .
    \label{mom-images}
\end{align}
This final expression for the two-point function closely resembles the generalized method of images introduced in Eq. \eqref{eq:GeneralizedMI}.
For each example considered in Section \ref{sec:applications}, we compute both the dispersion relation and the corresponding momentum-space thermal blocks.

In the theories that admit a perturbative expansion in terms of a small parameter (coupling constant, dimensional regularization parameter $\varepsilon$ etc.), we find perfect agreement between the Fourier transform of the momentum-space OPE and the result obtained from the dispersion relation.
This agreement holds not only at the level of the full correlator, but also block by block.
This is not surprising, given that the momentum-space correlator can itself be expressed in terms of the discontinuity of the position-space correlator (see Eq.(4.5) in~\cite{Manenti:2019wxs}). 

The distinctive feature of perturbative correlators is the fact that terms of the form $e^{-k}$ cannot appear: any such contribution would originate from diagrams that behave as exponentials of momenta, which would violate locality in quantum field theory.
As a consequence, there are no non-analytic corrections to the OPE in momentum space, explaining the exact match between the momentum-space OPE and the full correlator reconstructed via the generalized method of images.

When the correlator is considered at the non-perturbative level, corrections of the type $e^{-k}$ may appear both by resummations of diagrams or as renormalon/instanton corrections.
These contributions do not allow to have a full control over the momentum space OPE of the correlator, as shown explicitly in Section \ref{sec:applications} for correlators in two dimensions. 

Nonetheless, we emphasize that the exact knowledge of the correlator provided by the generalized method of images offers a constructive way to compute the exponential corrections to the OPE blocks.
As previously noted, the method of images applied to the inversion of each block $g_{\text{dr}}^{(\mathcal{O})}(r,w)$ can be interpreted as a KMS-symmetric inversion of the contribution from the operator $\mathcal{O}$.
In perturbation theory, this quantity coincides — upon Fourier transformation — with the momentum-space OPE block associated with $\mathcal{O}$. 
This observation naturally suggests a non-perturbative extension: we propose that the Fourier transform of Eq.\eqref{eq:GeneralizedMI}, applied to a single block, defines the momentum-space OPE block associated with the operator $\mathcal{O}$ beyond perturbation theory.
Explicitly, this leads to the following non-perturbative definition of the momentum-space OPE block:
\begin{equation}\label{eq:momspaceopenew}
    \tilde{f}_{\Delta,J}(\omega_n, \vec{k}) = \int_0^1 \mathrm{d}\tau \, e^{i \omega_n \tau} \int_{\mathbb{R}^{d-1}} \mathrm{d}^{d-1} \vec{x} \, e^{i \vec{k} \cdot \vec{x}} \left( \frac{1}{2} \sum_{m = -\infty}^\infty g_\text{dr}^{(\mathcal{O})}(\tau - m, x)+ g_\text{arcs}^{(\mathcal O)}(\tau,x) \right) \,,
\end{equation}
where $g_\text{arcs}^{(\mathcal O)}(\tau,x)$ is the contribution of $g_\text{arcs}$ of the operator $\mathcal O$ ($J_{\mathcal O}<J_\star$), $\omega_n = 2\pi n$, and where $\tilde{f}_{\Delta,J}$ is not defined as the Fourier transform of the thermal block in the original OPE, but matches it in perturbative theories.

\subsubsection{Momentum space and the  \texorpdfstring{$\tau$}{\texttwoinferior}-plane}

In Eq.~\eqref{eq:dispcomplextau}, we showed that thermal two-point functions at zero spatial separation can be decomposed into generalized free field correlators.
In this way, the correlator admits a block decomposition where each block is individually KMS invariant.
This structure is particularly convenient, as it allows us to perform a discrete Fourier transform of each block separately.

Consider the block associated with an operator of conformal dimension $\Delta$
\begin{equation}
    g^{(\Delta)}(\tau) = a_{\Delta} \Big[ \zeta_H(2\Delta_\phi - \Delta, \tau) + \zeta_H(2\Delta_\phi - \Delta, 1 - \tau) \Big] \ .
\end{equation}
The discrete Fourier transform of this function yields
\begin{equation}
    \tilde{g}^{(\Delta)}_n = a_{\Delta} \, \frac{\pi \, \sec\left[ \frac{\pi(2\Delta_\phi - \Delta)}{2} \right]}{\Gamma(2\Delta_\phi - \Delta)} \, |\omega_n|^{2\Delta_\phi - \Delta - 1} \ .
\end{equation}
Summing over all contributions, we obtain an expression for the generic Fourier mode
\begin{equation}
    \tilde{g}_n = \sum_{\Delta} \tilde{g}^{(\Delta)}_n 
    = \pi \sum_{\Delta} a_{\Delta} \, \frac{\sec\left[ \frac{\pi(2\Delta_\phi - \Delta)}{2} \right]}{\Gamma(2\Delta_\phi - \Delta)} \, |\omega_n|^{2\Delta_\phi - \Delta - 1} \,.
\end{equation}
This equation is the exact analog of \eqref{eq:momspaceopenew} in the case where the correlator is evaluated at zero spatial separation.
Indeed, the Fourier coefficients computed here correspond precisely to the spatial-momentum ($\vec k$) integral of the momentum-space OPE blocks defined in~\eqref{eq:momspaceopenew}.
This procedure showcases the power of Eq.~\eqref{eq:dispcomplextau} and its equivalent form~\eqref{eq:momspaceopenew}, as it allows for the straightforward computation of observables — such as the low-frequency contribution to thermal correlators — that were previously difficult to access.
   
   Interestingly, if we perform the Fourier transform along the shifted imaginary $\tau$ direction, corresponding to real time, the Fourier transform is not discrete. Concretely we perform the Fourier transform in $t$, where $\tau = 1/2+ it$ (see Eq. (4.2) of \cite{Dodelson:2023nnr}). In this case, the result for a block associated with conformal dimension $\Delta$ corresponds to the two-sided Wightman correlator, given by:
   \begin{equation}\label{eq:momspace}
       \tilde g^{(\Delta)}(\omega)=\frac{\pi\, a_\Delta}{\Gamma (2\Delta_\phi - \Delta)}  \left[\coth \left(\frac{| \omega | }{2}\right)+\frac{\omega}{|\omega|}\right] | \omega | ^{2\Delta_\phi - \Delta-1} e^{-\frac{\omega}{2}}\ .
   \end{equation}
 This expression differs from the thermal (Matsubara) case due to the absence of periodicity in imaginary time. From this form we can extract the imaginary part of the retarded Green function in frequency space for each operator contributing to the thermal two-point function. It would be interesting to understand the connection of this formula with the quasi-normal modes in holographic theories, as discussed in \cite{Dodelson:2023nnr}. Note that in the limit $|\omega| \to \infty $, Eq. \eqref{eq:momspace} reproduces the known OPE limit \cite{Caron-Huot:2009ypo,Dodelson:2023nnr} \begin{equation}
    \tilde g^{(\Delta)}(\omega) \sim a_{\Delta} e^{-\frac{\omega}{2}} |\omega|^{2\Delta_\phi-\Delta-1} \ .
 \end{equation}
 It would be interesting to extend this formula for the two-point functions at non-zero spatial separation to obtain the Fourier transform $\tilde g(\omega,k)$.

\section{Applications}
\label{sec:applications}

This section is dedicated to applying the formulae and methodology developed in this work to concrete models.

\subsection{Generalized free theories}
\label{subsec:GeneralizedFreeTheories}

The simplest case to study is the generalized free scalar theory.
This theory features a fundamental scalar field $\phi$ of scaling dimension $\Delta_\phi$, satisfying the equation of motion
\begin{equation}
    \Box^{\frac{d}{2}-\Delta_\phi} \, \phi = 0.
\end{equation}
The theory is non-local unless $\Delta_\phi = d/2 - 1$, which corresponds to the standard free scalar field theory.
The OPE expansion of two fundamental fields 
can be schematically represented as follows:
\begin{equation}
    \phi \times \phi = \mathds{1} + [\phi \phi]_{n,J} \ ,
    \label{eq:opeGFF}
\end{equation}
where $[\phi \phi]_{n,J}$ denotes the double-twist operators, of the form $\phi \Box^n \partial_{\mu_1} \cdots \partial_{\mu_J} \phi$, with conformal dimension $\Delta = 2\Delta_\phi + 2n + J$.

This theory is one of the few exactly solvable models at finite temperature.
Since it is non-interacting, the thermal two-point function is given by the method of images:\footnote{For the sake of clarity, we remark the conceptual difference between Eq.~\eqref{exactGFF}, where the sum is over images of the free propagator, and Eq.~\eqref{eq:GeneralizedMI}, where the sum is over images of the dispersion relation.}
\begin{equation}
    g(\tau,x) = \langle \phi(\tau,x) \phi(0,0) \rangle_\beta = \sum_{m = -\infty}^\infty \frac{1}{[(\tau + m)^2 + x^2]^{\Delta_\phi}} \ .\label{exactGFF}
\end{equation}

\paragraph{The complex \texorpdfstring{$\tau$}{tau}-plane.}
We first test Eq.~\eqref{eq:dispcomplextau}.
Odd-spin double-twist operators do not contribute to the discontinuity due to parity symmetry; using the identity
\begin{equation}
    \zeta_H(-2n, \tau) + \zeta_H(-2n, 1 - \tau) = 0 \ , \qquad \forall\, n\in\mathbb{Z}^{>0} \ ,
    \label{eq:VanishingIdentity}
\end{equation}
we conclude that even-spin double-twist operators do not contribute as well, since GFF is non-interacting and anomalous dimensions do not arise.
We conclude that the identity operator is the only operator contributing to the discontinuity.
The formula \eqref{eq:dispcomplextau} becomes
\begin{equation}
    g(\tau) = \zeta_H(2\Delta_\phi, \tau) + \zeta_H(2\Delta_\phi, 1 - \tau) \ ,
\end{equation}
matching the known result from \eqref{exactGFF}
\begin{equation}
    g(\tau) = \sum_{m = -\infty}^\infty \frac{1}{|\tau + m|^{2\Delta_\phi}} = \zeta_H(2\Delta_\phi, \tau) + \zeta_H(2\Delta_\phi, 1 - \tau) \ .
\end{equation}

\paragraph{The non-zero spatial separation.}
As observed in the previous paragraph, the identity operator is the only contribution in the OPE regime\footnote{With a slight abuse of notation, here and in what follows we use the notation $\operatorname{Disc}[\mathcal{O}]$ to indicate the contribution to the discontinuity given by the thermal block associated with the operator $\mathcal{O}$ in the OPE.}
\begin{equation}
    \Disc [\mathds{1}] = -2i \sin(\pi \Delta_\phi)\, r^{2\Delta_\phi} \left(\frac{r}{w'} - 1\right)^{-\Delta_\phi} (1 - rw')^{-\Delta_\phi} \Theta\left[\text{Re}\left(\frac{r}{w'}\right) - 1\right] \ .
\end{equation}
This contribution must be integrated against the kernel in Eq. \eqref{eq:DRw}.
Although a full analytic integration is challenging, one can perform a power series expansion in $r$ and reconstruct the result term by term. Switching to the coordinates $(z, \zb)$, the outcome is
\begin{equation}\label{eq:gdrfree}
    g_{\text{dr}}(z, \zb) = \frac{1}{(z - 1)^{\Delta_\phi} (\zb - 1)^{\Delta_\phi}} + \frac{1}{(z + 1)^{\Delta_\phi} (\zb + 1)^{\Delta_\phi}}\ .
\end{equation}
As expected, this expression is not KMS invariant. The OPE coefficients extracted from this correlator do not match those of the full thermal correlator.
In fact, by expanding \eqref{eq:gdrfree} in thermal blocks, one recovers the same result obtained from the Lorentzian inversion formula (LIF) applied to the discontinuity of the identity block, as detailed in Appendix \ref{app:LIF}.

\begin{table}[t]
    \centering
    \caption{Thermal OPE coefficients in the $4d$ free scalar theory extracted by expanding the exact correlator, using the dispersion relation (DR), which is shown to match the result of the Lorentzian inversion formula (LIF) in Appendix \ref{app:LIF}, and generalized method of images (GMI). Only even-spin operators contribute.}
    \renewcommand{\arraystretch}{1.5}
    \begin{tabular}{|c|c|c|}
    \hline
    $\mathcal O$ & $a_{\mathcal O}^\text{(exact)}$ = $a_{\mathcal O}^\text{(GMI)}$  & $a_{\mathcal O}^\text{(DR) }$  \\
    \hline
    $\mathds{1}$ & 1 & 0  \\
    \hline
    $[\phi \phi]_{0,J}$ & $2\zeta(2 + J)$ & 2  \\
    \hline
    \end{tabular}
    \label{tab:drFree}
\end{table}

\paragraph{The generalized method of images.}
We now wish to test the formula \eqref{eq:GeneralizedMI}.
For simplicity, we focus on the $4d$ free scalar theory, with $\Delta_\phi = 1$.
We first perform the sum over the images to obtain
\begin{equation}\label{eq:39}
   \frac{1}{2} \sum_{m = -\infty}^\infty g_{\text{dr}}(z - m, \zb - m)= - \frac{\pi}{(z - \zb)} \left[ \cot\left( \pi z \right) - \cot\left( \pi \zb \right) \right] \ .
\end{equation}
This expression already encodes the correct clustering behaviour at infinite distance, hence we conclude that the sum over images already solves the thermal bootstrap problem without any arc correction.

One may ask whether the usual method of images, Eq. \eqref{exactGFF}, coincides with the generalized method of images, Eq. \eqref{eq:39}: the two methods can be shown to be equivalent, since in the case of GFF, the dispersion relation can be reshaped into the sum over the images of the free propagator, by using the relation
\begin{equation}
    \sum_{m = -\infty}^\infty \frac{1}{(z - m + n)^{\Delta_\phi} (\zb - m + n)^{\Delta_\phi}}
    =
    \sum_{m = -\infty}^\infty \frac{1}{(z - m)^{\Delta_\phi} (\zb - m)^{\Delta_\phi}} \ ,
\end{equation}
for any integer $n$, by a simple change of summation variable.

\paragraph{The momentum space interpretation.}
We compute the contributions of the momentum space conformal blocks to the momentum space correlator by using Eq.~\eqref{eq:momentumblocks}.
Out of all the operators that are contributing to the OPE \eqref{eq:opeGFF}, only the identity operator has a non-zero contribution in momentum space:
\begin{align}
    \tilde f_{0,0}(\omega_n, \vec{k})= \frac{4\pi^{2}}{\left(k^2+\omega_n^2\right)}\ .
\end{align}
In particular, exponential corrections $e^{-k}$ are not present, since the theory is free.
As a check, we transform the $4d$ momentum-space correlator back to position space, expecting to recover the exact known result.
To do so we perform the four-dimensional inverse Fourier transformation on flat space $\mathbb R^4$, followed by a Poisson resummation 
to recover the result on $S_{\beta}^1\times\mathbb R^3$.
The result is

\begin{equation}
     g(\tau,x)=\frac{\pi}{2 x}\left[\coth (\pi( x+i  \tau) )+ \coth (\pi  (x-i  \tau) )\right]\ ,
\end{equation}
which coincides with \eqref{eq:gdrfree} when translated into $z,\zb$ coordinates.

\subsection{\texorpdfstring{$\mathrm O(N)$}{\texttwoinferior} model}
\label{subsec:ONModel}

We discuss here the $\mathrm{O}(N)$ model, which has a number of interesting physical applications and is considered one of the simplest models to study; for a broad review of the subject (at zero temperature), see \cite{kleinert2001critical,Henriksson:2022rnm,Giombi:2012ms} and references therein.
We follow two approaches: the $\veps$-expansion around $d=4$, and the large $N$ theory, which can be solved in $3d$.

The model can be defined starting from the Ginzburg–Landau action
\begin{equation}\label{eq:GLON}
   \mathcal{A} = \int \text{d}^{d}x  \left[\frac{1}{2}(\partial_\mu \phi_i)^2 + \frac{\lambda}{4!} (\phi_i \phi_i)^2\right] \ ,
\end{equation}
where $i = 1, \ldots, N$, and the upper critical dimension is $d = 4$.
In practice, this means that a possible definition of the $\mathrm{O}(N)$ fixed point is as the infrared fixed point of the Lagrangian in Eq. \eqref{eq:GLON}, in the range $2 < d < 4$.
In $d = 4$, the only fixed point is the free one ($\lambda = 0$), while in $d = 4 - \varepsilon$ the Wilson--Fisher fixed point occurs at $\lambda_\star \sim \varepsilon$~\cite{Wilson:1971dc}.
For this reason, the theory can be studied perturbatively using an expansion in $\varepsilon$.
We carry out this analysis at finite temperature in Section~\ref{sec:pertON}.

Far from $d = 4$, the theory becomes strongly-coupled, and in general it is very difficult to predict finite-temperature quantities in the physical dimension $d = 3$.
For $N = 1, 2, 3$, numerical results from bootstrap are available in~\cite{Iliesiu:2018zlz,Barrat:2024fwq}.
A major simplification arises in the limit $N \to \infty$, where the action~\eqref{eq:GLON} becomes
\begin{equation}\label{eq:LNlag}
    \mathcal{A} = \int \text{d}^{3}x  \left[\frac{1}{2}(\partial_\mu \phi_i)^2 + \frac{1}{2}\sigma (\phi_i \phi_i)-\frac{\sigma^2}{4 \lambda  }\right] \ ,
\end{equation}
with $\sigma$ the Hubbard–Stratonovich field. The critical point is located at $\lambda \to \infty$, so the last term is suppressed at large $N$.
This theory is free and massive, with $\langle \sigma \rangle_{\beta}$ playing the role of an effective mass.
We discuss this limit at finite temperature in Section~\ref{sec:largeN}.

\subsubsection{\texorpdfstring{$\veps$}{\texttwoinferior}-expansion}
\label{sec:pertON}


In this section we bootstrap the two-point function $\langle\phi \phi\rangle_\beta$ at first order in the parameter $\veps$.
This operator does not receive an anomalous contribution at this order and the scaling dimension corresponds to the free point
\begin{equation}
    \Delta_\phi = 1 - \frac{\veps}{2} + O (\veps^2)\ .
    \label{eq:ScalingDimPhi}
\end{equation}
One can perform an $\veps$-counting of the two and three-point functions to convince ourselves that the only operators relevant at first order are the identity $\mathds{1}$, the higher-spin currents $[\phi \phi]_{0,J}$ and the operator $[\phi \phi]_{1,J}$.\footnote{See for instance Section 2.1 of \cite{Gimenez-Grau:2022ebb} for a detailed counting.}
The latter classically corresponds to $\phi \partial^J \Box \phi$: at the free point, i.e., $d = 4$, equations of motions makes them zero vectors in the Hilbert space of the theory.
Nonetheless the equations of motions for the $\mathrm O(N)$ model gives $\phi \partial^J \Box \phi\sim \lambda \phi \partial^J \phi^3 \sim  \varepsilon \phi \partial^J \phi^3$, recalling that the critical value of the quartic coupling is proportional to $\varepsilon$. Because of equations of motion, operators of the type $[\phi \phi]_{n,J}$ for $n >1$ will only appear at higher order in $\varepsilon$.

The last ingredient that we need to consider are the anomalous dimensions.
Due to the fact that OPE coefficients for $[\phi \phi]_{1,J}$ start at order $O(\veps)$, we only need to consider the corrections to the operators of $[\phi \phi]_{0,J}$.
In fact, only the operator $\phi^2$ (corresponding to $[\phi \phi]_{0,0}$) receives an anomalous contribution at this order, meaning that most of the terms in \eqref{eq:dispcomplextau} and \eqref{eq:GeneralizedMI} will simply vanish \cite{kleinert2001critical,Henriksson:2022rnm}.

Note that it is also possible to calculate the correlator directly through Feynman diagrams.
We perform this analysis in Appendix \ref{subsec:ComparisonToDiagrammaticCalculation} to compare with our bootstrap result.

\paragraph{The complex $\tau$-plane.}
We first test Eq. \eqref{eq:dispcomplextau}.
Due to the $\veps$-counting above, the two-point function in complex $\tau$-plane consists of only two contributions:
\begin{align}
    g(\tau)=g^{(\mathds{ 1})}(\tau)+g^{(\phi^2)}(\tau)\ .
\end{align}
The identity contribution is simple and is again given by
\begin{equation}
    g^{(\mathds{1})}(\tau) = \zeta_H\left(2-\varepsilon,\tau\right)+\zeta_H\left(2-\varepsilon,1-\tau\right)  \ .
\end{equation}
The contribution of $\phi^2$ is new and corresponds to 
\begin{equation}
   g^{(\phi^2)}(\tau)=
   \zeta_H\left(- \veps \gamma_{\phi^2},\tau\right)+\zeta_H\left(-  \veps \gamma_{\phi^2},1-\tau\right) = - \veps\, a_{2}^{(0)} \gamma_{\phi^2} \log \left[\frac{ \csc(\pi \tau)}{2} \right]+O(\varepsilon^2) \ ,
\end{equation}
where $a_{2}^{(0)}$ corresponds to the coefficient $a_2$, i.e., $a_\Delta$ with $\Delta = 2$, in the $4d$ free theory, while $\gamma_{\phi^2}$ refers to the anomalous dimension of the operator $\phi^2$ at first order in $\varepsilon$, which is known to be \cite{kleinert2001critical,Sachdev:2011fcc}
\begin{equation}
    \gamma_{\phi^2} = \frac{N+2}{N+8}\label{eq:on_anomolous} \ .
\end{equation}
The correlator obtained differs from the free correlator in $d=4-\varepsilon$ as 
\begin{equation}
    g(\tau)-g_\text{free}(\tau)=- \veps\, a_{2}^{(0)} \gamma_{\phi^2} \log \left[\frac{ \csc(\pi \tau)}{2} \right]+O(\varepsilon^2) \ .
\end{equation}
This matches exactly the diagrammatic computation presented in Appendix \ref{subsec:ComparisonToDiagrammaticCalculation}.

\paragraph{Non-zero spatial separation.}
Eq. \eqref{eq:gdrfree} already presents the contribution of the identity block to $g_{\text{dr}}$ for any value of $\Delta_\phi$ and any spacetime dimension.
Separating the first order in $\varepsilon$, we have 
\begin{equation}
    g_{\text{dr}}^{(\mathds{1})}
    =\frac{\varepsilon(z \zb+1)}{\left(z^2-1\right)
   \left(\zb^2-1\right)}
    \log \left[\frac{(z-1)(\zb-1)}{z \zb}\right]+ \frac{\varepsilon}{\left(z+1\right)
   \left(\zb+1\right)}\left[\tanh ^{-1}(z)+ \tanh ^{-1}(\zb) \right]\ .
\end{equation}

The only new contributions come from $\phi^2$.
The calculation of the discontinuity is straightforward:
\begin{equation}
    \operatorname{Disc}[\phi^2] = i \pi \, \gamma_\phi \,   z  \zb \, \Theta (\text{Re}(\zb)-1) \ .
\end{equation}

The integration against the kernel is again non-trivial but can be performed, as in the free case, switching to $(w, r)$ coordinates and expanding order by order in $r$, from which we can guess the full closed form.
The result is
\begin{equation}
    g_{\text{dr}}^{(\phi^2)} = \frac{1}{2}  \veps\, a_{\phi^2}^{(0)} \gamma_{\phi^2} \left[\log(1-z^2)+\log(1-\zb^2)\right] \ ,
\end{equation}
where $a_{\phi^2}^{(0)}$ is the thermal OPE coefficient at order $O(\varepsilon^0)$.
We can expand these results in thermal blocks, and obtain the OPE coefficients given in Table \ref{tab:drepsilon}. 

Let us further observe that by solving the KMS condition \eqref{eq:KMSCondition}, one also finds that
\begin{equation}
    a_{\phi^2}^{(0)} = \frac{2}{1-\gamma_{\phi^2}} \ .
\end{equation}
This equation is however not consistent with the previous findings: $a_{\phi^2}^{(0)}$ is a free theory quantity that should not depend on $\gamma_{\phi^2}$.
We interpret this result as a consequence of the fact that contributions to $\phi^2$ appear in the arcs, and therefore it has to be treated separately.

Note furthermore that the quantity $a_{\phi^2}^{(0)}$ is scheme-dependent, being the constant part of the correlator and of the sum of images. To work around this scheme dependence, it is best to extract this OPE coefficient from a correlator in which $\phi^2$ does not contribute as the constant term, for instance $\vev{\phi^2 \phi^2}_\beta$. From
\begin{equation}
    \vev{\phi^2 (\tau,x) \phi^2 (0,0)}^{(0)}_\beta = 2 (\vev{\phi (\tau,x) \phi (0,0)}^{(0)}_\beta)^2\,,
\end{equation}
and
\begin{equation}
    \frac{f_{\phi\phi\phi^2}^{(0)}}{f_{\phi^2\phi^2\phi^2}^{(0)}} = \frac{1}{2}\,,
\end{equation}
we find
\begin{equation}
    a_{\phi^2}^{(0)}
    =
    8 \zeta_2\,.
\end{equation}

\begin{table}[t]
    \centering
    \caption{Thermal OPE coefficients at order $\mathrm{O}(\veps)$ in the OPE between the two fundamental scalars in the free theory computed by expanding the exact correlators in blocks, using the dispersion relation (DR), generalized method of images (GMI), which matches the results of the Lorentzian inversion formula (LIF) by computing the discontinuity in the OPE (see Appendix \ref{app:LIF}). In the latter cases the only contribution to the discontinuity is given by the identity contribution. It is given for granted that $J \in 2\mathbb N$: odd spin operators have zero one-point functions.  The operator $\phi^2$ is not included but it is discussed in the main text.}
    \renewcommand{\arraystretch}{1.5}
    \begin{tabular}{|c| c| c|}
        \hline
        $\mathcal O$ & $a_{\mathcal O}^\text{(exact)} = a_{\mathcal O}^\text{(GMI)}$  &  $a_{\mathcal O}^\text{(DR)}$  \\  \hline
        $\mathds{1}$ & 1 & 0  \\ \hline
        $[\phi \phi]_{0,J>0}$ & $-2 \zeta'_{2+J}-\frac{1}{J} a_{\phi^2}^{(0)}\gamma_{\phi^2}\zeta_J$ & $-\frac{1}{J}a_{\phi^2}^{(0)}\gamma_{\phi^2}$  \\ \hline
        $[\phi \phi]_{1,J}$ & $\frac{1}{J+2}\zeta_{2+J} a_{\phi^2}^{(0)}\gamma_{\phi^2}$ & $\frac{1}{J+2}a_{\phi^2}^{(0)}\gamma_{\phi^2}$  \\ \hline
    \end{tabular}
    \label{tab:drepsilon}
    \end{table}
    
\paragraph{The generalized method of images.}
We can now perform the sum over images of the dispersion relation as prescribed in Eq. \eqref{eq:GeneralizedMI}. 
We are only interested in $g(\tau)-g_\text{free}(\tau)$ since the free part can be exactly solved in $4-\varepsilon$ dimensions.
We perform the sum:
\begin{equation}\label{eq:sumoverphi2}
    \frac{1}{2}\sum_{m = -\infty}^\infty g_\text{dr}^{(\phi^2)}(z-m,\zb-m) = \frac{\veps}{4}
    a_{\phi^2}^{(0)} \gamma_{\phi^2}
    \sum_{m = -\infty}^\infty \left[\log(1-(z-m)^2)+\log(1-(\zb-m)^2)\right]\ ,
\end{equation}
which is strictly speaking diverging on the real axis.
It can however be obtained as a result of an analytic continuation. 
By using the fact that
\begin{equation}
    \log(1 + (x-m)) = -\left. \frac{\partial}{\partial s}\frac{1}{(1+x-m)^s}\right|_{s \to 0 } \ ,
\end{equation}
we find that the following sum can be regularized via (Hurwitz) $\zeta$-function regularization, and the real part reads
\begin{equation}
    \sum_{m = -\infty}^\infty  \log(1 + (x-m)) = -\log\left[ \frac{\csc(\pi x)}{2}\right]\ .
    \label{eq:RegularizedSum}
\end{equation}
Using this result, it is easy to show that the sum \eqref{eq:sumoverphi2} gives the exact same correlator as in  \eqref{eq:perturbativeepislonres}, apart from a constant 
\begin{equation} \label{eq: finalcorrzz}
    g(z,\zb)-g_\text{free}(z,\zb)
    =
    \frac{\veps}{2}\sum_{m = -\infty}^\infty g_\text{dr}^{(\phi^2)}(z-m,\zb-m) =
     -\frac{\veps}{2}a_{\phi^2}^{(0)} \gamma_{\phi^2}\log\left[ \frac{\csc(\pi z) \csc(\pi \zb)}{4}\right]\,.
\end{equation}
This constant can be understood as the contribution of the arc coming from the block of $\phi^2$ and cannot be constrained by KMS, as it corresponds to a constant term.

\paragraph{The momentum space interpretation.} 
Similarly to the dispersion relation, at the order $O(\veps^0)$ only the identity operator has a non-vanishing contribution.
At the order $O(\veps)$, it was computed in \cite{Manenti:2019wxs} that both identity and the $\phi^2$ operators contribute:
\begin{align}
    \tilde f_{0,0}(\omega_n,k)&=4^{\Delta_\phi} \pi^{2-\frac{\veps}{2}} \left(\frac{1}{\Gamma \left(\Delta_\phi\right)}\right) \frac{1}{\left(k^2+\omega_n^2\right)^{\Delta_\phi}}\ , \\
    \tilde f_{2+\gamma_{\phi^2},0}(\omega_n,k)&=4^{2\Delta_\phi+\veps}\gamma_{\phi^2} \pi^{2-\frac{\veps}{2}} \left(\frac{ \Gamma \left(2\Delta_\phi+\veps\gamma_{\phi^2}\right)}{\Gamma \left(-\frac{\veps}{2}\gamma_{\phi^2}\right)}\right)\frac{1}{\left(k^2+\omega_n^2\right)^{2\Delta_\phi+\veps}\gamma_{\phi^2}}\ .
\end{align}
Therefore the inverse Fourier transform back to position space gives the contribution 
\begin{align}
    f_{0,0}(\tau,x) =&\frac{\pi}{2 x}\left[\coth (\pi( x+i  \tau) )+ \coth (\pi  (x-i  \tau) )\right]\notag\\
    &+\frac{\veps}{4}\sum _{m=-\infty }^{\infty } \frac{\log \left[(\tau -m)^2+x^2\right]+\gamma_E -2-\log \pi}{(\tau -m)^2+x^2} +O(\veps^2)\ ,
\end{align}
while the $\phi^2$ operator contributes as follows
\begin{equation}
    f_{2+\gamma_{\phi^2},0}(\tau,x)= \frac{\veps}{4} \gamma_{\phi^2} \sum_{m=-\infty}^\infty \Big( \log \left[(\tau -m)^2+r^2\right]-2\log 2+2\gamma_E \Big) +O(\veps^2) \ .
\end{equation}
We perform the Poisson resummation by using the $\zeta$-function regularization and we match the result coming from the perturbative computation and generalized method of images as given in Eq.~\eqref{eq:GeneralizedMI}.

\subsubsection{Large \texorpdfstring{$N$}{\texttwoinferior}}
\label{sec:largeN}

We now study the large $N$ limit of the $\mathrm{O}(N)$ model at finite temperature.
The relevant operators in the OPE between two fundamental scalars are the double-twist $[\phi \phi]_{n,\ell}$ and the operators $\sigma^m$,\footnote{Due to the equations of motion, we have $\partial^2 \phi_i \sim \sigma \phi_i$. As a consequence, degeneracies in the conformal dimension may arise between additional operators and those in the families $[\phi \phi]_{n,\ell}$ and $\sigma^m$. For instance, certain operators in the family $[\phi_i \sigma \phi_i]_{0,\ell}$ can be degenerate with operators in the family $[\phi \phi]_{n,\ell}$.} where the spectrum in $d = 3$ is given by
\begin{equation}\label{eq:scalon3}
    \Delta_\phi
    =
    \frac{1}{2} + O\left(\frac{1}{N}\right) \ ,
    \qquad
    \Delta_\sigma
    =
    2 + O\left(\frac{1}{N}\right) \ .
\end{equation}
Moreover, any anomalous dimension of composite operators is suppressed in $1/N$. In the following, we will make use of the golden ratio $\varphi=\frac{1+\sqrt{5}}{2}$.

\paragraph{The complex $\tau$-plane.}
The first step is to understand how the two-point function can be reconstructed in the complex $\tau$-plane. We aim to apply Eq.~\eqref{eq:dispcomplextau} to the spectrum indicated above.
It is useful to recall the identity
\begin{equation}
    \zeta_H(2\Delta_\phi - \Delta, \tau) + \zeta_H(2\Delta_\phi - \Delta, 1 - \tau) = \sum_{n = -\infty}^\infty |\tau + n|^{\Delta-2\Delta_\phi} \ ,
\end{equation}
from which we can compute first the sum over operators, and then the sum over images (i.e., over $n$ in the equation above).
This strategy turns out to be particularly convenient in our case.

Let us consider the operators $\sigma^m$, which are the only ones contributing to the discontinuity.
We obtain
\begin{equation} \label{eq: summ}
    g(\tau) - \kappa = \sum_{n = -\infty}^\infty \sum_{m=0}^{\infty} a_{\sigma^m} |\tau + n|^{2m - 1} = \sum_{n = -\infty}^\infty \frac{\cosh(m_\text{th} |\tau + n|)}{|\tau + n|} \,,
\end{equation}
where the thermal coefficients are given by $a_{\sigma^m} = m_\text{th}^{2m} / \Gamma(2m + 1)$ \cite{Iliesiu:2018fao} and the thermal mass is $m_\text{th} = \log \varphi^2$ \cite{Sachdev:1992py}. Notice that in the computation above we identified $\sigma^0=\mathds{1}$.

Having computed the sum over operators, we now consider the sum over images.
This sum is clearly divergent and needs to be regularized.
To do so, we split the hyperbolic cosine into exponentials, perform the sum, and then analytically continue in the exponential variable, rendering the sum convergent.
Following this procedure, one can verify that the final result is
\begin{equation} \label{eq: betaeuler}
    g(\tau) = \text{B}_{1/\varphi^2}(\tau, 0) + \text{B}_{1/\varphi^2}(1 - \tau, 0) \ , 
\end{equation}
where the incomplete Euler B-function appears. This result is a reformulation of the result presented in~\cite{Iliesiu:2018fao,Marchetto:2023xap}. Notice that the equation \eqref{eq: betaeuler} should be considered only in the strip $0< \text{Re}(\tau) < 1$, since this assumption was needed to perform the sum in Eq. \eqref{eq: summ}.
    
\paragraph{Non-zero spatial coordinates.}
We now consider the case of non-zero spatial coordinates.
To do so, we start from the thermal blocks corresponding to the operators $\sigma^m$ and compute the outcome of the dispersion relation when inverting a single operator.
This gives
\begin{equation}
    g_\text{dr}^{(\sigma^m)}(z,\zb) =\frac{m_\text{th}^{2m}}{\Gamma(2m+1)}\left\lbrace \frac{1}{\left[(1-z)(1-\zb)\right]^{\frac{1}{2}-m}}+\frac{1}{\left[(1+z)(1+\zb)\right]^{\frac{1}{2}-m}}\right \rbrace \ .
\end{equation}
The contribution above can be resummed together with the identity contribution.
This results in
\begin{multline}
    g_{\text{dr}}(z,\zb)
    =
    \frac{1}{\sqrt{(1-z)(1-\zb)}} \coth\left(m_\text{th} \sqrt{(1-z)(1-\zb)}\right)  \\
    +\frac{1}{\sqrt{(1+z)(1+\zb)}} \coth\left(m_\text{th} \sqrt{(1+z)(1+\zb)}\right) \ .
    \label{eq:zzlargeN}
\end{multline}
The expansion of the coefficients above gives the large spin contribution to the operators in the OPE. In order to complete the result we need to consider the sum over generalized images.
    
\paragraph{The generalized method of images and momentum space.}
Applying the generalized method of images to the result of Eq.~\eqref{eq:zzlargeN} requires regularization, as already observed in the case of the $\mathrm{O}(N)$ model in the $\varepsilon$-expansion.
The regularization appears to be more subtle in the present case.
To verify the correctness of the result, it is more efficient to compute the Fourier transform of the images of the individual blocks, and to show that they reproduce the correct massive propagator, with mass given by $m_\text{th}$.

For a single term corresponding to the family $\sigma^m$, we have
\begin{multline}
    g^{(\sigma^m)}(z,\zb)
    =
    \frac{m_\text{th}^{2m}}{2\Gamma(2m+1)} \sum_{n = -\infty}^\infty  \Bigg[\frac{1}{\left((1 - z - n)(1 - \zb - n)\right)^{\frac{1}{2} - m}}  \\
    + \frac{1}{\left((1 + z - n)(1 + \zb - n)\right)^{\frac{1}{2} - m}} \Bigg] \ .
\end{multline}
This corresponds to the standard method of images for a generalized free field of dimension $\Delta_\phi=1/2 - m$.
Retaining the appropriate prefactors, the Fourier transform of this expression is 
\begin{equation}
    \tilde{g}^{(\sigma^m)}(\omega_n, k) =4\pi (-1)^{m+1} \frac{ m_\text{th}^{2m}}{(k^2 + \omega_n^2)^{m+1}} \ .
\end{equation}
This coincides with the corresponding block computed using the momentum-space OPE.

In fact, the Fourier transform of the thermal blocks is non-vanishing only for the identity operator and the operators $\sigma^m$, whose scaling dimensions are given 
by $\Delta_{\sigma^m}=2m$.
As expected those operators are the same ones contributing to the discontinuity.
The corresponding Fourier blocks are~\cite{Manenti:2019wxs}
\begin{equation}\label{eq:largeNblocks}
    \tilde{f}_{\mathds{1}}(\omega_n,k) = \frac{4\pi}{k^2 + \omega_n^2} \ , \quad 
    \tilde{f}_{\sigma^m}(\omega_n,k) = \frac{4\pi \cos(\pi m) \, \Gamma(2m + 1)}{(k^2 + \omega_n^2)^{m+1}} \ .
\end{equation}
It is now possible to resum all blocks explicitly, and the final result reads
\begin{align}\label{eq:momptopN}
    \tilde{g}(k, \omega) = \frac{4\pi}{k^2 + \omega^2 + m_\text{th}^2} \ .
\end{align}
This is manifestly the correct propagator, consistent with the Lagrangian in Eq.~\eqref{eq:LNlag}, since $\langle \sigma \rangle_\beta=m_{\text{th}}^2$.
Performing the inverse Fourier transform to position space yields
\begin{align}
    \frac{1}{(2 \pi)^3}\int_0^\infty \text{d}k \, k^2 \int_0^\pi \text{d}\theta \, \sin\theta \, e^{-i k r \cos\theta} \, \tilde{g}(\omega_n,k) 
    = \frac{e^{-m_\text{th} r}}{r} \ .
\end{align}
The full thermal two-point function in position space is therefore given by
\begin{align}\label{eq:proplargeN}
    g(\tau, x) = \sum_{m = -\infty}^\infty \frac{ e^{-m_\text{th} \sqrt{(\tau - m)^2 + x^2}}}{\sqrt{(\tau - m)^2 + x^2}} \ ,
\end{align}
which matches the exact result obtained in~\cite{Iliesiu:2018fao}.
As expected, because the $\mathrm{O}(N)$ model at large $N$ is a quasi-free theory, the momentum-space OPE is sufficient to reconstruct the full thermal correlator.
The dispersion relation reproduces precisely the same result. 
Since the clustering condition is satisfied, note that we also have $g_{\text{arcs}} = 0$, i.e., the problem is fully solved by the generalized method of images. 

The thermal mass of the system -- namely, the rate of decay of the correlator along spatial directions at large distances -- is encoded in the discontinuity.
In this explicit example it arises from the contribution of the operators $\sigma^m$, whose resummation produces precisely a mass term, which can be identified to correspond to the thermal mass.
The mechanism is shown explicitly in Equations~\eqref{eq:largeNblocks}--\eqref{eq:momptopN}.

Let us take this example as an opportunity to emphasize that the function $g_{\text{arcs}}$, defined in \eqref{eq:GeneralizedMI}, is not necessarily equal to the missing low-spin contribution in the dispersion relation before applying the method of images. 
Indeed, as we have shown here, $g_{\text{arcs}} = 0$.
However, in this case the arc contribution appearing in \eqref{eq:DRw} is non-zero, as explicitly demonstrated in~\cite{Iliesiu:2018fao, Petkou:2018ynm}: it contains contributions from scalar operators of the form $\sigma^m$.

\subsection{\texorpdfstring{$2d$}{\texttwoinferior} CFTs}
\label{subsec:2dCFTsAndBTW}

It is interesting to test the same methodology of analytic bootstrap also for the case of Virasoro primaries in two spacetime dimensions.
As it is well-known, the two-point functions of identical Virasoro primaries can be derived exactly as a consequence of a conformal map between the plane and the cylinder.
This map is anomalous: in fact, the only non-vanishing thermal one-point functions are the ones of the Virasoro vacuum module since these are the ones which are affected by the conformal anomalies.
For a primary of conformal weights $(h,\hb)$, we have
\begin{equation}\label{eq:2dcorr}
    \vev{\phi(z,\zb) \phi(0,0)}_\beta  = \left(\frac{\pi}{\beta}\right)^{h+\hb} \csc^{h}\left(\frac{\pi}{\beta}z \right)\csc^{\hb}\left(\frac{\pi}{\beta}\zb \right)\ .
\end{equation}
We consider here the case of a scalar primary such that $h = \hb = \Delta_\phi/2$.
The operators in this OPE that develop non-zero one-point functions sit in the vacuum module, meaning that they are such that $ \Delta \ge J$, with $\Delta \in 2\mathbb N$ and $J \in 2 \mathbb  N$.
In the following, we focus on the cases in which $\Delta_\phi = 1, 2$.

It is useful to notice that the thermal blocks defined in general dimension in \eqref{eq:ThermalBlocks} are not well-defined for $d= 2$.
This is due to the fact that the OPE coefficients contain a factor $\Gamma(\nu)$, which is diverging when $\nu = 0$, i.e., $d= 2$.
On the other hand, the blocks contain the Gegenbauer polynomial $C_{J}^{(0)}(\eta)$ which is zero and compensates the infinity of the OPE coefficient.
Therefore, it is convenient to redefine the blocks and OPE coefficients such that
\begin{equation}
    f_{\Delta,J}(z,\zb) = \lim_{\nu \to 0}\frac{J!}{(\nu)_J} z^{\frac{\Delta}{2}-\Delta_\phi}\zb^{\frac{\Delta}{2}-\Delta_\phi} C_{J}^{(\nu)} \left(\frac{z+\zb}{2 \sqrt{z \zb}}\right)\ ,
\end{equation}
and
\begin{equation}
    a_{\mathcal O}
    =
    \frac{f_{\phi \phi \mathcal O} b_{\mathcal O}}{2^J c_{\mathcal O}} \ .
\end{equation}
With a small abuse of notation, we use the same notation for blocks and OPE coefficient in $d = 2$ as in the $d>2$ cases of the previous sections, keeping in mind this redefinition.

\subsubsection{Virasoro primaries with \texorpdfstring{$\Delta_\phi = 1$}{\texttwoinferior}}
\label{subsubsec:VirasoroPrimariesWithDeltaphi1}

We start by studying the case $\Delta_\phi = 1$.

\paragraph{The complex $\tau$-plane.}
As a first step, we would like to use \eqref{eq:dispcomplextau} to compute the correlator in the complex $\tau$-plane.
All the operators in the OPE, being the operators in the vacuum module, have dimensions $\Delta = 2+\mathbb N$, and since odd-spin operators do not contribute, all the dimensions corresponding to non-zero one-point functions are associated with even conformal dimensions.
As a consequence of \eqref{eq:VanishingIdentity}, we find that only the identity operator contributes to the correlator.
This leads straightforwardly to
\begin{equation}
    g(\tau) = \zeta_H(2,\tau)+ \zeta_H(2,1-\tau)+\kappa  = \pi^2 \csc^2(\pi \tau) +\kappa \ ,
    \label{eq:2dCocircular1}
\end{equation}
which perfectly matches the exact result in \eqref{eq:2dcorr} with $\kappa = 0$.

\paragraph{Non-zero spatial separation.}
In two spacetime dimensions, operators with twist zero, i.e., $\Delta = J$, contribute to the discontinuity in the dispersion relation \eqref{eq:DRw}.
The contribution of each single block is
\begin{equation}
    g_\text{dr}^{(J)}(z,\zb)
    =
    2^{1-J} a_{J} \frac{(1+z \zb) (1-z \zb)^J}{z \zb\left(1-z^2\right)
    \left(1-\zb^2\right)} \ .
\end{equation}
In order to sum over all these contributions, we need the thermal OPE coefficients $a_J$.
In this case, we have access to the exact result and in particular\footnote{Notice that all twists appear in the thermal block expansion of \eqref{eq:2dcorr} for $\Delta_\phi=1$.
Using our dispersion relation, we only need one family ($\Delta = J$) to reconstruct them all.}
\begin{equation}
    a_J = - 2^{J+1} \Li_J (-1) \ .
\end{equation}
Summing over all the contributions, we conclude that
\begin{equation}
    g_\text{dr} (z,\zb)
    =
    \frac{2 \pi   \left(1 -z^2\zb^2\right)}{z \zb\left(1-z^2\right) \left(1-\zb^2\right)} \csc (\pi  z \zb) \ .
\end{equation}
This expression can be expanded in blocks as usual, from which we find that $a_{\mathds{1}} = 0$ while all the other OPE coefficients organize in the following sequence:
\begin{equation}
    a_{\Delta,J} =-4 \, \Li_{\frac{\Delta-J}{2}}(-1) \ ,
\end{equation}
which can be verified to be the correct large-spin behavior of all the OPE coefficients that can be extracted from \eqref{eq:2dcorr}.

\begin{table}[t]
    \centering
    \caption{Thermal OPE coefficients in the $2d$ Virasoro primary ($\Delta_\phi = 1$) two-point function extracted by expanding the exact correlator, using the dispersion relation (DR), and generalized method of images (GMI). Only even-spin operators in the vaccum module contribute.}
    \renewcommand{\arraystretch}{2}
    \begin{tabular}{|c|c|c|}
    \hline
    $\mathcal O$ & $a_{\mathcal O}^\text{(exact)}$ = $a_{\mathcal O}^\text{(GMI)}$  & $a_{\mathcal O}^\text{(DR) }$ \\
    \hline
    $\mathds{1}$ & 1 & 0  \\
    \hline
    $\mathcal O_{\Delta,J}$ & $4\  \Li_{\frac{\Delta-J}{2}}(-1) \Li_{\frac{\Delta+J}{2}}(-1)$ & $-4\  \Li_{\frac{\Delta-J}{2}}(-1)$  \\
    \hline
    \end{tabular}
    \label{tab:dr2d1}
\end{table}

\paragraph{The generalized method of images.}
In order to reproduce \eqref{eq:2dcorr}, our prescription would be to sum over all the images of the output of the dispersion relation.
Here for simplicity we limit ourselves to two limits ($z = \zb$ and $z = -\zb$) and show that they correspond to the expectation \eqref{eq:2dcorr}.

The first case corresponds to the limit in which the two operators are placed on the same thermal circle, i.e., at zero spatial separation.
It is easy to show that only the sum of images over the identity block gives a non-zero result we get
\begin{equation}
    g(z,z) = \pi^2 \csc^2(\pi z) \ ,
\end{equation}
which matches \eqref{eq:2dCocircular1}.

The other limit is $z=-\zb = i x$, which corresponds to the case in which the time distance is zero but the spatial distance is finite.
In this case, we have
\begin{align}
    g(i x,-i x)
    &=
    \sum_{m = -\infty}^\infty g_\text{dr}(i x,-ix) \notag \\
    &=\frac{1}{2}\sum_{m= -\infty}^\infty \left(\frac{1}{x^2+(m-1)^2}+\frac{1}{x^2+(m+1)^2}\right) \notag \\
    &\phantom{=\ }-\sum_{\Delta \ge 2}2^\Delta\Li_{\Delta}(-1) \sum_{m= -\infty}^\infty\left(\frac{(m-1)^\Delta}{x^2+(m-1)^2}+\frac{(m+1)^\Delta}{x^2+(m+1)^2}\right) \ ,
\end{align}
where we already neglected terms which are zero after analytic continuation.
We find
\begin{equation}
    g(i x,-i x)  = \pi^2 \text{csch}^2(\pi x) \ ,
\end{equation}
which matches \eqref{eq:2dcorr} when $h = \hb = 1/2$ and $\zb \to -z = -i x$.
    
\paragraph{Momentum space interpretation.}
In momentum space, the only non-trivial contributions come from the identity operator and the operators obeying the condition $\Delta=J$.
In this case, the Fourier thermal blocks read
\begin{align}
    \tilde f_{0,0}(\omega_n,k)&=-\frac{1}{2} \pi  \left[\log \left(k^2+\omega_n ^2\right)+2 \gamma_{E} -2 \log 2\right]\,, \\
    \tilde f_{\Delta,\Delta}(\omega_n,k)&= 2^{\Delta-1} \pi \Gamma (\Delta) \frac{1}{(k+i \omega_n)^{\Delta}}\,.
\end{align}
Summing over all these contributions by inserting the correct OPE coefficients only gives us the momentum space OPE contributions:
\begin{align}
    \tilde{g}_{\text{OPE}}(\omega_n,k)=-2\tilde f_{0,0}(\omega_n,k)-2\sum_{\Delta=2}^\infty \text{Li}_{\Delta}(-1)\tilde f_{\Delta,\Delta}(\omega_n,k)\;.
\end{align}
We can perform the sum over all twist-zero operators above. The sum over these operators diverges, nonetheless it is possible to perform the Borel transformations and Laplace transform back the result: this procedure results in a regularization of the sum over the twist zero operators.

We conclude that the momentum space OPE contribution of the two-point function is given by
\begin{multline} \label{eq:2dpert1} 
    \tilde{g}_{\text{OPE}}(\omega_n,k)=\pi  \left[\log \left(k^2+\omega_n ^2\right)+2 \gamma_E -2 \log 2+1\right]\\-\frac{i(k+2 \pi i n)}{4} \psi ^{(1)}\left(\frac{1-n}{2}+\frac{i k}{4 \pi }\right)-\frac{i(-k+2 \pi in)}{4} \psi ^{(1)}\left(\frac{1-n}{2}-\frac{i k}{4 \pi }\right)\ .
\end{multline}
On the other hand, the Fourier transform of the full thermal two-point function given in \eqref{eq:2dcorr} is
\begin{equation}
    \tilde{g}(\omega_n,k)=-\pi \Bigg[\psi^{(0)}\left(\frac{1+n}{2}-\frac{ik}{4\pi}\right)+\psi^{(0)}\left(\frac{1+n}{2}+\frac{ik}{4\pi}\right)\Bigg] \ , \label{eq:2dfull}
\end{equation}
which also appears in \cite{Son_2002}. 
While Eq. \eqref{eq:2dpert1} does not capture the non-perturbative corrections at small $k$, those are captured by the Fourier transform of the generalized method of images in Eq. \eqref{eq:2dfull}. In contrast, the momentum-space thermal blocks clearly miss these contributions, as expected.

Furthermore, let us comment on the holographic interpretation of these correlators. For example, the poles of the two-point function in the complex $\omega$-plane are directly related to the quasi-normal modes of the black hole—in particular, to those of the black brane limit of the BTZ black hole \cite{Dodelson:2023vrw}. By locating the poles, one finds that the quasi-normal modes are given by \cite{Son_2002}
\begin{equation}
    \Omega_n = \pm k - 2\pi i (n + 1) \, ,
\end{equation}
with $n \in \mathbb{N}_0$.

\subsubsection{Virasoro primaries with  \texorpdfstring{$\Delta_\phi = 2$}{\texttwoinferior}}
\label{subsubsec:VirasoroPrimariesWithDeltaphi2}

We study here the two dimensional case with $\Delta_\phi = 2$.

\paragraph{The complex $\tau$-plane.}
Let us apply \eqref{eq:dispcomplextau} to compute the correlator in the complex $\tau$-plane.
All the operators in the OPE have dimensions $\Delta = 2+\mathbb N$ and, since odd spin operators do not contribute, all the dimensions corresponding to non-zero one-point functions are associated with even conformal dimensions.
Therefore we have that in this case only the identity operator and the stress tensor ($\Delta = J = 2$) contribute to the correlator.
This results in
\begin{align}
    g(\tau) &= \zeta_H(4,\tau)+ \zeta_H(4,1-\tau)+ a_2 \big[ \zeta_H(2,\tau)+ \zeta_H(2,1-\tau)\big] +\kappa \notag \\
    &=  \pi ^2 a_2 \csc ^2(\pi  \tau )+\frac{1}{3} \pi ^4 \left[\cos (2 \pi  \tau )+2\right] \csc ^4(\pi  \tau)+\kappa \ .
\end{align}
From the exact correlator one can read that $a_2 = 2 \pi^2/3$, which simplifies the correlator to
\begin{equation}
    g(\tau)= \pi^4 \csc^4(\pi \tau) + \kappa \ ,
    \label{eq:2dCocircular2}
\end{equation}
matching the result in \eqref{eq:2dcorr} for $\kappa = 0$.
     
\paragraph{Non-zero spatial separation.}
As for $\Delta_\phi=1$, there are operators with zero twist in $d= 2$ ($\Delta = J$) that need to be considered.
We also have, in principle, a new trajectory of operators of twist $\Delta-J= 2$.
Luckily, these operators vanish in the thermal OPE, as it can be confirmed by direct computation from the exact result \eqref{eq:2dcorr}, and for this reason also for $\Delta_\phi = 2$ we only have to care about the operators of zero twist.
The contribution of each single block is given by
\begin{multline}
    g_\text{dr}^{(J)}(z,\zb)=
    a_{J}\frac{2 (1-z \zb)^{J-2}}{(1-z^2)^2(1-\zb^2)^2} \Big[1+z^2+\zb^2-(J+2)z \zb(1-z^2)(1-\zb^2) \\+(1-J)z^4 \zb^4+z^2 \zb^2 (1+J)(z^2+\zb^2)-(J+6)z^2\zb^2 \Big]\ .
\end{multline}
Also in this case it is possible to resum over all the spins: the result gives the large-spin limit of the exact result given in Table \ref{tab:dr2d2}.

\begin{table}[t]
    \centering
    \caption{Thermal OPE coefficients in the $2d$ Virasoro primary ($\Delta_\phi = 2$) two-point function extracted by expanding the exact correlator, using the dispersion relation (DR) and generalized method of images (GMI). Only even-spin operators in the vacuum module contribute.}
    \renewcommand{\arraystretch}{2}
    \begin{tabular}{|c|c|c|}
    \hline
    $\mathcal O$ & $a_{\mathcal O}^\text{(exact)}$ = $a_{\mathcal O}^\text{(GMI)}$  & $a_{\mathcal O}^\text{(DR) }$  \\
    \hline
    $\mathds{1}$ & 1 & 0  \\
    \hline
    $\mathcal O_{\Delta,J}$ & $(\Delta-J -2) (\Delta+J -2) \Li _{\frac{\Delta-J }{2}}(1) \Li_{\frac{\Delta+J }{2}}(1)$ & $2 J (\Delta-J-2) \Li_{\frac{\Delta-J }{2}}(1)$  \\
    \hline
    \end{tabular}
    \label{tab:dr2d2}
\end{table}

\paragraph{The generalized method of images.}
 Trying to compensate with explicit corrections in $1/n^J$, as shown in Appendix \ref{app:KMScompensator}, is very inefficient because of the double truncation in the number of corrections and twist.
We will not consider this possibility here and instead simply consider the sum over images of the dispersion relation.
As in the case $\Delta_\phi = 1$, we only consider the two limits in which the two operators are on the same thermal circle and in the limit in which the two operators are only spatially separated.

In the limit $z = \zb$, we have only two non-zero contributions after analytic continuation.
Those are the operators corresponding to $\Delta = 0$ and $\Delta = 2$, corresponding respectively to the identity operator and the stress tensor.
After resumming, we find
\begin{equation}
    g(z,\zb) = \pi^4 \csc^4(\pi z) \ ,
\end{equation}
which perfectly matches Eq. \eqref{eq:2dCocircular2}.

In the zero spatial direction limit, i.e., $z = - \zb = i x$, we perform the same steps as in Section \ref{subsubsec:VirasoroPrimariesWithDeltaphi1}, we conclude that
\begin{equation}
    g(i x,-i x) = \pi^{4}\text{csch}^4 (\pi x) \ .
\end{equation}
It is easy to show that this also matches the exact result \eqref{eq:2dcorr} in the appropriate limit.

\paragraph{Momentum space interpretation.}
Exactly as in the case of $\Delta_\phi=1$, we compute
\begin{align}
    \tilde f_{0,0}(\omega_n,k)&=\frac{1}{8} \pi  \left(k^2+\omega ^2\right) \left[\log \left(k^2+\omega_n ^2\right)+2 \gamma_E -2-2 \log 2\right]\ , \\
    \tilde f_{\Delta,\Delta}(\omega_n,k)&= 2^{\Delta-3}\pi \Gamma (\Delta-1) \left(k^2+\omega ^2\right) (k+i \omega )^{-\Delta}\ .
\end{align}
Note that the Fourier blocks of twist-two operators also have non-vanishing blocks in momentum space.
The OPE coefficients vanish however, and therefore we do not take them into account.
Summing over all the relevant contributions by using Borel transform, we obtain
\begin{align}
    \tilde{g}_{\text{OPE}}(\omega_n,k)=\tilde f_{0,0}(\omega_n,k)+\sum_{\Delta=2}^\infty a_{\Delta,\Delta}\tilde f_{\Delta,\Delta}(\omega_n,k) \ ,
\end{align}
with the OPE coefficients given in Table \ref{tab:dr2d2}.
Similarly to the $\Delta_\phi = 1$ case, the sum over an infinite number of operators is divergent. However, it is possible to regularize the sum by performing a Borel transform followed by a Laplace transform of the Borel sum.

Comparing to the exact result computed in \cite{Son_2002},
\begin{equation}
    g(\omega_n,k) = \pi \frac{\omega_n^2 + k^2}{4} \left[ \psi^{(0)}\left(1 + \frac{\omega_n}{4\pi} - \frac{i k}{4\pi} \right) + \psi^{(0)}\left(1 + \frac{\omega_n}{4\pi} + \frac{i k}{4\pi} \right) \right] \,,
\end{equation}
from which it is again possible to locate the poles and thus compute the quasi-normal modes \cite{Son_2002}
\begin{equation}
    \Omega_n = \pm k - 2\pi i (n + 2) \,,
\end{equation}
we observe once more that the non-perturbative corrections relevant at small momenta are missing.

\subsubsection{Virasoro primaries with \texorpdfstring{$\Delta_\phi = -2/5$}{Delta\_phi = -2/5} (Lee--Yang theory)}
\label{subsubsec:VirasoroPrimariesLeeYang}

We now study the two-dimensional case with $\Delta_\phi = -2/5$.
This theory is non-unitary and some operators have negative scaling dimensions.\footnote{In non-unitary models, the reality of the spectrum is not guaranteed.
However, this model -- as well as many other minimal models~\cite{Belavin:1984vu,Lencses:2022ira,Fonseca:2001dc} -- enjoys $\mathcal P\mathcal T$ symmetry, which ensures that all conformal dimensions are real (see \cite{Bender:2007nj,Bender:1998gh} and references therein).}
The correlator we study is the (Virasoro) primary-primary two-point function of the only non-trivial primary field in the Lee--Yang model \cite{Fisher:1978pf,Cardy:1985yy}, i.e., the minimal model $\mathcal{M}(2,5)$.
This example shows that the formulas presented in this paper do not rely on unitarity.
For simplicity, we present the case of zero spatial distance. 

\paragraph{The complex $\tau$-plane.}
Let us apply \eqref{eq:dispcomplextau} to compute the correlator in the complex $\tau$-plane.
All operators in the OPE, except for the identity, have dimensions $\Delta = 2+\mathbb{N}$. Moreover, since odd-spin operators do not contribute, all operators with non-zero one-point functions have even integer dimensions. However, because the external operator has non-integer dimension $\Delta_\phi$, all exchanged operators have a non-vanishing discontinuity.

Since all operators contribute to the discontinuity, we need as input the full set of OPE coefficients, which can be extracted from the exact two-point function.\footnote{To obtain this result, one may use
\begin{equation}
    \log\!\left(\frac{\sin x}{x}\right) = \sum_{n\ge 1} \frac{5}{4}\,c_n\,x^{2n}\,,
\end{equation}
and then employ the definition of the complete Bell polynomials.}
They read
\begin{equation}
    a_{\Delta = 2n}
    = \frac{1}{n!}\,\mathrm{Bell}_n\!\bigl(1!\,c_1,2!\,c_2,\ldots,n!\,c_n\bigr),
    \qquad
    c_n = (-1)^n \frac{4}{5}\,\frac{2^{2n-1} B_{2n}}{n (2n)!}\,,
\end{equation}
where $B_n$ are Bernoulli numbers and $\mathrm{Bell}_n$ are the complete Bell polynomials, defined as 
\begin{equation}
    \mathrm{Bell}_n(x_1,\ldots,x_n)
    = n!\sum_{1 j_1+\cdots+n j_n = n}\,
    \prod_{i=1}^{n}\frac{x_i^{j_i}}{(i!)^{j_i}\,j_i!}\,.
\end{equation}
The two-point function, according to \eqref{eq:dispcomplextau}, is
\begin{equation}\label{eq:lab}
    g(\tau) = \sum_{\Delta} a_{\Delta} 
    \left[
        \zeta_H(2\Delta_\phi-\Delta,\tau )
        + 
        \zeta_H(2\Delta_\phi-\Delta,1-\tau )
    \right]
    + \kappa \, ,
\end{equation}
where we set $\beta = 1$. 

The sum over $\Delta$ runs over the identity operator ($\Delta = 0$, $a_{0}=1$) and over the operators in the (Virasoro) vacuum module, namely $T,\, TT,\, \ldots$, so that $\Delta = 2n$. 
Although it is difficult to perform the resummation of \eqref{eq:lab} analytically, the series is convergent. Indeed, the coefficients satisfy the large-$\Delta$ asymptotics
\begin{equation}
    a_{\Delta=2n}
    \overset{n\to\infty}{\sim}
    \frac{\pi^{-2n}}{\Gamma(-4/5)}\, n^{-9/5} \, .
\end{equation}

\begin{figure*}[t]
\centering
   \includegraphics[width=0.8\textwidth]{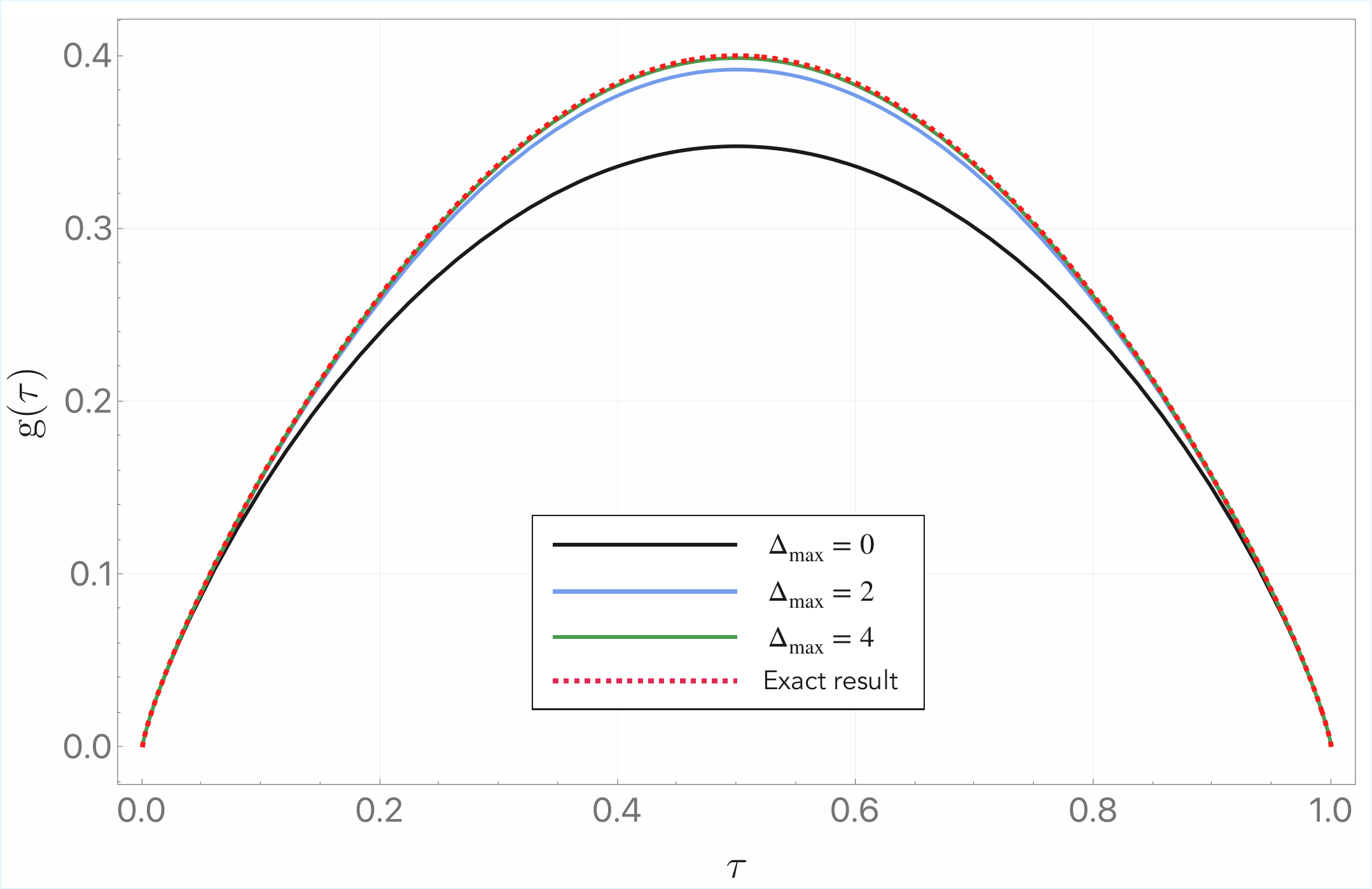}

  \caption{ Two-point function of the (Virasoro) primary field in the two-dimensional Lee--Yang model. 
The dashed red curve denotes the exact result obtained from the conformal map between the plane and the cylinder. 
The solid curves show the truncated expansion in GFF blocks of the form \eqref{eq:lab}, for different truncation levels: 
$\Delta_{\rm max}=0$ (black), 
$\Delta_{\rm max}=2$ (blue), and 
$\Delta_{\rm max}=4$ (green). }
  \label{Fig:LY}
\end{figure*}

Instead of attempting an analytic resummation, in Fig.~\ref{Fig:LY}
we compare the full correlator with the truncated sum obtained from
\eqref{eq:lab} by restricting to $\Delta \le \Delta_{\rm max}$, for various
values of $\Delta_{\rm max}$. 
Note that even for relatively small $\Delta_{\rm max}$ the truncated
sum already provides an excellent approximation to the exact correlator.

In the comparison, one must specify how the constant term $\kappa$ is fixed. 
In the present case this is straightforward, since a constant contribution is not compatible with the OPE. 
We therefore choose $\kappa$ such that the constant term in the OPE vanishes.
As expected, the correlator does not diverge in the limit $\tau \to 0$ 
(equivalently, $\tau \to \beta$). 
This behavior is consistent with the non-unitarity of the theory and with the fact that $\Delta_\phi = -\tfrac{2}{5} < 0$.


%
%

%

\section{A thermal semi-analytic bootstrap}\label{eq:hyb}

In this section, we explore the connection between the analytic bootstrap techniques proposed in this work and the numerical bootstrap results obtainable through the method discussed in \cite{Barrat:2024fwq}.
We show that the formula \eqref{eq:dispcomplextau} can be used to iteratively extract corrections to the asymptotics of heavy operators' OPE coefficients.
Moreover, we show that the same formula results in correlators that accurately reproduce Monte Carlo estimations.

\subsection{Heavy operators from dispersion relation} \label{ssec: heavy}

The novel analytical understanding of thermal two-point functions presented in this work can be used to improve the numerical bootstrap procedure discussed in \cite{Barrat:2024fwq}.
One of the main obstacles to increase the precision lies in the approximation of the tail of heavy operators.
This is crucial in order to study correlators of heavier external operators or simply to increase the precision of the current results.
In \cite{Marchetto:2023xap}, the authors argued that the large $\widetilde{\Delta}$ asymptotic of the thermal OPE coefficients, under suitable assumptions, read
\begin{equation}\label{eq:tauberian}
    a_{\widetilde{\Delta}} \overset{{\widetilde{\Delta}} \gg 1}{\sim} \frac{2 {\widetilde{\Delta}}^{2\Delta_\phi-1}}{\Gamma(2\Delta_\phi)}  \left(1+\frac{c_1}{{\widetilde{\Delta}}}+
    \sum_{\ell} \frac{c_{\ell}}{{\widetilde{\Delta}}^{\alpha_\ell}}\right)\ ,
\end{equation} 
where the coefficients $c_1, \lbrace  c_\ell \rbrace$ and the powers $\lbrace  \alpha_\ell \rbrace$ are unknown and theory-dependent. Although the leading term already provides a rather good approximation \cite{Marchetto:2023xap}, numerical bootstrap procedures require at least the first correction, parametrized by $c_1$. 
More refined approximations are required to achieve higher-precision results. We will shortly demonstrate that the leading behavior and its corrections can be systematically recovered from the dispersion relation in Eq.~\eqref{eq:dispcomplextau}. Before doing so, let us recall the physical intuition and interpretation underlying this result.

\paragraph{Channel duality and the KMS condition.}
Consider the thermal two-point function \( g(\tau) \) in the regime \( \tau \sim 0 \), where the identity operator dominates the OPE. The KMS condition reads \( g(\tau) = g(1-\tau) \), implying that the identity-dominated limit \( \tau \sim 0 \) is mapped to the regime \( \tau \sim 1 \) in the crossed channel. Explicitly, this leads to the relation
\begin{equation}\label{eq:ee}
    \frac{1}{\tau^{2\Delta_\phi}} \sim \frac{1}{(1-\tau)^{2\Delta_\phi}} \sum_{\Delta} a_\Delta\, (1-\tau)^{\Delta} \ .
\end{equation}
It is evident that the right-hand side does not exhibit any poles at \( \tau = 0 \) arising from individual conformal blocks. Therefore, the only way to reproduce the pole generated by the identity contribution on the left-hand side is through the cumulative effect of an infinite number of heavy operators in the crossed channel. This illustrates that the identity is KMS-dual to an infinite tower of heavy operators.

To obtain a quantitative estimate of how these heavy operators contribute, one may rewrite the right-hand side of Eq.~\eqref{eq:ee} as an integral over a spectral density of OPE data, $\rho(\Delta)$, weighted by the coefficients \( a_\Delta \). For a precise formulation, we refer to~\cite{Marchetto:2023xap}. The relevant Laplace transform
\begin{equation}
   \rho(\Delta) \propto \int_0^\infty \text{d} \tau \ \frac{\tau^{-\Delta-1}}{(1-\tau)^{2\Delta_\phi}} \ ,
\end{equation}
predicts that this density grows asymptotically as \(\rho(\Delta)\sim \Delta^{2\Delta_\phi - 1} \). Upon computing the prefactors and imposing appropriate analyticity conditions, one recovers Eq.~\eqref{eq:tauberian}.

This argument makes the channel duality manifest and connects the identity operator to an infinite set of heavy states. However, it is not intended as a rigorous derivation; for a detailed and formal treatment, including the connection with Tauberian theorems,\footnote{For further applications of Tauberian theorems to bootstrap we refer to \cite{Mukhametzhanov:2018zja,Pal:2022vqc,Ganguly:2019ksp,Pal:2019zzr,Pal:2023cgk,Pal:2025yvz,vanRees:2024xkb,Qiao:2017xif}.} see~\cite{Marchetto:2023xap}. Finally, it is natural to expect that subleading corrections to Eq.~\eqref{eq:tauberian} arise from the contribution of light operators on the left-hand side of Eq.~\eqref{eq:ee}. This expectation is indeed correct. Demonstrating it, however, requires more sophisticated techniques than the ones of~\cite{Marchetto:2023xap}. We will show that these corrections can be systematically obtained using the dispersion relation in Eq.~\eqref{eq:dispcomplextau}.

\paragraph{From dispersion relation to heavy operators.}
The Hurwitz $\zeta$-function admits a Taylor expansion around $\tau=0$:
\begin{equation}\label{eq:zetaexp}
    \zeta_H(s,\tau)+\zeta_H(s,1-\tau) = \frac{1}{\tau^s}+\sum_{k=0}^{\infty} \left[1+(-1)^k\right]\binom{s+k-1}{k}\zeta(k+s) \tau^{k} \ ,
\end{equation}
where $s = 2\Delta_\phi-\Delta$.
Observe that we simply recovered the OPE of a GFF correlator with external dimension given by $s$: the expansion \eqref{eq:zetaexp} is given by operators of classical\footnote{The anomalous dimensions could be recovered by studying the \textit{pole-shifting} through the inversion formula, as in \cite{Iliesiu:2018fao}.} conformal dimension $\widetilde{\Delta} = 2\Delta_\phi+ k$.
From now on, we will use $\widetilde{\Delta}$ as a variable, instead of $k$.
Moreover, only the terms proportional to even powers of $\tau$ contribute, as a consequence of the parity symmetry $\tau \to -\tau$.
It makes therefore sense to compare the OPE coefficient $a_{\widetilde{\Delta}}$ and the generic coefficient of \eqref{eq:zetaexp} in the limit $\widetilde{\Delta} \to \infty$.
We notice that the asymptotic of the two-point function can be approximated by summing of even integer dimensional operators, corresponding classically to the double twist, the associated OPE coefficients can be expanded in terms of the inversion of the light operators.
Furthermore, the exponential corrections are completely disentangled and subleading with respect to the polynomial corrections in $1/\Delta$.
Therefore, in the following, we will drop them.\footnote{They can be easily reintroduced by multiplying each term in the sum by $\zeta(\widetilde{\Delta}- \Delta)$.}
In the large $\widetilde{\Delta}$ limit we find that the contribution of the operator $\Delta$ to the heavy operator of dimensions $\widetilde{\Delta}$ is given by
\begin{equation} \label{eq:contr}
    a_{\widetilde{\Delta}}^{(\Delta)} \overset{\widetilde{\Delta} \gg 1}{\sim}a_{\Delta}\frac{2\widetilde{\Delta} ^{2 \Delta_\phi -\Delta-1}}{ \Gamma (2 \Delta_\phi -\Delta)} \left[1+\frac{c_1^\Delta}{ \widetilde{\Delta} }+\frac{c_2^\Delta}{\widetilde{\Delta} ^{2}}+\dots \right] \ ,
\end{equation}
where the correction coefficients are given by
\begin{align}
    c_1^\Delta&=\frac{(\Delta+2 \Delta_\phi ) (\Delta-2 \Delta_\phi +1)}{2} \label{eq:cdelta1}\ , \\
    c_2^\Delta&=\frac{(\Delta-2 \Delta_\phi +2) \left(3 \Delta^2+ \Delta (12\Delta_\phi +1)+2 \Delta_\phi  (6 \Delta_\phi -1)\right) (\Delta-2 \Delta_\phi +1)}{24} \label{eq:cdelta2}\ .
\end{align}

The asymptotic \eqref{eq:contr} should be interpreted as the correction to the Tauberian leading order for the double-twist OPE cofficients \eqref{eq:tauberian} obtained by inverting the operator (or operators, in case of degeneracy) of conformal dimension $\Delta$ appearing in the OPE.
The main contribution, as expected, comes from the inversion of the identity operator, and it reads\footnote{In the final stage of this work, we became aware of \cite{NewPaperc}, where the authors also computed some of the corrective terms in Eq.~\eqref{eq:iden}: our results and theirs perfectly match. We thank Matthew Dodelson for the interesting conversation, and the authors of \cite{NewPaperc} for sharing their results with us.}
\begin{equation} \label{eq:iden}
    a_{\widetilde{\Delta}} \overset{\widetilde{\Delta} \gg 1}{\sim}\frac{2\widetilde{\Delta} ^{2 \Delta_\phi-1}}{ \Gamma (2 \Delta_\phi)} \left[1-\frac{\Delta_\phi(2 \Delta_\phi -1)}{\widetilde{\Delta}}+\frac{\Delta_\phi(\Delta_\phi -1)(2 \Delta_\phi -1)(6 \Delta_\phi -1) }{6\, \widetilde{\Delta} ^{2}}+\dots \right]+ \dots \ .
\end{equation}
Interestingly, the inversion of the identity operator correctly reproduces the Tauberian asymptotic derived in \cite{Marchetto:2023xap}.
This section therefore provides a derivation of this asymptotic, under the assumption of analyticity and Regge boundedness of the correlator in the complex $\tau$-plane. 
From this perspective, it is clear that the result arises from the inversion of operators from the $t$-channel to the $s$-channel.
In fact, the result of~\cite{Marchetto:2023xap} can be effectively interpreted as the inversion of the identity operator.
In this work, we generalize that result by promoting it — through the analytic techniques developed in this paper — to the inversion of arbitrary operators, making stronger the idea of the channel duality between light and heavy sectors.

 Thanks to the closed form of the thermal two point function in complex $\tau$ plane given in \eqref{eq:dispcomplextau} we can improve the tail of the heavy operators further by computing corrections coming from the other light operators in the OPE. If the light operators have conformal dimensions $0<  \Delta_1<\ldots$, the corrections take the following form: 
  \begin{align}
      \label{eq:full}
    a_{\widetilde{\Delta}} &
    \overset{\widetilde\Delta \gg 1}{\sim}
    2\frac{\widetilde \Delta^{2\Delta_\phi-1}}{\Gamma(2\Delta_\phi)}\left(1+\frac{c_1^{0}}{\widetilde \Delta }+\frac{c_{2}^{0}}{\widetilde\Delta^2}+\ldots \right)  \notag\\ &
    \phantom{\overset{\Delta \gg 1}{\sim}\ }
    +\sum_{\Delta_{i} \neq 0} 2a_{\Delta_i} \frac{\widetilde\Delta^{2\Delta_\phi-\Delta_i-1}}{\Gamma(2\Delta_\phi-\Delta_i)}\left(1+\frac{c_1^{\Delta_i}}{ \widetilde\Delta }+\frac{c_2^{\Delta_i}}{\widetilde \Delta^2 }+\ldots \right)\ ,
  \end{align} 
 such that for each light operator contribution the coefficients of the order-by-order corrections are defined in \eqref{eq:cdelta1} and \eqref{eq:cdelta2}. This gives us a better understanding of the dominant contributions in terms of the light operators of the theory.
 This expression can be used to bootstrap thermal OPE coefficient as in~\cite{Barrat:2024fwq}: as a proof of concept, in the following section, we make use of Eq.~\eqref{eq:dispcomplextau} to compute the two-point function $\langle \epsilon \epsilon\rangle_\beta$ of $3d$ Ising.

It was pointed out in~\cite{Iliesiu:2018fao} that the inversion of individual operators can only generate operators of the double-twist families $[\phi\phi]_{n,J}$.
However in general theories (e.g., in the $3d$ Ising model) we expect other families of operators to appear, for instance $[\eps \eps]$ or even multi-twist $[\phi\phi\phi\phi]$.
The reason why these other families do not appear in the asymptotics lies in the fact that the sum of operators and the inversion procedure do not commute.
In order to see these families one could first resum a family, for instance $[\phi\phi]_{0,J}$, and then invert the result.
This procedure is explained in detail in~\cite{Iliesiu:2018fao} and the formalism presented here can be adapted to this case.

In~\cite{Buric:2025anb}, it was pointed out that certain operators, such as multi-stress tensors, may have alternating sign OPE coefficients, naively rendering the expansion in Eq.~\eqref{eq:full} inconsistent. However, by applying the dispersion relation~\eqref{eq:dispcomplextau} — for instance, in the case $\Delta_\phi = 3/2$ — by using the OPE coefficients for the multi-stress tensor sector given in the paper, one recovers the result given in Eq.~(2.24) of~\cite{Buric:2025anb}. As discussed in that work, the corresponding solution exhibits unexpected poles in the complex $\tau$-plane.\footnote{These unexpected poles are holographically associated with the bouncing singularity~\cite{Ceplak:2024bja}.}  These poles should not appear in physical correlators and indicate that some consistency condition, necessary to derive the dispersion relation, is violated. The authors suggests a procedure to remove by hand these poles, but this has no direct analogue within the dispersion relation~\eqref{eq:dispcomplextau} itself. Let us take this opportunity to point out that Eq.~\eqref{eq:dispcomplextau} requires, as input, the discontinuity of the two-point function. However, not every choice of input is consistent: new, emergent analytic structures -- such as the extra poles appearing in Eq.~(2.24) of~\cite{Buric:2025anb} -- may arise. This imposes nontrivial constraints on the admissible input of the dispersion relation which would be interesting to investigate in future studies.

\subsection{Correlators in the \texorpdfstring{$3d$}{\texttwoinferior} Ising model}
\label{subsec:CorrelatorsInThe3dIsingModel}

In Section \ref{subsubsec:AFormulaForg}, we derived a formula for the thermal two-point function of two scalar operators in the complex $\tau$-plane (see \eqref{eq:dispcomplextau}).
Although such formula completely fixes the kinematics of the correlator, it requires the dynamical information encoded in the OPE coefficients $a_{\Delta}$ as an input.
When studying strongly-coupled thermal CFTs, computing these coefficients is usually a formidable challenge, which can be tackled by using a numerical bootstrap approach \cite{Barrat:2024fwq}.
In this Section, we combine the formula \eqref{eq:dispcomplextau} and the numerical results of \cite{Barrat:2024fwq} to study correlation functions in the $3d$ Ising model, in particular the correlators $\langle \sigma(\tau)\, \sigma(0)\rangle_\beta$ and $\langle \epsilon(\tau)\, \epsilon(0)\rangle_\beta$ (see Table \ref{tab:3dIsingBotstrap} for a short summary of the light operators in this theory).

\begin{table}[t]
\centering
\renewcommand{\arraystretch}{1.25}
\begin{tabular}{|c|c|c|c|}
\hline
$\mathcal O$ & $\mathcal O_{\text{GL}}$ & $\Delta_{\mathcal{O}}$~\cite{Kos:2016ysd,Reehorst:2021hmp,Chang:2024whx} & $a_{\Delta_\mathcal O}$ \cite{Barrat:2024fwq} \footnotesize{(from $\langle\sigma \sigma \rangle_{\beta}$)} \\
\hline
$\sigma$ & $\phi$ & 0.518148806(24) & 0 \\ \hline
$\epsilon$ & $\phi^2$ & 1.41262528(29) & 0.75(15) \\ \hline
$T_{\mu \nu}$ & $T_{\mu \nu}$ & 3 & 1.97(7) \\ \hline
$\epsilon'$ & $\phi^4$ & 3.82951(61) & 0.19(6) \\ \hline
$C'_{\mu \nu \rho \sigma}$ & $\phi\, \partial_\mu \partial_\nu \partial_\rho \partial_\sigma \phi$ & 5.022665(28) & / \\ \hline
$T'_{\mu \nu}$ & $\phi\, \partial_\mu \partial_\nu \Box \phi$ & 5.50915(44) & / \\ \hline
\end{tabular}
\caption{Light operators in the $3d$ Ising model, their classical (Ginzburg–Landau) counterparts ($\mathcal O_{\text{GL}}$), conformal dimensions, and thermal OPE coefficients (when available and considering the $\langle\sigma(\tau) \sigma(0) \rangle_{\beta}$ correlator).}
\label{tab:3dIsingBotstrap}
\end{table}

In applying the formula \eqref{eq:dispcomplextau} to strongly-coupled theories, and in particular to the $3d$ Ising model, we are faced with a problem: since the conformal dimensions appearing in the OPE are non-integers, all of them will contribute to the discontinuity and hence have to be summed over in the formula.
Usually only the coefficients for the first light dimensions are explicitly known (see Table \ref{tab:3dIsingBotstrap}).
However, close to $\tau/\beta \sim 0$, the dominant contributions come from the lightest operators, which also dominate the limit $\tau/\beta \sim 1$ thanks to KMS invariance. 
Therefore, even though in the $3d$ Ising model (and other strongly-coupled theories) we cannot perform the full sum over the conformal dimensions appearing in the OPE as prescribed in~\eqref{eq:dispcomplextau}, it is still meaningful to truncate the sum to the first few light operators.
We expect this approximation to hold well close to $\tau/\beta \sim 0$ and  $\tau/\beta \sim 1$.

\paragraph{The correlator \texorpdfstring{$\langle \sigma \sigma \rangle_\beta$}{\texttwoinferior}: consistency checks.}
In this paragraph, we start by studying the correlator $\vev{\sigma(\tau)\, \sigma(0)}_\beta$. The first light operators appearing in the OPE are, schematically,
\begin{equation} \label{eq: ss ope}
    \sigma \times \sigma \sim \mathds{1}+\epsilon+T^{\mu \nu}+\epsilon'+\dots \ .
\end{equation}
All the operators appearing on the right-hand side of the OPE \eqref{eq: ss ope} contribute to the discontinuity.
Their thermal OPE coefficients were bootstrapped in \cite{Barrat:2024fwq} and are listed in Table \ref{tab:3dIsingBotstrap}, along with their conformal dimensions.

Following the line of thought described above, the correlator can be approximated by truncating the sum in Eq.~\eqref{eq:dispcomplextau} and plugging the dynamical data:
\begin{equation}\label{eq:sigmasigmaapprox}
    \langle \sigma(\tau)\, \sigma(0)\rangle_\beta \approx \hspace{-0.3em} \sum_{\mathcal O \in \lbrace\mathds{1},\, \epsilon,\, T,\, \epsilon' \rbrace} \hspace{-0.3em} \frac{a_{\Delta_{\mathcal \mathcal O}}}{\beta^{2\Delta_{\mathcal \sigma}}} \left [\zeta_H\left(2\Delta_\sigma - \Delta_{\mathcal O}, \frac{\tau}{\beta}\right) + \zeta_H\left (2\Delta_\sigma - \Delta_{\mathcal O}, 1 - \frac{\tau}{\beta}\right)\right] + \frac{\kappa}{\beta^{2\Delta_\sigma}} \,.
\end{equation}
As discussed in Section \ref{sec:AnalyticToolsForThermalCorrelators}, the only unfixed parameter of the dispersion relation in the complex $\tau$-plane is an additive constant $\kappa$.
If we were considering all the operators contributing to the discontinuity in the sum above, this constant could be fixed by clustering; however, the truncation of the sum spoils the limit $\tau \to i\infty$.
To fix $\kappa$ we require that the OPE does not contain any constant term: the procedure yields an estimated value $\kappa \approx -55.86 \beta^{2\Delta_\sigma}$.

\begin{figure*}[t]
\centering
   \includegraphics[width=0.8\textwidth]{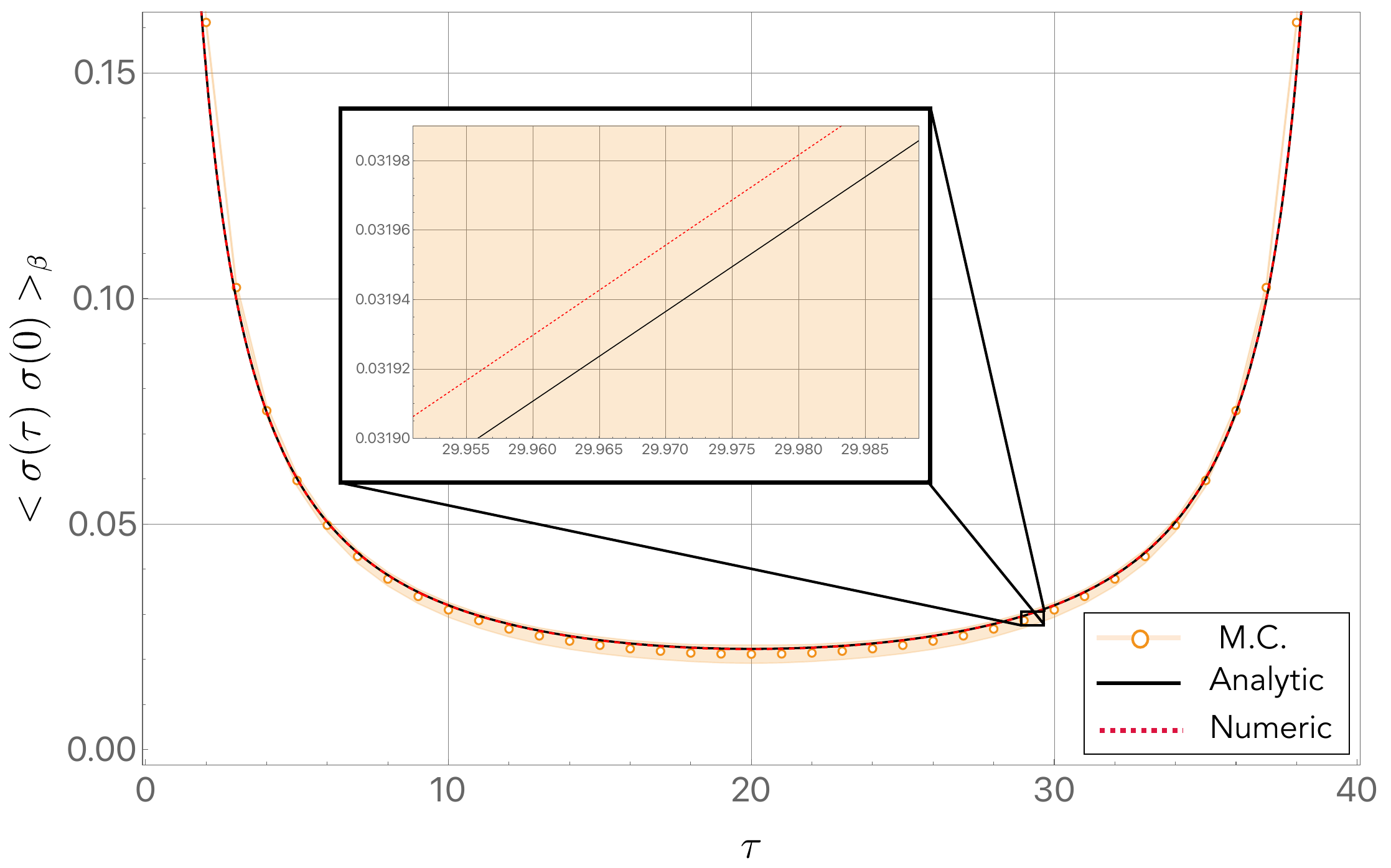}

  \caption{ Two point function $\vev{\sigma(\tau) \sigma(0)}_\beta$ in the limit in which $\vec x = 0$. The black continuous line is obtained by using the dispersion relation in the complex $\tau$-plane  and by inputting the thermal OPE coefficients from \cite{Barrat:2024fwq}.
  The dashed curve represent directly the OPE result of \cite{Barrat:2024fwq} while the points are the result of a Monte Carlo simulation on a lattice $500\times 500 \times 40$ where the inverse temperature is given by $\beta = 40$.}
  \label{Fig:3dIsingDR}
\end{figure*}

The result is reported in Fig.~\ref{Fig:3dIsingDR}.
In the plot, we compare the approximation obtained from the truncated version of Eq. \eqref{eq:sigmasigmaapprox} and the numerical evaluation of~\cite{Barrat:2024fwq}. This procedure will clearly decrease the numerical error since the number of Tauberian correction can be arbitrarily high without adding new unknown coefficient.
The relative discrepancy between the two curves is of order $10^{-3}$, in agreement with the level of accuracy to which the numerical results in~\cite{Barrat:2024fwq} can be trusted.
Notably, the largest deviation occurs in the region $\tau \in (1/2,1)$, which is precisely the region where we expect the corrections to the Tauberian approximation used in~\cite{Barrat:2024fwq} to become more significant.
The motivation behind the discrepancy lies in the amount of corrections employed in each case to refine the leading Tauberian approximation: in \cite{Barrat:2024fwq}, the authors made use of only the first correction appearing in Eq.~\eqref{eq:full}, of order $1/\widetilde{\Delta}$.
Moreover, the coefficient was considered to be an unknown variable.
In this work, we instead made use of the formula \eqref{eq:dispcomplextau}, which already encodes an infinite amount of corrections, each one with a predetermined coefficient (see Section \ref{ssec: heavy} for details).

For completeness, we also compare our results to Monte Carlo simulations, finding very good agreement.
Details on the numerical implementation and simulation parameters are provided in Appendix~\ref{app:MC}.
It is important to note that the two-point function extracted from Monte Carlo simulations is not unit-normalized. As a result, we need to redefine the operator $\sigma$ in Eq.~\eqref{eq:sigmasigmaapprox} so that its zero-temperature two-point function matches the normalization used in the Monte Carlo data.
This mismatch amounts to a single overall multiplicative constant, which can be fixed by fitting. We find that this normalization constant is approximately $C_\sigma \approx 0.305$.

Note that one can also adopt the method presented in \cite{Barrat:2024fwq} to implement further corrections in the tail of heavy operators, as shown in Eq.~\eqref{eq:iden}. This is equivalent to comparing the result of \cite{Barrat:2024fwq} with the approximation in \eqref{eq:sigmasigmaapprox}, and indeed, this procedure is consistent with the numerical results for thermal OPE coefficients shown in Table \ref{tab:3dIsingBotstrap}. This confirms the findings of \cite{Barrat:2024fwq}. Interestingly, however, these results do not match the Casimir amplitude computed via Monte Carlo in \cite{Bulgarelli:2025riv}. It would be worthwhile to investigate this discrepancy in more detail, for instance by comparing additional observables such as the thermal two-point function with higher precision.

\paragraph{The correlator \texorpdfstring{$\langle \epsilon\epsilon\rangle_\beta$}{\texttwoinferior}: new result.}
To produce a numerical approximation of the correlator $\vev{\epsilon(\tau)\epsilon(0)}_\beta$ using the methods of \cite{Barrat:2024fwq}, we need to address the issue that the external dimension $\Delta_\epsilon$ is larger than $\Delta_\sigma$.
While $\vev{\sigma \sigma}_\beta$ only required a single correction to the Tauberian leading behaviour, the $\vev{\epsilon \epsilon}_\beta$ correlator requires more corrections to obtain the same level of precision.\footnote{As argued in \cite{Marchetto:2023xap} and \cite{Barrat:2024fwq}, this can be seen heuristically by noticing that the contribution of the heavy tail to the thermal sum rules increases when the external dimension $\Delta_\phi$ increases as well, leading to the conclusion that correlators with higher external dimension require a more precise evaluation of the tail.
\emph{A posteriori}, this line of thought is justified by the fully corrected asymptotics \eqref{eq:iden} and \eqref{eq:full}, where it can be seen that the corrections to the leading order are monotonic functions of $\Delta_\phi$.}

\begin{figure}[t]
    \centering
    \hspace{-2.5em}
    \includegraphics[width=0.8\textwidth]{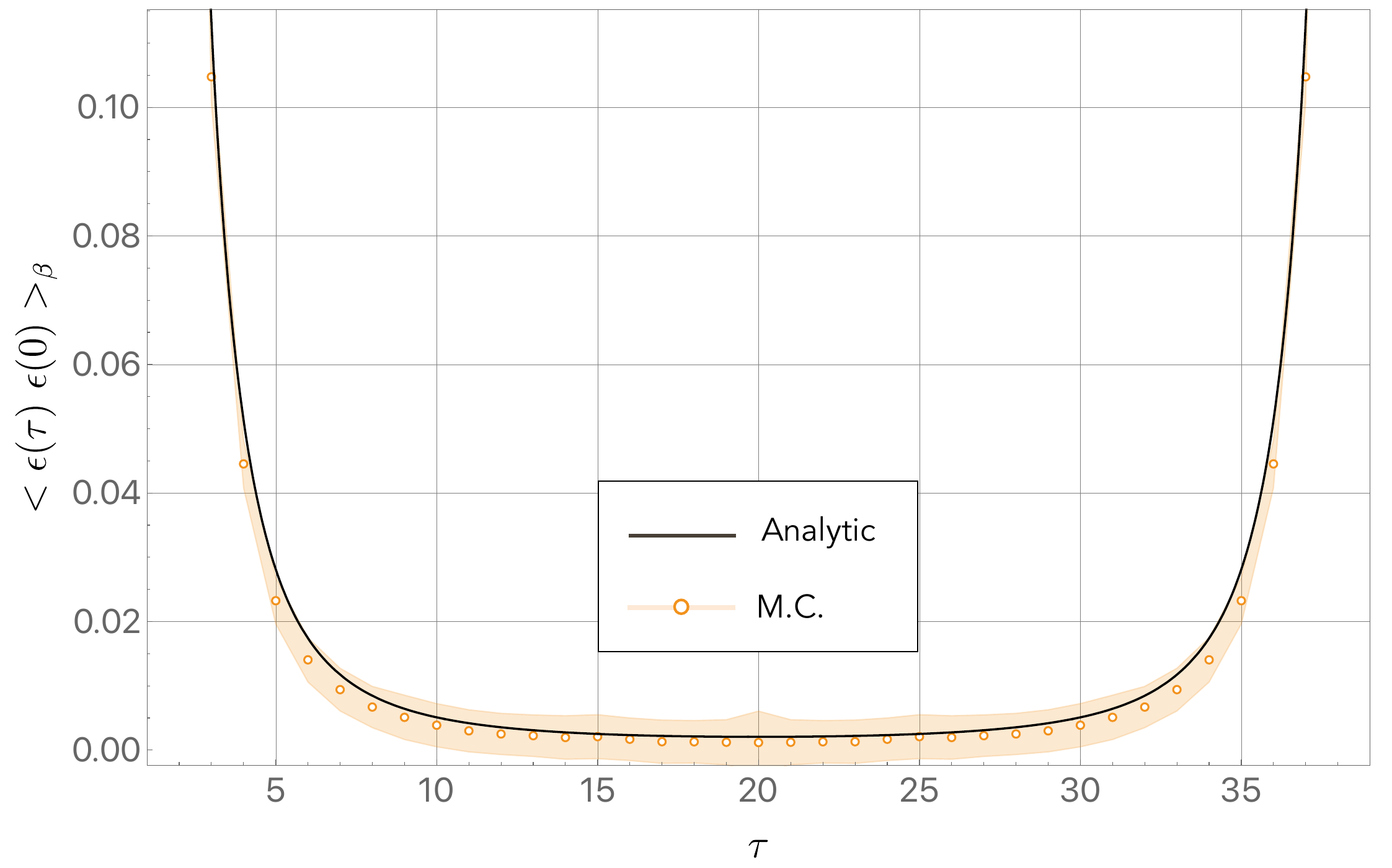}
    \caption{Two-point function $\vev{\epsilon(\tau)\epsilon(0)}_\beta$ computed by inverting the contributions of $\mathds{1}$, $\epsilon$, and $T_{\mu \nu}$ using the dispersion relation in the complex $\xi$-plane.
    The result is compared with Monte Carlo simulations on a $500\times 500 \times 40$ lattice, corresponding to inverse temperature $\beta = 40$.}
    \label{fig:3dIsingee}
\end{figure}

The dispersion relation \eqref{eq:dispcomplextau} overcomes this limitation since, as we pointed out in the previous paragraph, it already encodes an infinite number of corrections.
We can therefore adopt the same strategy employed for the $\vev{\sigma \sigma}_\beta$ correlator and plug the bootstrap results of~\cite{Barrat:2024fwq} in the formula.
First we convert the OPE coefficients extracted out of $\vev{\sigma \sigma}_\beta$ (indicated by $a_{\Delta_\mathcal{O}}^{\langle\sigma \sigma \rangle}$) into those relevant for $\vev{\epsilon \epsilon}_\beta$ (indicated by $a_{\Delta_\mathcal{O}}^{\langle\epsilon \epsilon\rangle}$). Specifically, we use the relation
\begin{equation}
    a_{\Delta_\mathcal O}^{\langle\epsilon\epsilon \rangle}  =  \left(\frac{f_{\mathcal O \epsilon \epsilon}}{f_{\mathcal O \sigma \sigma}}\right)a_{\Delta_\mathcal O}^{\langle\sigma\sigma \rangle} \ .
\end{equation}
This yields the estimates
\begin{equation}
    a_{\Delta_\epsilon}^{\langle\epsilon\epsilon \rangle}  = 1.09(22) \ , \qquad
    a_{\Delta_T}^{\langle \epsilon\epsilon \rangle}  = 5.37(19) \ .
\end{equation}
Using this information, we approximate the thermal two-point function as
\begin{equation} \label{eq: appee}
  \langle \epsilon(\tau) \epsilon(0)\rangle_\beta \approx \hspace{-0.75em} \sum_{\mathcal O \in \lbrace\mathds{1},\, \epsilon,\, T \rbrace} \hspace{-0.75em} \ \   \frac{a_{\Delta_{\mathcal O}}}{\beta^{2\Delta_{\epsilon}}} \left[ \zeta_H\left(2\Delta_\epsilon - \Delta_{\mathcal O}, \frac{\tau}{\beta}\right) + \zeta_H\left(2\Delta_\epsilon - \Delta_{\mathcal O}, 1 - \frac{\tau}{\beta}\right) \right] + \frac{\kappa}{\beta^{2\Delta_\epsilon}}\ .
\end{equation}
As before, the dispersion relation is ambiguous up to an additive constant $\kappa$.
To fix it, we match the result with the OPE-expanded correlator, truncated to the contributions from $\mathds{1}$, $\epsilon$, and $T_{\mu\nu}$ in the region $\tau/\beta \lesssim 0.05$, where the OPE series is rapidly convergent.
One could check that including additional operators (using large $\Delta$ asymptotic to estimate their OPE coefficients) does not significantly alter the result.
The procedure yields an estimated value $\kappa \approx 1.81$ .

In Fig.~\ref{fig:3dIsingee}, we plot the function \eqref{eq: appee} and compare it to the outcome of Monte Carlo simulations, observing again very good agreement.
As in the $\vev{\sigma \sigma}_\beta$ case, a multiplicative normalization constant is required.
This constant, which amounts to a redefinition of the $\epsilon$ operator, is determined by fitting the Monte Carlo data, yielding $C_\epsilon \approx 5.04$.

\section{Conclusions and outlook}
\label{sec:conclusions}

In this paper, we developed an analytic framework to bootstrap thermal two-point functions based on dispersion relations, the operator product expansion, and the Kubo-Martin-Schwinger condition.
We showed that, in the periodic Euclidean time coordinate and at zero spatial separation, any thermal two-point function can be expanded in terms of generalized free-field correlators.
The coefficients of this expansion are given by certain thermal OPE coefficients.
This result, explicitly presented in Eq.~\eqref{eq:dispcomplextau}, follows from expressing the two-point function in terms of its discontinuity in the complex time plane.

We further investigated the dispersion relation at non-zero spatial separation introduced in~\cite{Alday:2020eua}, and its interplay with the OPE and the KMS condition.
We found that computing the discontinuity in the OPE regime captures only the leading polynomial behavior at large spin.
In contrast, non-polynomial corrections arise from contributions outside the OPE regime, in full agreement with the structure of large-spin perturbation theory described in~\cite{Iliesiu:2018fao}.
Moreover, we observed that the naive result from the dispersion relation within the OPE regime is not KMS invariant.
Motivated by the complex $\tau$-plane analysis, we proposed a prescription to restore KMS invariance and simultaneously reconstruct the low-spin contributions.
This involves summing over images of the dispersion relation result, leading to what we called the generalized method of images.
We demonstrated that this construction satisfies all thermal bootstrap axioms, except clustering, which must be verified on a case-by-case basis.
Additionally, we showed that if clustering holds, then the solution is unique — equivalently, arc contributions vanish.

We illustrated the applicability of our method in several examples.
In the case of the free scalar theory, the generalized method of images reduces to the standard one, effectively reproducing a sum over propagators.
We then studied the $\mathrm{O}(N)$ model in the $\varepsilon$-expansion and found perfect agreement with direct perturbative calculations.
Finally, we tested our results in exactly solvable strongly coupled regimes, namely the large $N$ limit of the $\mathrm{O}(N)$ model and in two-dimensional models, reproducing the known results.

By combining analytic and numerical techniques — in particular the method proposed in~\cite{Barrat:2024fwq} — we believe that substantial progress can be made on the thermal bootstrap program.
Using the dispersion relation in the complex time direction, we rederived the Tauberian estimate for the contribution of heavy operators in the OPE, originally obtained through different means in~\cite{Marchetto:2023xap}.
Furthermore, our method allows the computation of the full tower of corrections to the leading heavy-state contribution, including the coefficient of the first correction used in~\cite{Barrat:2024fwq}.
As an application, we combined the numerical approach of~\cite{Barrat:2024fwq} with our analytic dispersion relation to compute the thermal two-point function $\langle\epsilon \epsilon\rangle_\beta$ in the $3d$ Ising model, finding excellent agreement with Monte Carlo simulations.

The thermal bootstrap program remains rich with open questions, both at the conceptual level and in terms of applications to physically relevant models.
We highlight below several promising directions for future work:
\begin{itemize}
    \item[$\star$] The most immediate application of the methods presented in this paper is to extend them to a wider range of physically relevant theories.
    It would be clearly valuable to improve the result of \cite{Barrat:2024fwq} at the level of numerical precision of the result for the $\mathrm O(N)$ model in three dimensions \cite{NewPaperc}.
    In the space of CFTs, a special role is played by holographic theories, which are dual, via the AdS/CFT correspondence, to black holes and black branes in AdS space.
    Among these, $\Nm = 4$ super Yang-Mills theory — dual to gravity in AdS$_5 \times S^5$ — stands out as a particularly compelling target.
    Understanding its behavior at finite temperature remains one of the central goals of the thermal bootstrap program~\cite{NewAnalytic}.
    Holographic models also present precise structures of two-point function as a function of the quasi-normal modes of the black hole in AdS \cite{Dodelson:2022yvn,Dodelson:2023vrw,Dodelson:2023nnr,Dodelson:2024atp,Dodelson:2025rng,Bhattacharya:2025vyi}.
    As we have highlighted in the simple case of BTZ black holes, the quasi-normal modes of the black hole can be the outcome of a proper thermal bootstrap problem.

    \item[$\star$] One of the main challenges in thermal bootstrap is the lack of positivity of thermal OPE coefficients, which complicates the derivation of rigorous numerical bounds.
    While existing studies~\cite{Iliesiu:2018fao,Barrat:2024fwq} have circumvented this by developing non-positivity-based methods, it would be highly desirable to find intrinsically positive observables — such as the hydrodynamic moments discussed in~\cite{Dodelson:2023vrw} — or to reformulate the thermal bootstrap as a semi-definite programming problem.

    \item[$\star$] This work has focused on conformal field theories at criticality.
    However, many observables of physical interest arise in theories deformed away from the critical point.
    Extending the non-perturbative methods developed here to non-conformal theories at finite temperature, or to relevant deformations of CFTs, would open new avenues for understanding finite-temperature dynamics in realistic models~\cite{future1}.
    Furthermore the non-perturbative study of QFTs at finite temperature could give further insides on the order or disorder phases at high temperature which are extremely important for the understanding of the thermodynamic of QFTs  \cite{Chai:2020onq,Komargodski:2024zmt,Hawashin:2024dpp,Han:2025eiw}.
    
    \item[$\star$] In~\cite{Barrat:2024aoa}, a bootstrap framework was proposed for line defects wrapping the thermal circle.
    These include, for instance, Polyakov loops in gauge theories, which serve as order parameters for confinement/deconfinement transitions and are closely related to the Kondo effect in low-energy physics \cite{Polyakov:1978vu,Affleck:1995ge}.
    Extending the techniques developed in this paper to include such line defects would be a significant step forward and would enable the study of setups such as the magnetic line in $\mathrm{O}(N)$ models \cite{Cuomo:2021kfm,Gimenez-Grau:2022czc,Gimenez-Grau:2022ebb,Bianchi:2022sbz,Hu:2023ghk} and the supersymmetric Wilson line in $\Nm=4$ SYM \cite{Giombi:2017cqn,Liendo:2018ukf,Giombi:2018qox,Ferrero:2021bsb,Cavaglia:2021bnz,Barrat:2021tpn,Artico:2024wut} at finite temperature.
    Many aspects are expected to carry over to correlators of defect operators \cite{future2}.
    
    \item[$\star$] Most of the equations and results derived in this work concern two-point functions of scalar operators in bosonic theories.
    It would be highly valuable to generalize this framework to include fermionic operators, enabling the study of theories such as the chiral Gross–Neveu model and the Gross–Neveu–Yukawa model.
    Recent progress in this direction can be found in~\cite{David:2023uya}.

    \item[$\star$] Finally, it would be of great interest to study finite-temperature systems at finite volume. Holographically, placing the theory on $S^1 \times S^{d-1}$ corresponds to exploring a broader class of black hole solutions in AdS.
    The $\mathrm{O}(N)$ model at large $N$ was recently studied on $S^1 \times S^2$ in~\cite{David:2024pir}.
    Extending this analysis to finite $N$ and formulating a bootstrap approach in finite volume — along the lines developed in this paper — would represent a valuable extension of the thermal bootstrap program.
    Furthermore, the generalized method of images resembles certain formulae that appeared in literature in holographic contexts at finite volume \cite{Brigante:2005bq,Belin:2025nqd}.
    It would be interesting to understand the connection between these formulae and the techniques of this paper. 
\end{itemize}

\section*{Acknowledgements}
 It is a pleasure to thank Lorenzo Bianchi, Davide Bonomi, Ilija Burić, Michele Caselle, Simon Caron-Huot, Matthew Dodelson, Ignacio Haraya, Manuela Kulaxizi, Dalimil Mazáč, Andrei Parnachev, Tassos Petkou, and Subir Sachdev for very useful discussions.
 JB and EP are supported by ERC-2021-CoG - BrokenSymmetries 101044226. AM, DB, and EP have benefited from the German Research Foundation DFG under Germany’s Excellence Strategy – EXC 2121 Quantum Universe – 390833306.
 EM and EP’s work is funded by the German Research Foundation DFG – SFB 1624 – “Higher structures, moduli spaces and integrability” – 506632645.
 AM, DB, EM, EP would like to express special thanks to the organizers of the Program ``Black hole physics from strongly coupled thermal dynamics'' and to the Simons Center for Geometry and Physics (SCGP) for the hospitality and support in the final stage of this project. 
  
 Part of the results contained in this work are presented in the Ph.D. thesis of AM \cite{Miscioscia:2025pjh}.

\appendix

\section{Constraining \texorpdfstring{$g_{\text{arcs}}(\tau)$}{\texttwoinferior}}
\label{app:pefconst}

The goal of this Appendix is to prove that a function \( f(w) \), with \( w \in \mathbb{C} \), satisfying the following conditions:
\begin{itemize}
    \item[$\star$] \( f(w) \) is an entire function;
    \item[$\star$] \( f(w) = f(w+1) \);
    \item[$\star$] \( |f(w)| < e^{|w|} \);\footnote{This condition can be easily generalized to \( |f(w)| < C  e^{|w|} \) by appropriately redefining the function \( f \).}
\end{itemize}
must be constant, i.e., \( f(w) = f(0) \in \mathbb{C}\).

The proof proceeds as follows. Consider the function
\begin{equation}
    h(w) = \frac{f(w) - f(0)}{\sin(\pi w)} \ .
\end{equation}
Note that \( h(w) \) is entire, since the potential poles introduced by the sine function in the denominator are canceled by the zeroes of the numerator at the same points. Furthermore, the function \( h(w) \) is periodic, i.e., \( h(w) = h(w + 2n) \) for all \( n \in \mathbb{Z} \), and satisfies \( |h(w)| \to 0 \) as \( \text{Im}(w) \to\pm \infty \) for any fixed real number \( \text{Re}(w)\). Therefore, \( h(w) \) is bounded in the complex plane. By Liouville’s theorem, any entire and bounded function must be constant. Hence, \( h(w) = \lambda \in \mathbb{C} \), and we conclude that
\begin{equation}
    f(w) - f(0) = \lambda \, \sin(\pi w) \ .
\end{equation}
However, the only way this expression can be compatible with the growth bound \( |f(w)| < e^{|w|} \) is if \( \lambda = 0 \), since \( \sin(\pi w) \) grows exponentially along the imaginary axis. Therefore, we conclude that
\begin{equation}
    f(w) = f(0) \ ,
\end{equation}
which means that \( f \) is constant.

Note that this statement holds for any period, by a simple redefinition of the function \( f \) and the variable \( x \). Furthermore, the assumption \( |f(w)| < e^{|w|} \) is crucial: without it, counterexamples can be easily constructed. For instance, the function
\begin{equation}
    f(w) = \sin (2 \pi w) \ ,
\end{equation}
satisfies all the conditions except the last one, as it obeys \( |f(w)| \le  e^{2\pi |w|} \) instead of \( |f(w)| < e^{|w|} \).

\section{Comments and details on the thermal Lorentzian inversion formula}\label{app:LIF}

Dispersion relations and the Lorentzian inversion formula (LIF) are based on the same set of assumptions and, in fact, encode equivalent information~\cite{Alday:2020eua}. In this Appendix, we review how to recover the result of the dispersion relation from the Lorentzian inversion formula in two cases: the free scalar theory and the $\mathrm{O}(N)$ model in the $\varepsilon$-expansion.

\subsection{Free scalar theory}

As shown in~\cite{Iliesiu:2018fao}, the contribution of an operator $\mathcal{O}$ to the OPE coefficient of the $n=0$ subsector of double-twist operators $[\phi \phi]_{0,J}$ is given by
\begin{equation}\label{eq:LIFgenericOp}
    a^{\mathcal{O}}_{[\phi \phi]_{0,J}} = a_{\mathcal{O}} \left[1 + (-1)^J\right] \frac{K_J}{K_{J_{\mathcal{O}}}}\, S_{h_{\mathcal{O}} - \Delta_\phi, \Delta_\phi}(\hb) \ ,
\end{equation}
where
\[
    h_{\mathcal{O}} = \frac{\Delta_{\mathcal{O}} - J_{\mathcal{O}}}{2} \ , \qquad
    \hb = h + J \ , \qquad
    \nu = \frac{d}{2} - 1 \ ,
\]
and
\begin{equation}
    K_j = \frac{\Gamma(j+1)\Gamma(\nu)}{4\pi\, \Gamma(j + \nu)} \ , \qquad
    S_{\ell,\Delta}(\hb) = \frac{1}{\Gamma(-\ell)} \frac{\Gamma(\hb - \Delta - \ell)}{\Gamma(\hb - \Delta + 1)} 
    - \frac{B_{1/z_{\text{max}}}(\hb - \Delta - \ell, 1 + \ell)}{\Gamma(-\ell)\Gamma(1 + \ell)} \ .
\end{equation}

We now specialize to the free scalar theory, where $\Delta_\phi = d/2 - 1$, and consider the contribution from the identity operator $\mathcal{O} = \mathds{1}$. In this case, the inversion yields the exact OPE coefficients
\begin{equation}
    a_{[\phi \phi]_{J}} = 1 + (-1)^J \ .
\end{equation}
As anticipated, expanding the correlator in thermal blocks (see Eq.~\eqref{eq:gdrfree}) gives the same result as obtained from the LIF.

\subsection{\texorpdfstring{$\mathrm{O}(N)$}{\texttwoinferior} model in the \texorpdfstring{$\veps$}{\texttwoinferior}-expansion}

We have already computed the inversion of the identity operator. Its contribution is universal in all dimensions for a free scalar, since the spacetime dependence of the residues at $\Delta = 2\Delta_\phi + J$ vanishes when $\Delta_\phi = (d-2)/2$. This corresponds to the free scalar theory in $d$ dimensions. This observation is consistent with the exact result~\cite{Iliesiu:2018fao}:
\begin{equation}
    a_{[\phi\phi]_{0,J}} = 2\, \zeta(d - 2 + J) \ .
\end{equation}
In particular, the large-spin limit always yields
\begin{equation}
    a_{[\phi\phi]_{0,J}} \to 2 \quad \text{as } J \to \infty \ ,
\end{equation}
independently of the spacetime dimension.

The next operator to consider is $\phi^2$, since it acquires an anomalous dimension at order $\varepsilon$. We can again use Eq.~\eqref{eq:LIFgenericOp} to determine its contribution. The results are
\begin{equation}
    a_{[\phi \phi]_{0,J}}^{(\phi^2)} = - \left[1 + (-1)^J\right] \frac{a_{\phi^2}^{(0)}\, \gamma_{\phi^2}}{2J} \ , \qquad
    a_{[\phi \phi]_{1,J}}^{(\phi^2)} = \left[1 + (-1)^J\right] \frac{a_{\phi^2}^{(0)}\, \gamma_{\phi^2}}{2(J+2)} \ ,
\end{equation}
with
\begin{equation}
    a_{[\phi \phi]_{n,J}}^{(\phi^2)} = 0\,, \qquad \text{for } n \geq 2 \ .
\end{equation}
In deriving the expressions above, we used that $\gamma_{\phi^2} \propto \varepsilon$, so only the leading-order one-point function $a_{\phi^2}^{(0)}$ contributes at this order. Higher-order corrections to the one-point function would contribute only at higher orders in $\varepsilon$.
These results match exactly those obtained from the dispersion relation (see Table~\ref{tab:drepsilon}).

\section{Examples of numerical KMS compensation}
\label{app:KMScompensator}

We comment here on the strategy described in Section \ref{subsubsec:ANumericalApproach}, to reconstruct the out-of-OPE contribution, starting from the OPE and the dispersion relation.

\subsection{Free scalar theory}

In the free theory we know that the only operator contained in the arcs is the identity operator: its one-point function is simply the normalization of the scalar field $\phi$ and we can decide to unit-normalize its zero-temperature two-point functions so that $a_{\mathds{1}} = 1$.
Even after adding this contribution by hand, the function is not KMS invariant.
According to the discussion in Section \ref{subsubsec:ANumericalApproach}, the Ansatz for the correlator is
\begin{equation}\label{eq:freeansaz}
    g (z,\zb) = g_{\text{dr}}(z,\zb)+\frac{1}{z \zb}+ \sum_{n = 1}^\infty c_n g_\text{comp.}^{(n)}(z, \zb) \ ,  
\end{equation}
where the first term is simply the outcome of the dispersion relation, the second term the identity operator contribution, which in this case is the only term in the arcs, and the last term should be the contribution to the thermal OPE coefficients coming from the region out of the OPE regime.
The Ansatz for this contribution is given by
\begin{equation}
    a_{\mathcal O} = a_{\mathcal O}^{(\text{dr})} \left(1+\sum_{n = 0}^\infty \frac{c_n}{n^J}\right)\ .
\end{equation}
In the free scalar case, the only operators in the OPE are the higher-spin currents $[\phi \phi]_{0,J}$: the compensation at any order in $n$, i.e., in $1/n^J$, can be computed directly by summing over all the operators.
As a result we have 
\begin{equation}
   g_\text{comp.}^{(n)} =  \frac{n^2  \left(n^2+z \zb\right)}{\left(n^2-z^2\right)
   \left(n^2-\zb\right)} \ .
\end{equation}
In order to impose KMS, we need to truncate the Ansatz \ref{eq:freeansaz} to some $n_\text{max}$ and numerically find the coefficients $c_n$ by expanding the correlator both around $z, \zb \sim 0$ and $z,\zb \sim 1$.
The value of $n_\text{max}$ can be then modified to verify the stability of the solutions explicitly.
In this specific case, by considering $n_\text{max}$ large enough it is possible to notice that all the coefficients are rational (as expected from the analytic solution) and the exact value can be determined with the function \texttt{Rationalize} in \textsc{Mathematica}.

\begin{table}[t]
    \caption{Results for the first coefficients $\alpha_k$ for the corrections in spin for the case $\Delta_\phi = 1$, obtained from the KMS condition.}
    \centering
    \begin{tabular}{ccccccl}
        \hline
        $n_\text{max}$ & 2 & 3 & 4 & 5 & 10 & Exact results  \\ \hline
        $c_2$ & 0.614 & 0.486 & 0.501 & 0.500 & 0.500 & $1/2 = 0.5$ \\
        $c_3$ & - & 0.384 & 0.172 & 0.232 & 0.222 & $2/9 = 0.222 \ldots$ \\
        $c_4$ & - & - & 0.323 & 0.017 & 0.125 & $1/8 = 0.125$ \\ \hline
    \end{tabular}
    \label{tab:KMS_Free}
\end{table}

\begin{figure}[t]
\centering
\hspace{-3em} \includegraphics[width=110mm]{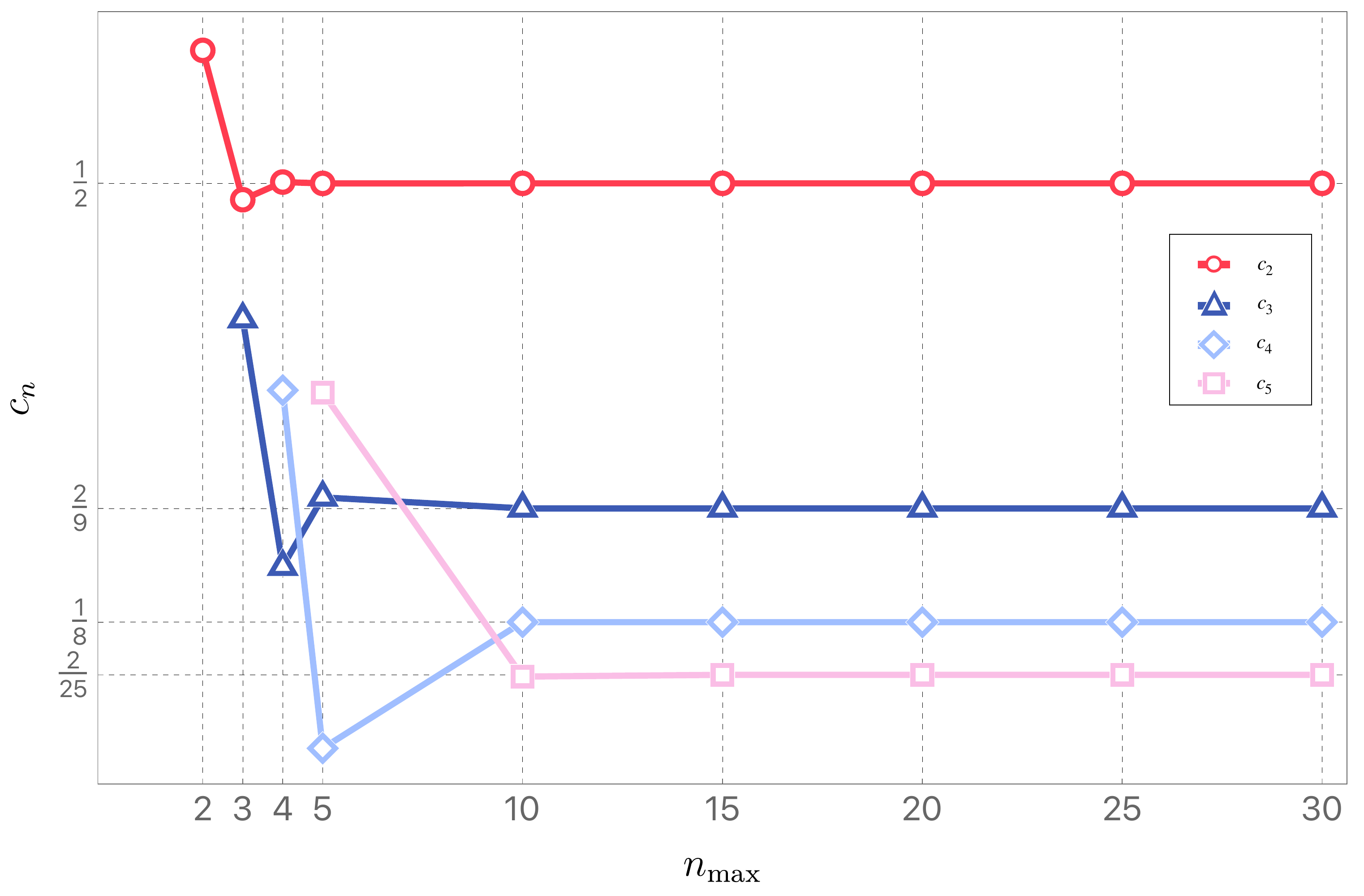}
    \caption{Coefficients of the first five corrections of Eq. \eqref{eq:freeansaz} as a function of $n_\text{max}$. Notice that when $n_\text{max}$ increases, the corrections stabilize rather quickly to the values that correctly re-establish KMS invariance of the corrected correlator.}
    \label{fig:cnvsmmax} 
\end{figure}

For the sake of clarity we show in Table \ref{tab:KMS_Free} and Fig.~\ref{fig:cnvsmmax} the evolution of the solutions for the first coefficients imposing KMS while varying the number of corrections $n_\text{max}$.
We observe a very fast convergence, in particular for the first corrections.
The reader can already notice a clear pattern, which is confirmed also from higher coefficients.
In particular,
\begin{equation}
    c_n = \frac{2}{n^2} \ .
\end{equation} 
In fact this fixes the OPE coefficients to be $a_{[\phi \phi]_{0,J}} = 2 \zeta(2+J)$, perfectly matching the exact results.

\subsection{\texorpdfstring{$\mathrm O(N)$}{\texttwoinferior} model in \texorpdfstring{$\varepsilon$}{\texttwoinferior}-expansion}\label{appendix:ONModelNumerical}

We now consider the $\mathrm{O}(N)$ model in the context of the $\veps$-expansion.
It is possible to subtract the free theory from the total correlation function:
\begin{equation}
    \hat g(z,\zb) = g(z,\zb)-g_\text{free}(z,\zb) \ ,
\end{equation}
where $g_\text{free}$ refers to the free two-point function in $d = 4-\varepsilon$.
Notice that since $g$ and $g_\text{free}$ are KMS invariant, also $\hat g$ is KMS invariant.
By using this definition, we can easily impose KMS only on the portion of the correlation function proportional to $\gamma_{\phi^2}$ since, at first order in $\veps$, this is what we expect for $\hat g$.
The identity operator and $\phi^2$ are operators in the arcs.
In principle, since $[\phi \phi]_{1,0}$ is also a scalar, it should be considered in the arcs, too: we will show that this is not necessary.
Said differently, its contribution to the arc is zero.
As at zero order, the Ansatz is the following:
\begin{equation}
    \hat a_{[\phi \phi]_{0,J}} = -\frac{1}{J}a_{\phi^2}^{(0)} \gamma_{\phi^2} \left(1+\sum_{n = 0}^\infty \frac{\hat c_n}{n^J}\right)\ , \hspace{0.5 cm} \hat a_{[\phi \phi]_{1,J}} = \frac{1}{(J+2)}a_{\phi^2}^{(0)} \gamma_{\phi^2} \left(1+\sum_{n = 0}^\infty \frac{\hat d_n}{n^J}\right)\ .
\end{equation}
In the formula above, the $\hat a_\Om$ refer to the OPE coefficients of $\hat g$.
The coefficients $\hat c_n$ and $\hat d_n$ can be computed numerically by imposing KMS.
Also in this case the corrections can be re-summed and the contribution of the compensator is
\begin{align}
    \hat g_\text{comp.}^{(n)}(z,\zb)
    &=
   \gamma_{\phi^2}\, a_{\phi^2}^{(0)} \hat c_n \frac{z \log
    \left(1-\frac{z^2}{n^2}\right)-\zb \log
    \left(1-\frac{\zb^2}{n^2}\right)}{2 (z-\zb)} \notag \\
    &\phantom{=\ }
    +  \gamma_{\phi^2} \, a_{\phi^2}^{(0)}  \hat d_n \, n^2 \frac{z \log
    \left(1-\frac{\zb^2}{n^2}\right)-\zb \log \left(1-\frac{z^2}{n^2}\right)+z
    \zb (\zb-z)}{2 (z-\zb)}\ .
\end{align}
The operators $\phi^2$ and $[\phi\phi]_{1,0}$ are left as part of the arcs and their contributions are added by hand with open coefficients $a_{\phi^2}$ and $a_{[\phi\phi]_{1,0}}$.
The contribution of $a_{\phi^2}$ cannot be constrained by KMS since this operator contributes as a constant to the correlator.
We show in Table \ref{tab:TableCompensatorON} the numerical values by using $n_\text{max} = 30$, from which we can guess the following pattern:
\begin{equation}
    \hat c_n = 1 \ , \hspace{1 cm} \hat d_n = \frac{1}{n^2} \ .
\end{equation}
Meanwhile, we compute the value of $a_{[\phi\phi]_{1,0}} = 0.3073 \dots $.
Those results indeed reproduce the results that can be obtained analytically and are summarized in Table \ref{tab:drepsilon}.

\begin{table}[t]
    \centering
    \caption{Values for the coefficients of the $\mathrm{O}(N)$ model in the $\veps$-expansion using the numerical KMS compensation.
    We used $n_\text{max} = 30$.}
    \renewcommand{\arraystretch}{1.5}
    \begin{tabular}{cccc}
        \hline
         $\hat c_2$ & $\hat c_3$ & $\hat c_4$ & $\ldots$ \\ \hline
         $1.0000$ & $1.0000$ & $1.0000$ & $\ldots$  \\
        \hline \hline 
        $\hat d_2$ & $\hat d_3$ & $\hat d_4$ & $\ldots$ \\ \hline 
        $0.2500$ & $0.1111$ & $0.06250$ & $\ldots$\\ \hline 
    \end{tabular}
    \label{tab:TableCompensatorON}
\end{table}

\section{\texorpdfstring{$\mathrm O(N)$}{\texttwoinferior} model in \texorpdfstring{$\varepsilon$}{\texttwoinferior}-expansion: diagrammatic computation}
\label{subsec:ComparisonToDiagrammaticCalculation}

In this appendix, we compute the correlator at next-to-leading order in the context of the $\veps$-expansion in the $\mathrm{O}(N)$ model, in order to compare it to the results of Section \ref{sec:pertON}.

\paragraph{The Feynman rules.} We first give the Feynman rules for the theory.
The action for the $\mathrm{O}(N)$ model is given by \eqref{eq:GLON}, which we consider at the Wilson--Fisher fixed point
\begin{equation}
    \frac{\lambda_\star}{(4\pi)^2} = \frac{3 \veps}{N+8} + O(\veps^2)\ .
\end{equation}
When the time dimension is compactified, the scalar propagator reads
\begin{equation}
    \ScalarPropagator\ = \delta^{ij} \frac{\Gamma(d/2-1)}{4 \pi^{d/2}} \sum_{m=-\infty}^\infty \frac{1}{((\tau - m)^2 + x^2)^{d/2-1}}\ ,
\end{equation}
while the insertion of a vertex brings the following integral and index contractions:
\begin{equation}
   \begin{tikzpicture}[scale=0.7, anchor=base, baseline]
        \node (a) at (0.25,0.60) {\footnotesize{$i$}};
        \node (b) at (0.25,-0.75) {\footnotesize{$j$}};
        \node (c) at (1.75,0.60) {\footnotesize{$k$}};
        \node (d) at (1.75,-0.75) {\footnotesize{$l$}};
        
        \draw[very thick] (0.5,0.5) -- (1.5,-0.5);
        \draw[very thick] (1.5,0.5) -- (0.5,-0.5);
        
        \draw[fill=white] (0.5,0.5) circle (2.5pt);
        \draw[fill=white] (1.5,0.5) circle (2.5pt);
        \draw[fill=white] (1.5,-0.5) circle (2.5pt);
        \draw[fill=white] (0.5,-0.5) circle (2.5pt);
        
        \filldraw[WCOrange] (1,0) circle (2.5pt);
    \end{tikzpicture} = - \frac{\lambda_{\star}}{4!} \left(\delta^{ij}\delta^{ij} + \delta^{ik}\delta^{jl} + \delta^{il}\delta^{jk}\right)
    \int \text{d}^{d-1} x \int_0^1 \text{d}\tau\ .
\end{equation}

\paragraph{The correlator.} We now calculate the correlator $\vev{\phi (0) \phi(\tau,x)}_\beta$.
In the bootstrap calculation of Section \ref{sec:pertON}, the operator is normalized such that
\begin{equation}
    \vev{\phi(0) \phi(x)} = \frac{1}{\tau^{2\Delta_\phi}}\ .
    \label{eq:Normalization}
\end{equation}
We therefore express the interacting part of the correlator as
\begin{equation}
    g(\tau,x) - g_\text{free}(\tau,x)
    =
    \frac{1}{n_\phi} Z_\phi^{-2} \left( \; \DiagramB + O(\lambda^2) \; \right)\ ,
\end{equation}
with $n_\phi$ the normalization constant ensuring \eqref{eq:Normalization}, while the diagram consists of non-normalized fields.
$Z_\phi$ is the usual wave-function renormalization factor, however at this order in $\lambda$ the operator does not renormalize:
\begin{equation}
    Z_\phi^{-2} = 1 + O(\lambda^2)\ .
\end{equation}
For our calculation, we only need the leading term of the normalization constant, which is simply
\begin{equation}
    \frac{1}{n_\phi} = 4 \pi^2 + O(\veps)\ .
\end{equation}
The diagram itself is explicitly given by the expression
\begin{align}
    \DiagramB
    &=
    - \frac{\lambda_{\star}}{6} (N+2) \left(\frac{\Gamma(d/2-1)}{4 \pi^{d/2}}\right)^3 \left( \sum_{m=-\infty}^\infty \frac{1}{m^{d-2}} \right) \notag \\
    &\times
    \sum_{n=-\infty}^\infty \sum_{n'=-\infty}^\infty
    \int \text{d}^{d-1} x' \int_0^1 \text{d}\tau' \frac{1}{((\tau-\tau'+n)^2+(x-x')^2)((\tau'+n')^2+x'^2))^{d/2-1}}\,.
\end{align}
Using the regularized sum
\begin{equation}
    \sum_{m=-\infty}^\infty \frac{1}{m^{\alpha}} = [1 + (-1)^{-\alpha}] \zeta_\alpha\ ,
\end{equation}
and shifting the $\tau$-integral as well as one of the sums, we obtain
\begin{align}
    \DiagramB
    &=
    - \lambda_{\star} (N+2) \zeta_{d-2} \frac{\Gamma^3(d/2-1)}{192 \pi^{3d/2}} \notag \\
    &
    \times \sum_{n=-\infty}^\infty
    \int \frac{\text{d}^d x'}{((\tau-\tau'+n)^2+(x-x')^2)(\tau'^2+x'^2))^{d/2-1}}\ .
\end{align}
Note that the integral is now simply equivalent to a two-point Feynman integral at zero temperature, which is well-known and given by \cite{tHooft:1978jhc}
\begin{equation}
    \int \frac{\text{d}^d x'}{\left[(x-x')^2 x'^2\right]^\alpha}
    =
    \pi^{d/2}\frac{ \Gamma(2\alpha-d/2)\Gamma^2(d/2-\alpha)}{\Gamma^2(\alpha) \Gamma(d-2\alpha)} \frac{1}{x^{2(2\alpha-d/2)}}\ .
\end{equation}
We now have the following exact result for the diagram in an arbitrary number of dimensions:
\begin{align}
    \DiagramB
    &=
    - \lambda_{\star} (N+2) \zeta_{d-2} \frac{\Gamma(d/2-1) \Gamma(d/2-2)}{192 \pi^{d}} \sum_{n=-\infty}^\infty \frac{1}{((\tau+n)^2+x^2)^{d/2-2}}\ .
\end{align}
Expanding this expression up to first order in $\veps$, we find
\begin{equation} \label{eq: cci}
    \DiagramB
    =
    \frac{\lambda_{\star} (N+2)}{576 \pi^2}
    \sum_{n=-\infty}^\infty
    \left[
    \frac{1}{\veps} + 12 \log A - \log 2 + \frac{1}{2} \log [(n+z)(n+\zb)] + O(\veps)
    \right]\ ,
\end{equation}
where $A=1.282427 \ldots$ is the Glaisher--Kinkelin constant, related to $\zeta'_2$ via
\begin{equation}
    \zeta'_2
    =
    \zeta_2 (\gamma_E + \log (2\pi) - 12 \log A)\ .
\end{equation}
Notice that in Eq. \eqref{eq: cci} we switched to the coordinates $z, \zb$ defined in \eqref{eq:zzbVariables}. Moreover, the divergent term corresponds to the renormalization of the thermal mass.\footnote{There are some subtleties related to the choice of scheme for the $\zeta$-sums. In particular the divergence appears to drop when using our regularization methods, since the sum over images of a constant evaluates to zero. The proper way to do it is to not evaluate the sum explicitly and perform the renormalization.}
Using the sum \eqref{eq:RegularizedSum}, we find that the interacting part of the correlator is given by\footnote{We thank Daniele Artico for pointing out a typo in this expression.} 
\begin{equation}
    g_\text{int} (z,\zb)
    =
    - \veps \frac{\pi^2}{6} \frac{N+2}{N+8}
    \log \left[\frac{\csc (\pi z) \csc (\pi \zb)}{4}\right]\,.
    \label{eq:perturbativeepislonres}
\end{equation}
The result in equation \eqref{eq:perturbativeepislonres} matches the bootstrap result provided in the main text.

It is worth noticing that the correlator consists of transcendental terms only with homogeneous weight $1$.
To conclude, observe that the contributions explicitly computed here are not the full order $ O(\veps)$: the theory lives in $d = 4-\veps$ and for this reason we have a contribution up to order $\veps^1$ coming from the free propagator:
\begin{align}
    a^\text{free}_{[\phi \phi]_{0,J>0}} &= 2 \zeta_{2+J} -2 \varepsilon \zeta'_{2+J}+O(\varepsilon^2) \,, \\
    a^\text{free}_{[\phi \phi]_{1,0}} &= 0\,.
\end{align}
which should be added to the interacting part.

\section{Details on Monte Carlo simulations}
\label{app:MC}

In this Appendix we briefly comment on the standard Monte Carlo \textsc{Metropolis} algorithm employed for the simulations in Section \ref{subsec:CorrelatorsInThe3dIsingModel}.
The code is available on \href{https://github.com/alemiscio/Monte_Carlo_Ising_3d.git}{GitHub}.

\paragraph{The \textsc{Metropolis} algorithm for the $3d$ Ising model.}
The $3d$ Ising model consists of a cubic lattice where each site $i$ hosts a spin variable $s_i = \pm 1$. The Hamiltonian of the system is given by
\begin{equation}
    H = -J \sum_{\langle i,j \rangle} s_i s_j+ h \sum_{i} s_i \ ,
\end{equation}
where the first sum runs over all the pairs of nearest-neighbor sites $\langle i,j \rangle$, and $J$, $h$ are the coupling constant (which we set to the critical value).

The \textsc{Metropolis} algorithm is employed in order to reproduce how a given sample spin configuration, randomly generated, thermalizes according to the Boltzmann distribution, which assigns to a given spin configuration $\{s_i\}$ the probability
\begin{equation}
    P(\{s_i\}) \propto \exp \big[-\beta H(\lbrace s_i \rbrace)\big] \ .
\end{equation}
The starting point for the algorithm is the choice of a lattice size.
In order to simulate the physics on the thermal manifold $S^{1}_{\beta}\times \mathbb{R}^{2}$, the number of sites on one of the axis must be much smaller than the one on the other two axis.
On top of that, boundary conditions must be taken into account.
After having generated an initial random spin distribution, the algorithm proceeds as follows:
\begin{enumerate}
    \item it randomly chooses a lattice site $i$;
    \item it computes the energy difference $\Delta E$ associated with flipping the spin: $s_i \to -s_i$; 
    \item it accepts the spin flip with probability
    \begin{equation}
        p = 
        \begin{cases}
            1 & \text{if } \Delta E \leq 0 \ , \\
            e^{-\beta \Delta E} & \text{if } \Delta E > 0 \ .
        \end{cases}
    \end{equation}
    \item it repeats the steps above for many iterations until the system reaches thermal equilibrium.
\end{enumerate}
Once the system is sufficiently thermalized, physical observables can be measured.

\paragraph{The code.}
On \href{https://github.com/alemiscio/Monte_Carlo_Ising_3d.git}{GitHub} one can find:
\begin{itemize}
    \item[$\star$] The file \texttt{TD}$\_$\texttt{Ising}$\_$\texttt{Monte}$\_$\texttt{Carlo.py}, which contains a number of functions used for the \textsc{Metropolis} algorithm as well as for computing magnetization and two-point functions of spin and energy operators. All the functions contained in this file are listed with a short explanation in the file \texttt{ising3d}$\_$\texttt{function}$\_$\texttt{description.txt};
     \item[$\star$]The file \texttt{Main}$\_$\texttt{3d.py}, which contains an example of how to use the functions of the previous file to simulate the $3d$ Ising model and save the spin configurations after thermalization;
    \item[$\star$]The file \texttt{Data}$\_$\texttt{Analysis}$\_$\texttt{3d.py}, which shows how to compute physical observables from the previously saved spin configurations.
\end{itemize}

\bibliography{./auxi/biblio.bib}
\bibliographystyle{./auxi/JHEP}

\end{document}